\documentclass[11pt]{article}
\usepackage{jheppub} 
\usepackage{slashed,bbm,units}
\usepackage{array,multirow,booktabs}
\usepackage[normalem]{ulem}
\usepackage[dvipsnames]{xcolor}
\usepackage{subcaption}
\usepackage{graphicx}
\usepackage{caption}
\graphicspath{{./}{./Figures/}}

\DeclareMathOperator{\Br}{Br}
\newcommand{\de}{\partial}
\newcommand{\g}{\gamma}
\newcommand{\cO}{\mathcal{O}}
\newcommand{\Lag}{\mathcal{L}}

\newcommand{\ie}{\emph{i.e.}}
\newcommand{\eg}{\emph{e.g.}}

\newcommand{\cgo}{C_{G_1}}
\newcommand{\cgt}{C_{G_2}}
\newcommand{\cgamma}{C_{\g\g}}
\newcommand{\cgoh}{C_{ah}}

\newcommand{\Ocgo}{\cO_{G_1}}
\newcommand{\Ocgt}{\cO_{G_2}}
\newcommand{\Ocgamma}{\cO_{\g\g}}
\newcommand{\Ocah}{\cO_{ah}}

\newcommand{\La}{\Lambda_a}

\newcommand{\aNLO}{{\sc\small MadGraph5\_aMC@NLO}}

\newcommand{\sA}{\sigma_{11}}
\newcommand{\sB}{\sigma_{22}}
\newcommand{\sC}{\sigma_{12}}

\newcommand{\rG}{r_{G}}

\definecolor{darkgreen}{rgb}{0.0,0.5,0.0}
\newcommand{\orange}{{\color{orange}orange}}
\newcommand{\blue}{{\color{blue}blue}}
\newcommand{\red}{{\color{red}red}}
\newcommand{\green}{{\color{darkgreen}green}}
\newcommand{\cyan}{{\color{Aquamarine}cyan}}
\newcommand{\purple}{{\color{Plum}purple}}

\title{ALP pair production at the LHC}

\author[a,b]{Ilaria Brivio,}
\author[a,b]{Simone Meoni,}
\author[b]{Davide Pagani}

\emailAdd{ilaria.brivio@unibo.it}
\emailAdd{simone.meoni3@unibo.it}
\emailAdd{davide.pagani@bo.infn.it}

\affiliation[a]{Dipartimento di Fisica e Astronomia, Università di Bologna, Via Irnerio 46, Bologna, I-40126 Italy}
\affiliation[b]{INFN, Sezione di Bologna, Via Irnerio 46, Bologna, I-40126 Italy}

\abstract{
We study axion-like particle (ALP) pair production at the LHC, investigating its sensitivity to the simultaneous presence of dimension-5 and dimension-6 ALP interactions. 
Focusing on the signature with four isolated photons, we analyze for the first time the non-resonant process $gg\to aa$, finding that it can constrain significantly ALP interactions, already at an integrated luminosity of $\unit[300]{fb^{-1}}$.
Particular attention is paid to the multidimensional nature of the ALP parameter space. 
To this end, we present a re-interpretation of a search for the Higgs-resonant process $gg\to h\to aa$ by the ATLAS Collaboration, recasting their results within a three-parameter space. 
We find that the multi-dimensional constraints resulting from both non-resonant and Higgs-resonant ALP pair production exhibit non-trivial features,
that are expected to extend to other searches in which the ALPs decay into Standard Model particles.

}

\date{}

\subheader{\hfill\rm COMETA-2026-27}

\begin{document}
\maketitle

\section{Introduction}
\label{sec:Intro}

Axion-Like Particles (ALPs) are arguably among the best-motivated classes of hypothetical particles beyond the Standard Model (BSM). They are defined as pseudo-Goldstone bosons (pGB) associated with the spontaneous breaking of global symmetries at high energy scales, a mechanism that appears in a wide variety of Standard Model (SM) extensions, see Refs.~\cite{DiLuzio:2020wdo,Choi:2020rgn,DiLuzio:2023lmd,Biekotter:2025fll,Arza:2026rsl} for recent reviews. The prototypical example is the QCD axion, predicted
in solutions to the strong CP problem~\cite{Peccei:1977hh,Peccei:1977ur,Weinberg:1977ma,Wilczek:1977pj}.  Further examples include the relaxion~\cite{Graham:2015cka},  the Majoron~\cite{Gelmini:1980re,Chikashige:1980qk,Chikashige:1980ui,Schechter:1981cv}, the axiflavon~\cite{Calibbi:2016hwq}, the flaxion~\cite{Ema:2016ops} and, more generally, flavons~\cite{Davidson:1981zd,Wilczek:1982rv}. pGBs also emerge in composite Higgs models~\cite{Katz:2005au,Contino:2011np} and in string theory~\cite{Witten:1984dg,Svrcek:2006yi}.

Their ubiquity in BSM theories is only part of the reason ALPs have attracted considerable interest in recent years: owing to their pGB nature, ALPs are expected to be much lighter than the other new states present in the UV theory. 
This makes them particularly attractive targets for direct searches and plausible candidates for the first BSM discovery. At the same time, the natural separation between the ALP mass scale and the scale of the UV sector ensures that their interactions can be described by an Effective Field Theory (EFT) containing only ALPs and SM fields as dynamical degrees of freedom~\cite{Georgi:1986df}.
This EFT has the great advantage of capturing the low-energy dynamics of \emph{any} ALP, absorbing all model-dependence and theoretical complexity into a small set of Wilson coefficients.

Depending on the ALP mass and on the process of interest, the ALP EFT contains a different set of SM degrees of freedom. In the context of LHC searches, the relevant EFT is formulated in terms of fields in $SU(3)_c\times SU(2)_L\times U(1)_Y$ representations~\cite{Georgi:1986df}.\footnote{The standard ALP EFT shares the field content of the SM EFT~\cite{Buchmuller:1985jz,Grzadkowski:2010es}, adopting a linear representation of the electroweak symmetry breaking. Adopting a  non-linear representation leads to an alternative ALP EFT~\cite{Brivio:2017ije} akin to the Higgs EFT (HEFT)~\cite{Buchalla:2013rka,Brivio:2016fzo}, with distinct phenomenological implications.}  
At the lower scales relevant for flavor physics, the ALP EFT is more appropriately constructed by extending with ALP interactions chiral perturbation theory or the weak effective theory~\cite{Buchalla:1995vs,Jenkins:2017jig}, which only contains light fermions and massless gauge bosons. At the even lower scales relevant for astrophysical and cosmological searches, the EFT can be reduced down to ALP couplings to photons, gluons and light leptons, in a theory which is only manifestly invariant under $SU(3)_c\times U(1)_{\rm em}$. The matching among these effective descriptions, together with the renormalization-group evolution of their Wilson coefficients, has been studied extensively in recent years~\cite{Bauer:2020jbp,Chala:2020wvs,Bonilla:2021ufe,Bauer:2021wjo,Bauer:2021mvw,DiLuzio:2023cuk,DasBakshi:2023lca,Bresciani:2024shu}.

A distinctive feature of ALP phenomenology is the enormous size of the available parameter space. Since both the ALP mass $m_a$ and characteristic scale $\La$ are essentially unconstrained \emph{a priori}, ALP searches encompass an exceptionally diverse set of experimental probes, including cosmological and astrophysical observations, dedicated laboratory experiments, flavor measurements, and high-energy collider searches. A regularly updated collection of these results can be found \eg\ at the public repository~\cite{AxionLimits}. 
Nearly all existing ALP searches focus on processes involving the production and, where relevant, the decay of a single ALP. This strategy is very sensible from the EFT perspective: since ALP interactions with SM fields first appear at dimension five, each additional ALP production or decay vertex increases the EFT order of the process, and therefore introduces an additional suppression. As a result, processes involving two or more ALPs are generally expected to occur at significantly lower rates than single-ALP ones.
For light ALPs, with masses below the GeV scale, this expectation is further reinforced by the stringent constraints already placed on their couplings by cosmological, astrophysical, and laboratory observations. These bounds require $\La\gtrsim\unit[10^4]{TeV}$, and in some cases considerably larger values~\cite{AxionLimits}. Consequently, the EFT suppression is expected to be so severe that processes with two ALPs in the final state are essentially unobservable in this region of parameter space. The situation changes, however, for ALP masses above the GeV, which is the regime relevant for collider experiments. Here, existing bounds are substantially weaker, opening the possibility that di-ALP production could provide a meaningful and potentially competitive probe of ALP interactions.

To our knowledge, the only di-ALP process that has been explored so far in the literature is the exotic Higgs decay $h\to aa$, mediated by the $haa$ vertex appearing at dimension-6 in the ALP EFT. This channel was first identified in Ref.~\cite{Bauer:2017ris}, and has subsequently been targeted by dedicated searches by the ATLAS and CMS Collaborations in 
$\g\g jj$~\cite{ATLAS:2018jnf}, 
$bb\mu\mu$~\cite{ATLAS:2021hbr}, 
$4\ell$~\cite{ATLAS:2021ldb,CMS:2021pcy}, $4\gamma$~\cite{CMS:2022fyt,ATLAS:2023ian,CMS:2026knm}, 
$4b$~\cite{CMS:2024zfv,ATLAS:2025rfm},
$bb\tau\tau$~\cite{ATLAS:2024vpj},
$\g\g\tau\tau$~\cite{ATLAS:2024nnm},
and
$4\tau$~\cite{ATLAS:2025qyn}
final states.

These analyses already set strong exclusion limits on the branching ratio $\Br(h\to aa\to X)$, reaching sensitivities of order $10^{-6} - 10^{-5}$ in the $4\ell$ and $4\gamma$ final states. These two channels provide the highest sensitivity owing to their remarkably clean experimental signatures, which allow for a significant background reduction that compensates for the small signal rates.
Since $\Br(h\to aa\to X)$ depends simultaneously on the $haa$ coupling in the Higgs decay and on the ALP couplings controlling its subsequent decays into SM particles, the translation of those bounds into constraints on the Wilson coefficients requires additional assumptions on these parameters.
Depending on the strategy adopted, existing $h\to aa$ searches have been interpreted as probing new-physics scales as large as $\La\sim \unit[10^7]{TeV}$~\cite{Bauer:2017ris,ATLAS:2023ian}, suggesting that the current sensitivity can already compete with single-ALP searches.

Motivated by these results, the goal of this work is twofold: 
on one hand, we perform the first study of direct non-resonant production of an ALP pair at the LHC, exploring its sensitivity to ALP interactions. On the other, we characterize such sensitivity in a multidimensional way, accounting for the presence of all the contributing Wilson coefficients simultaneously. 

For this first exploration, we focus on the process $gg\to aa\to 4\gamma$, without the intermediate Higgs resonance, which is expected to yield the largest production rate and a very clean experimental signature. As detailed in the next sections, this process depends on at least two dimension-5 Wilson coefficients (parameterizing  the $agg$ and $a\g\g$ interactions) and a dimension-6 one (giving a contact $aagg$ vertex). 
Rather than varying one parameter at a time, we explore the measurement sensitivity in the full three-dimensional space, highlighting characteristic features of the allowed regions and examining in detail how they arise from the interplay between different physical effects. We extend this approach to the $h\to aa\to 4\gamma$ process as well, by reinterpreting the bounds derived in Ref.~\cite{ATLAS:2023ian} within an analogous three-dimensional parameter space that includes both dimension-5 and 6 interactions.

While a complete global analysis of the ALP EFT lies beyond the scope of the present work, our study is intended as a step in that direction and, in particular, towards the inclusion of dimension-6 parameters in that context, providing also a first exploration of the interplay between single and double-ALP production processes.
Partial global analyses of ALP searches at colliders have been performed in recent years, focusing \eg\ on vector boson scattering~\cite{Bonilla:2022pxu} and multi-boson production processes~\cite{Esser:2025nmd}, top, di-jet and electroweak observables~\cite{Bruggisser:2023npd} and reinterpreting SMEFT global fits~\cite{Biekotter:2023mpd} through loop contributions of ALP couplings to dimension-6 SMEFT operators~\cite{Galda:2023qjx}. A comprehensive global analysis of ALP constraints is arguably desirable, particularly for ALP masses above a few GeV, as several couplings (including to electroweak bosons and heavy quarks) can in principle contribute simultaneously to ALP signals within this range. Nevertheless, this task is quite challenging due to the potentially complex dependence of the signal predictions on the ALP parameters. One of the goals of this work is to highlight examples of this feature, and to illustrate how it leads to non trivial shapes in the sampling of the parameter space. As most of these features are fundamentally due to the fact that the ALP is produced and decayed resonantly and that it has a finite lifetime, they are expected to extend to other ALP searches. 

Finally, this work aims at providing one of the first phenomenological studies of dimension-6 ALP interactions beyond the $haa$ operator, for ALPs with masses above a few GeV. The structure of the ALP EFT beyond dimension-5 has been explored recently from the theoretical point of view, for instance by examining how the approximate shift-invariance of the ALP Lagrangian behaves at higher orders~\cite{Bonnefoy:2022rik}, constructing a Hilbert series to determine the number of independent parameters~\cite{Grojean:2023tsd} and exploring the structure of the EFT with on-shell amplitude techniques~\cite{Bertuzzo:2023slg,Bresciani:2025ojh}. The role of quadratic interactions, and particularly of the dimension-6 $a^2F_{\mu\nu}F^{\mu\nu}$ interaction, in (light) QCD axion models was also examined in Refs.~\cite{Beadle:2023flm,Kim:2023pvt}.

This work is organized as follows: we first give an overview of the $pp\to aa\to 4\gamma$ process and its main physical features in Sec.~\ref{sec:overview}. Section~\ref{sec:theo} provides a more precise definition of the theoretical framework and a summary of existing constraints on the parameters of interest. 
Section~\ref{sec:modeling} contains detailed descriptions of the modeling of the ALP pair production and decay processes. The numerical results are presented in Sec.~\ref{sec:results}, with Sec.~\ref{sec:discussion} summarizing our main findings. In Sec.~\ref{sec:conclusion} we give the conclusions and outlook.


\section{Hadroproduction of two ALPs decaying into photons}
\label{sec:overview}

\begin{figure}[t]
    \centering
    \begin{subfigure}[b]{0.3\textwidth}
        \centering
        \includegraphics[width=\textwidth]{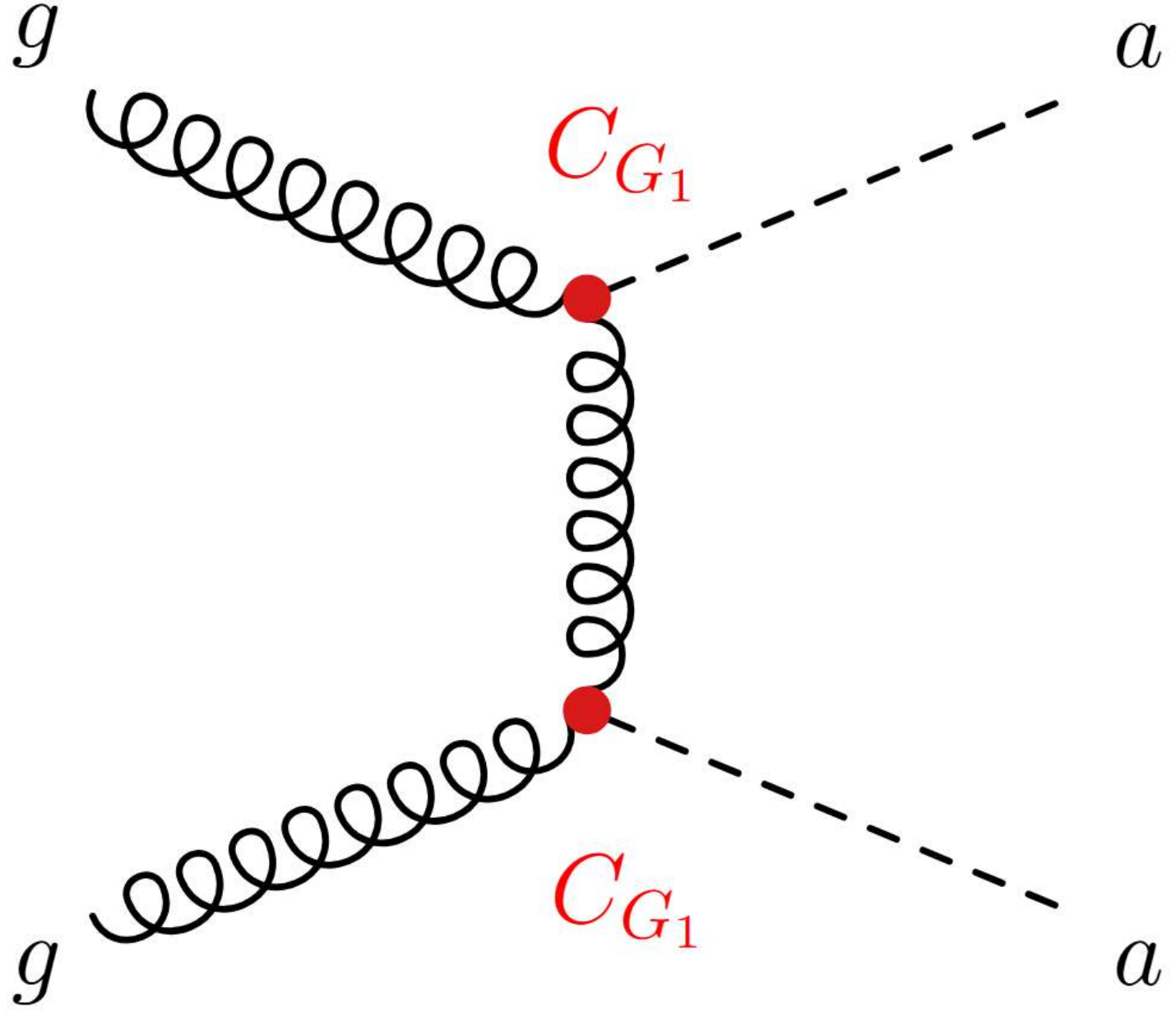}
        \caption{}
        \label{fig:t_channel}
    \end{subfigure}
    \hfill
    \begin{subfigure}[b]{0.3\textwidth}
        \centering
        \includegraphics[width=\textwidth]{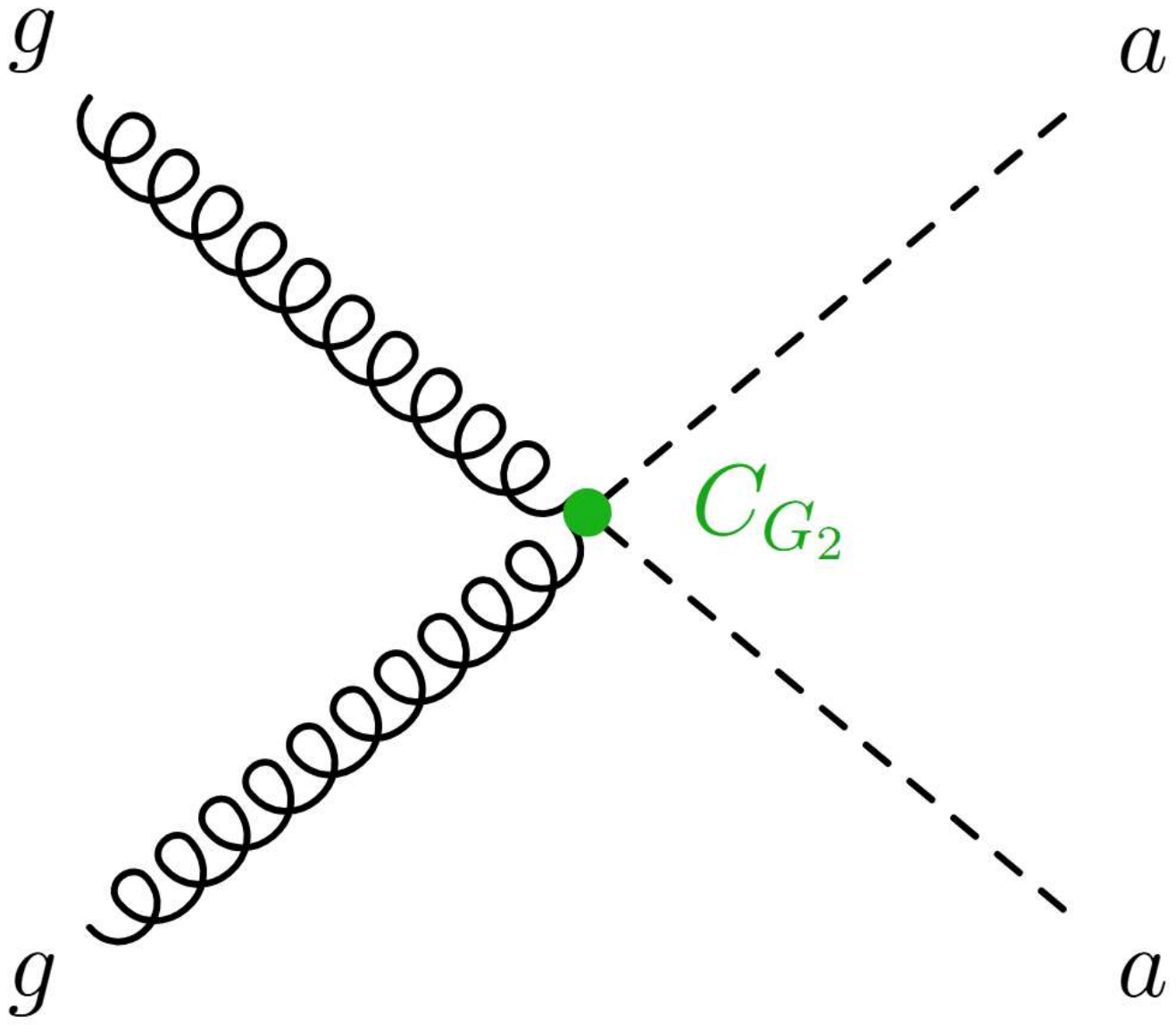}
        \caption{}
        \label{fig:contact}
    \end{subfigure}
    \hfill
    \begin{subfigure}[b]{0.3\textwidth}
        \centering
        \includegraphics[width=\textwidth]{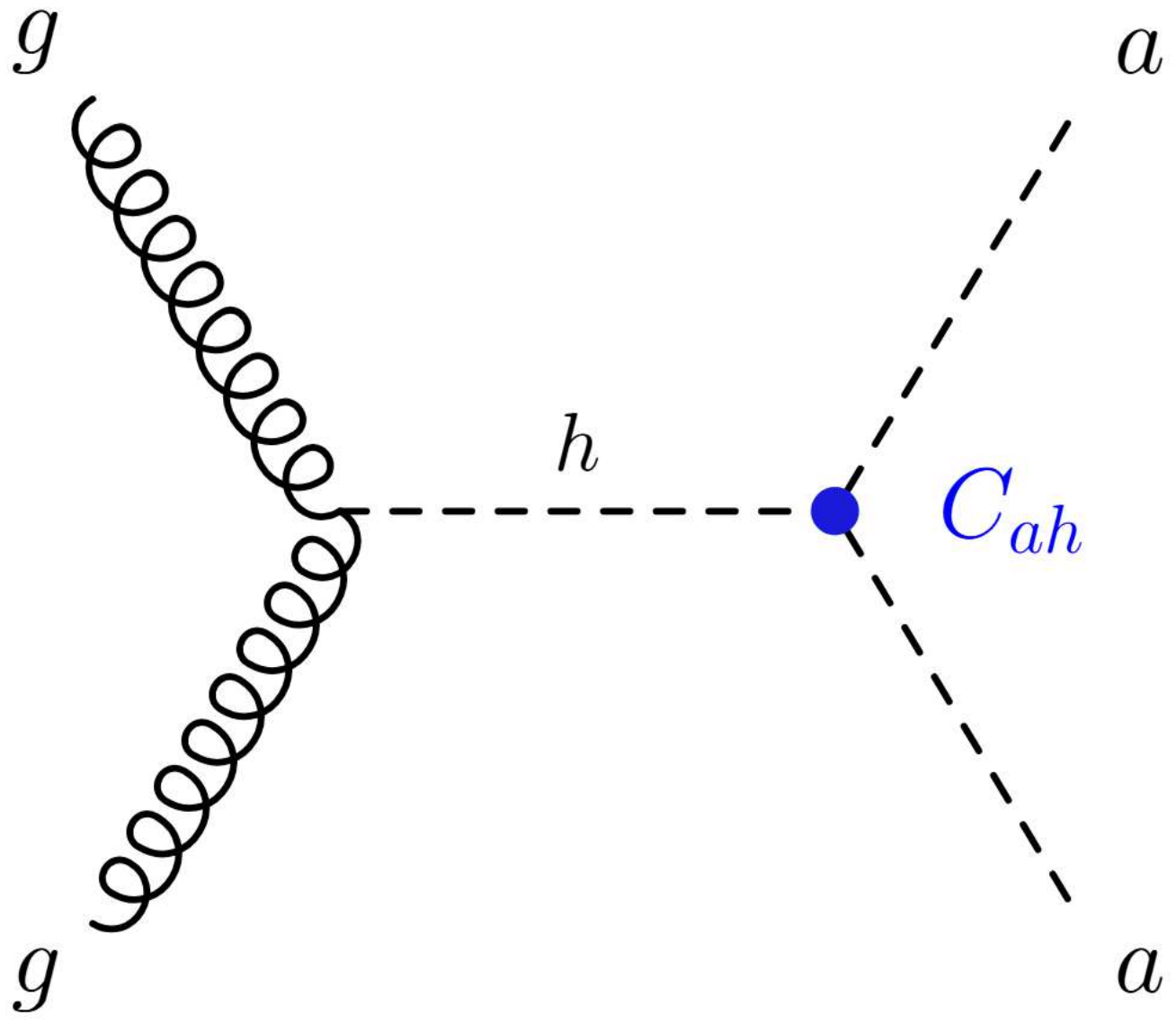}
        \caption{}
        \label{fig:gghaa}
    \end{subfigure}
    \caption{Representative Feynman diagrams contributing to $gg \to aa$, in the non-resonant (a), (b) and Higgs-resonant (c) channels.}
    \label{fig:gg_to_aa}
\end{figure}

As our main focus is the ALP pair production process at the LHC, we first introduce the relevant parameters and observables in this section, deferring a more detailed presentation of the theoretical framework to Sec.~\ref{sec:theo}.

\paragraph{Production.}
The dominant production mode for a \emph{single} ALP with a mass of the order of the electroweak (EW) scale or smaller at the LHC is via the gluon--gluon ($gg$) initial state \cite{Bauer:2017ris}: the parton-distribution function (PDF) of the gluon is by far dominant at small Bjorken-$x$ and the $q\bar q$-initiated mode is further suppressed by a factor $m_q/\La$ weighing the  quark--ALP coupling, where $m_q$ is the mass of the quark and $\La$ is the new-physics scale suppressing ALP interactions in the usual EFT parameterization, see Sec.~\ref{sec:theo}.
The $gg\to a$ process is realized at tree level via the $agg$ vertex originating from the dimension-5 operator 
\begin{equation}
\label{eq:Ocg1}
\Ocgo= - g_s^2\frac{a}{\La} G_{\mu\nu}^b \widetilde{G}^{b,\mu\nu}\,.
\end{equation}
This amplitude is of order $\La^{-1}$ and, clearly, via crossing symmetry, it can also induce the decay $a\to gg$. In Eq.~\eqref{eq:Ocg1} $b$ is the color index, $g_s$ is the strong coupling constant, $G_{\mu\nu}^b$ is the gluon field strength and $\widetilde{G}^{b}_{\mu\nu}\equiv\frac{1}{2}\epsilon_{\mu\nu\rho\sigma}{G}^{b,\rho\sigma}$ its dual, with $\epsilon_{\mu\nu\rho\sigma}$ being the Levi-Civita tensor.

The associated production of \emph{two} ALPs is also induced by the $\Ocgo$ operator via a $t$-  (see Fig.~\ref{fig:t_channel}) or $u$-channel diagram  for the $gg\to aa$ process, which features two $agg$ vertices. The contribution of such diagrams to the amplitude is of order  $\La^{-2}$ and at this order additional contributions originate from the dimension-6 operator
 \begin{equation}
 \label{eq:Ocg2}
\Ocgt = - g_s^2 \frac{a^2}{\La^2} G_{\mu\nu}^b G^{b,\mu\nu}\,,
\end{equation}
giving the contact vertex $ggaa$ in Fig.~\ref{fig:contact}. 

The production of an ALP pair at the LHC could also be mediated by a resonant Higgs boson
\begin{equation}
    gg\to h\to aa\,,
\end{equation}
as depicted in Fig.~\ref{fig:gghaa}. In this case, the relevant interaction is the $haa$ vertex generated by the operator~\cite{Draper:2012xt,Bauer:2017ris,Bauer:2018uxu,Bauer:2022rwf}
 \begin{equation}
    \Ocah=\dfrac{(\partial_\mu a)(\partial^\mu a)\phi^\dagger \phi}{\La^2}\,, \label{eq:Ocah}
\end{equation}
where $\phi$ is the SM Higgs doublet. This gives again an amplitude of order $\La^{-2}$. Thus, from the point of view of the ALP EFT, all diagrams in Fig.~\ref{fig:gg_to_aa} enter at the same order.\footnote{Due to the presence of an EW loop in the $gg\to h \to aa $ amplitude, the Higgs-mediated contribution is formally part of the one-loop EW corrections to the $gg\to aa$ amplitude.  However, it features a resonance and the corresponding  squared amplitude contributes only at NNLO EW accuracy and higher orders. Thus, we can safely treat the two channels as separate processes. }
As will be discussed in Sec.~\ref{sec:res}, the interference with the tree-level diagrams~\ref{fig:t_channel},~\ref{fig:contact}
is negligible  in most of the parameter space, so the two production channels can be safely treated separately. We will refer to diagrams~\ref{fig:t_channel},~\ref{fig:contact} and diagram~\ref{fig:gghaa}  as inducing respectively ``non-resonant'' and ``Higgs-resonant'' $aa$ production.
The two modes will clearly differ in the kinematic properties and in the targeted ALP mass windows, as the latter is only present for $m_a<m_h/2\simeq 62.5~{\rm GeV}$ and yields an ALP pair with invariant mass close to the Higgs mass $m_{aa}\simeq m_h$, while the former is not subject to either constraint. 
For definiteness, we will consider ALP masses up to 1~TeV for the non-resonant channel, as the analysis sensitivity worsens significantly at large ALP masses, due to the steep decrease of the gluon--gluon luminosity. As shown in Sec.~\ref{sec:nonresresults}, for $m_a\gtrsim \unit[1]{TeV}$ the bounds on the ALP couplings lie partially outside the region of validity of the narrow-width approximation (NWA), which we adopt in our calculation. 
A detailed analysis of the two production channels is presented in Sec.~\ref{sec:prod}.

The diagrams in Fig.~\ref{fig:gg_to_aa} are the only ones contributing to $gg\to aa$ production, up to higher-order corrections to the two modes, which we will neglect. 
We will not consider other double-ALP production channels in this work, for the following reasons:
in the case of $q \bar q \to aa$ production, the tree-level amplitude involves ALP-fermion couplings, which typically carry $m_q/\La$ suppressions (see Sec.~\ref{sec:theo}). Vector-boson-fusion and $Vaa$ associated production modes are possible, but they lead to signatures different from the one considered in this work, due to the presence of additional forward jets and a $W$ or $Z$ boson, respectively. Moreover, they are suppressed respectively by two and one power of $\alpha_{\rm EW}$. Finally,  the ``Higgs-resonant'' amplitude may feature an additional diagram  with a $a^3$ self-interaction. However, since the ALP is a pseudo-scalar, this coupling would violate CP, which is a scenario that we do not consider in this work.

\paragraph{Decay.}
Depending on the ALP mass and decay width, the production of an ALP pair can lead to various final states including missing energy and SM particles. In this work, we focus on the channel with four isolated photons
\begin{equation}
 p p \;(\to h) \to a a \to \g\g\g\g \,,
 \end{equation}
where each ALP decays into a $\gamma\gamma$ pair. 
This choice represents a natural playground for a first study of $aa$ production at colliders, for various reasons: first of all, in the absence of specific constraints on the UV modeling (see \eg\ photophobic ALPs~\cite{Craig:2018kne}), the decay rate to photons is naturally among the dominant ones for ALP masses in the GeV--TeV range. Moreover, the signature is very clean, it allows us to reproduce a realistic signal selection via simple kinematic cuts and, as discussed in Sec.~\ref{sec:cuts}, the corresponding SM background is negligible even at the HL-LHC. 
Finally, the ATLAS and CMS Collaborations have already presented searches for $pp\to h\to aa$ in the four-photon final state~\cite{ATLAS:2015rsn,CMS:2022xxa,CMS:2022fyt,ATLAS:2023ian,CMS:2026knm}, which provide a blueprint for a realistic analysis.

The ALP decay to photons proceeds at tree level through the $a\gamma\gamma$ vertex induced by the dimension-5 operator
\begin{equation}
\Ocgamma = - e^2\frac{a}{\La} F_{\mu\nu} \widetilde{F}^{\mu\nu}\,, \label{eq:Ocgamma}
\end{equation} 
where $F_{\mu\nu}$ is the photon field strength and $\widetilde{F}_{\mu\nu}\equiv\frac{1}{2}\epsilon_{\mu\nu\rho\sigma}{F}^{\rho\sigma}$ is its dual.
The relevant physical quantity is the branching ratio 
\begin{equation}
\Br(a\to \g\g) = \frac{\Gamma_{a\to\g\g}}{\Gamma_a^{\rm tot}}\,,
\label{eq.BR_a_gamgam_def}
\end{equation}
which, through the denominator, is in principle sensitive to \emph{all} the open decay channels. For this work, we will adopt a minimalistic approach, accounting only for $a\to\g\g, a\to gg$ and an irremovable combination\footnote{For $m_a\geq 2m_W$, gauge invariance requires the presence of at least one of the latter three channels, see Sec.~\ref{sec:decay}.} of $a\to Z\g, WW,ZZ$ decays. 
This choice avoids the introduction of additional free parameters affecting the signal normalization, to which our analysis would not be sensitive. The potential dependence on ALP-fermion and ALP-EW couplings can in principle be reintroduced a posteriori.
Our treatment of ALP decays is illustrated in Sec.~\ref{sec:decay}.

In the non-resonant case, we find that
the signal rate at small $m_a$ (order of a few GeV) is considerably suppressed by the selection criteria applied to final-state photons and it is highly sensitive to the specific definition of these cuts.
To avoid this problem, we restrict to the safe range $m_a\ge 30~{\rm GeV}$.
Extending our analysis to lower values of $m_a$ is in principle possible, at the cost of dealing with reduced sensitivity and complications in the signal identification, which is beyond the scope of this work. 
These issues do not affect our analysis of the Higgs-resonant channel, because the recasting of the bounds from Ref.~\cite{ATLAS:2023ian} can be performed without reproducing the experimental signal selection. On the other hand, a potential issue emerges due to the introduction of an ALP coupling to gluons, which opens up hadronic ALP decay channels: at low masses, a proper modeling of these decays requires matching the ALP EFT defined in Sec.~\ref{sec:theo} to a chiral Lagrangian, see \eg\ Refs.~\cite{Aloni:2018vki,Bauer:2020jbp,Bauer:2021mvw}. As this treatment is beyond the scope of this work, we restrict the Higgs-resonant analysis to $m_a\geq \unit[3]{GeV}$.  

\paragraph{Parameter space.} 
Overall, the signals considered in this work depend on 5 free parameters: the ALP mass $m_a$ and the four Wilson coefficients $C_i$ associated with the effective operators defined in this section.  Consistently with the notation for the operators, these coefficients are denoted as
\begin{equation} 
\cgo\,,\cgt\,,\cgamma\,, \cgoh\,.
\end{equation}
The mass will be taken in the range
 \begin{align}
     3~{\rm GeV} \le &~m_a \le 62.5~{\rm GeV}
     &
     &\text{for Higgs-resonant production}\,, \label{eq:masswindowres}
    \\
    30~{\rm GeV} \le &~m_a \le 1~{\rm TeV}
    &
    &\text{for non-resonant production}\,, \label{eq:masswindownonres}
\end{align}
for the reasons mentioned above and further discussed in Sec.~\ref{sec:cuts}. In the spirit of providing an unbiased estimate of the measurement's sensitivity, the Wilson coefficients will be left free to take values up to $10^2$, which is the maximal range consistent with basic requirements EFT validity and unitarity (see Sec.~\ref{sec:discussion}).
They enter the predictions with a specific pattern:
\begin{itemize}
\item the non-resonant production $gg\to aa$ depends only on $\cgo$ and $\cgt$. 
\item the Higgs-resonant production $gg\to h \to aa$ depends only on $\cgoh$. 
\item the ALP lifetime and branching ratio into $\g \g$ depend only on $\cgamma$ and $\cgo$.
\end{itemize}
This apparently simple setup can lead to non-trivial phenomenological consequences. In particular, the fact that $\cgo$ modifies both the production cross section and the branching ratio, entering \emph{at the denominator} of the latter, enables significant cancellation effects. Intuitively, pushing $\cgo$ to large values will simultaneously boost the production and suppress the decay, potentially resulting into an unconstrained direction. This condition, which is analyzed more precisely in Secs.~\ref{sec:nonres} and~\ref{sec:nonresresults}, is actually paradigmatic of all prompt ALP searches and marks a sharp difference compared \eg\ to SMEFT analysis.

Another relevant feature emerges in the limit where both $\cgo$ and $\cgamma$ are very small: in this case, the ALPs become sufficiently long-lived to escape the detector, as described in Sec.~\ref{sec:FSDE}. As a consequence, searches for $4\gamma$ signals can only place upper bounds outside this region. Notably, this phenomenon happens to dominate the sensitivity of the $pp\to h\to aa$ analyses presented in Refs.~\cite{Bauer:2017ris,ATLAS:2023ian}. Those results are re-examined in Sec.~\ref{sec:resresults} by relaxing the conditions $\cgo=0$ and $\cgoh=1$ implicitly imposed in the original derivations.

\section{Theoretical framework}
\label{sec:theo}
The ALP is introduced as a generic pseudoscalar particle whose interactions are parametrized by an effective Lagrangian~\cite{Georgi:1986df,Choi:1986zw} 
\begin{equation}
 \Lag_{\rm ALP} = \frac{1}{2}\de_\mu a \de^\mu a -\frac{m_a^2}{2}a^2 
 + \Lag_{5}
 +\Lag_6 + \dots
\end{equation}
where $\Lag_d$ contains a complete and non-redundant basis of dimension-$d$ operators involving at least one ALP insertion. As explained in the previous section, we are interested in going up to $d=6$, which is the first order at which interactions with two ALPs appear, and we will restrict ourselves to CP even operators. The operator basis for $\Lag_5$ is well-established
\begin{align}
\label{EQ:dimension_five_full_Lagrangian}
\Lag_5 &= \cgo \Ocgo + C_W \cO_W + C_B \cO_B + \sum_\psi \mathbf{C}_{\psi,ij} \cO_{\psi,ij}\,,
\end{align}
where $\Ocgo$ was defined in Eq.~\eqref{eq:Ocg1} and
\begin{align}
\cO_W &= -g^2 \dfrac{a}{\La} W^{I}_{\mu\nu} \widetilde{W}^{I,\mu\nu} \,,
&
\cO_B &= -g^{\prime 2}  \dfrac{a}{\La} B_{\mu\nu} \widetilde{B}^{\mu\nu}\,,
&
\cO_{\psi,ij} &= \dfrac{\partial_{\mu}a}{\La}  \,\bar{\psi}_{i} \gamma^{\mu} \psi_{j} \,.
\end{align}
Here $W_{\mu\nu}^I$ and  $B_{\mu\nu}$ are the field strengths of the  $SU(2)_L$ and $U(1)_Y$ gauge symmetries respectively, $\widetilde{W}^I_{\mu\nu}$ and  $\widetilde{B}_{\mu\nu}$ are their duals, and $g$ and $g'$ are the corresponding coupling constants.
$I$ is a $SU(2)_L$ index, while $i$ and $j$ are flavor indices. The sum in Eq.~\eqref{EQ:dimension_five_full_Lagrangian}  runs over the five fermionic fields of the SM $\psi = q_L,u_R,d_R,\ell_L,e_R$. As pointed out in Refs.~\cite{Chala:2020wvs,Bauer:2020jbp,Bonilla:2021ufe}, not all flavor entries are independent, as baryon- ($B$) and lepton-number ($L$) transformations allow the removal of two linear combinations of the diagonal ones. Assuming CP conservation, all Wilson coefficients are  real-valued, and the fermionic $\mathbf{C}_\psi$ are symmetric matrices in flavor space.
Upon EW symmetry breaking (EWSB), the couplings to weak gauge bosons from Eq.~\eqref{EQ:dimension_five_full_Lagrangian} can be recast into 
\begin{align}
\label{eq:pheno_ew_Lagrangian}
\mathcal{L}_{5,{\rm EW}}&=
\cgamma \Ocgamma 
-C_{Z\gamma}\, \dfrac{2ge}{c_w}\dfrac{a}{\La}F_{\mu\nu}\widetilde{Z}^{\mu\nu}
-C_{ZZ}\, \dfrac{g^2}{c_w^2}\dfrac{a}{\La}Z_{\mu\nu}\widetilde{Z}^{\mu\nu}
-C_{WW}\, 2g^2 \frac{a}{\La}W_{\mu\nu}\widetilde{W}^{\mu\nu}\,,
\end{align}
with $\Ocgamma$ defined in Eq.~\eqref{eq:Ocgamma} and
\begin{equation}
\label{eq:pheno_coupling}
\cgamma=C_{W}+C_{B}\,,\quad C_{Z \g}= c_w^2C_{W}-s_w^2C_{B}\,,\quad C_{ZZ}=c_{w}^4 C_{W}+s_w^4 C_{B}\,,\quad  C_{WW}=C_{W}\,.
\end{equation}
Here $s_w \equiv \sin(\theta_w)$ and $c_w = \cos(\theta_w)$,  with  $\theta_w$ the weak angle and $e=gs_w$ the electromagnetic coupling.

As anticipated in Sec.~\ref{sec:overview},
the structure of the fermionic operators $\cO_\psi$ is such that, when introduced in an amplitude, the result is always proportional to the fermion masses. This can be seen, for instance, by noting that through integration by parts and equations of motion $\cO_{d,ij} = - i (a/\La) (\bar q_L \phi Y_d)_i d_{R,j} + \text{h.c.} $ and analogously for the others, see \eg~Ref.~\cite{Bonilla:2021ufe} for the complete expressions. The proportionality to the Yukawa couplings represents a significant suppression for all fermions except the top quark, which however only enters through radiative corrections in the processes under study. For this reason, the $\cO_\psi$ operators will be neglected in the rest of this work.

As customary in the literature, the Lagrangian in Eq.~\eqref{EQ:dimension_five_full_Lagrangian} is approximately invariant under  $a(x)\mapsto a(x) + c$ transformations, with $c$ being a constant. This property, typically referred to as shift symmetry, is reminiscent of the Peccei-Quinn invariance of QCD axion models and characterizes the pseudo-Goldstone nature of the ALP.
To be more precise, the shift symmetry is explicitly broken by the introduction of the mass term $m_a\neq 0$ and it is also broken, albeit at a non-perturbative level, by the operators $\Ocgo,\cO_W$ and $\cO_B$ entering at $d=5$, see Refs.~\cite{Chala:2020wvs,Bauer:2020jbp,Bonilla:2021ufe,Bonnefoy:2022rik,Grojean:2023tsd} for recent discussions. 
When selecting the relevant $d=6$ operators, one should then specify whether further breakings of shift invariance are retained in $\Lag_6$ or not. To this end, let us first note that a complete basis of $d=6$ operators with and without shift symmetry was proposed in Ref.~\cite{Grojean:2023tsd}, which presented bases up to $d=8$. Taking this as a reference, the operators that could potentially contribute to $pp\to aa$ at Born level\footnote{We consider the $gg\to h \to aa$ process as mediated by the effective $ggh$ interaction.} are $\Ocgt$ and $ \Ocah$, defined in Eqs.~\eqref{eq:Ocg2},~\eqref{eq:Ocah} and discussed in Sec.~\ref{sec:overview}, plus 
\begin{align}
    \cO_{au\phi,ij}^{(6)} &= \frac{a^2}{\La^2} (\bar q_{L,i} \tilde\phi) u_{R,j}\,,
    &
    \cO_{ad\phi,ij}^{(6)} &= \frac{a^2}{\La^2} (\bar q_{L,i} \phi) d_{R,j}\,,
\end{align}
which can contribute to $ q \bar q\to aa$ via ALP-fermion interactions that do not scale with the fermion masses.\footnote{The same is true for certain flavor entries of the $d=5$ fermionic operators ${\cO_{au\phi,ij}^{(5)} = i (a/\La) (\bar q_{Li}\tilde\phi) u _{Rj}}$ (and analogously $\cO_{ad\phi}^{(5)}$) that manifestly break the shift symmetry, see Refs.~\cite{Chala:2020wvs,Bauer:2020jbp,Bonilla:2021ufe}. }  
The remaining $d=6$ operators induce interactions of two ALPs with EW bosons or interactions with three or more ALPs, so they do not contribute to the process we are interested in.

Among the potentially relevant operators, only $\Ocah$ preserves exactly the shift symmetry, and it is therefore retained. Of the remaining terms, we choose to retain $\Ocgt$ and discard the fermionic $O_{a\psi\phi}^{(6)}$, \ie\ we take
\begin{align}
  \Lag_6 &= \cgt \Ocgt + \cgoh \Ocah\,.
\end{align}
This choice is motivated by the fact that the fermionic operators $\cO_{a\psi\phi}^{(6)}$ manifestly break the shift symmetry at a perturbative level, and do not give tree-level contributions to non-resonant $gg\to aa$, which is the main focus of our study.
On the other hand, $\Ocgt$ gives relevant contributions to this process, and the way it breaks the shift symmetry is arguably similar to that of the $a^2 F_{\mu\nu}F^{\mu\nu}$ operator, which was recently shown in~Ref.~\cite{Beadle:2023flm} to be induced in QCD axion models. This suggests that the presence of $\Ocgt$ can hardly be ruled out in ALP scenarios that already contain some source of shift-symmetry breaking. Moreover, since we consider ALP masses up to 1~TeV, such effects can be potentially large.
Overall, retaining it in the Lagrangian can be regarded as a conservative choice.

As the operator $\Ocah$ contains the Higgs doublet, we illustrate explicitly how it induces interactions of the physical Higgs boson. For completeness, we include in this discussion the ALP-Higgs coupling 
\begin{align}
    -\lambda_{ah}\, a^2\phi^\dagger \phi\,,\label{eq:Ocahp}
\end{align}
as well, which is present in the $d=4$ Lagrangian if the shift symmetry is not imposed.
We therefore consider the terms
\begin{align}
  \Lag_{\rm ALP} \supset 
  \frac{1}{2}\de_\mu a\de^\mu a \left(1+ 2\cgoh \frac{\phi^\dag \phi}{\La^2}\right)-
  \frac{m_a^2}{2}a^2 \left(1 + 2\lambda_{ah} \frac{\phi^\dag \phi}{m_a^2}\right) \,.
\end{align}
Upon EWSB, $\phi^\dag \phi = (v+h)^2/2$, yielding:
\begin{align}
   \Lag_{\rm ALP}&\supset
   \frac{1}{2}\de_\mu a\de^\mu a \left(1+ \cgoh \frac{v^2}{\La^2}\right)-
  \frac{1}{2}a^2 \left(m_a^2 + \lambda_{ah}v^2\right) 
  \\
  &+\frac{\cgoh v}{\La^2}\de_\mu a\de^\mu a h
  -\lambda_{ah} v a^2 h
  +\frac{\cgoh }{2\La^2}\de_\mu a\de^\mu a h^2
  -\frac{\lambda_{ah}}{2} a^2 h^2\,.
\end{align}
The kinetic term of the ALP can be brought to canonical normalization (up to higher-order corrections) via the field redefinition $a\mapsto a(1 - \cgoh v^2/2\La^2)$, leading to
\begin{align}
   \Lag_{\rm ALP}&\supset
   \frac{1}{2}\de_\mu a\de^\mu a-
  \frac{1}{2}a^2 \left(m_a^2 + \lambda_{ah}v^2\right) \left(1-\cgoh\frac{v^2}{\La^2}\right)
  \label{eq.redefined_ma}
  \\
  &+\frac{\cgoh v}{\La^2}\de_\mu a\de^\mu a h
  -\lambda_{ah}\left(1-\cgoh\frac{v^2}{\La^2}\right) v a^2 h
  + (h^2 \text{ terms}) +\cO(\La^{-4})\,.
  \label{eq.redefined_aah}
\end{align}
From the second term in Eq.~\eqref{eq.redefined_ma} we see that the physical ALP mass can be expressed as
\begin{align}
m_{a,{\rm phys}}^2 = \left(m_a^2 + \lambda_{ah}v^2\right) \left(1-\cgoh\frac{v^2}{\La^2}\right) + \cO(\La^{-4})\, .
\end{align}
Because the ALP mass is a free parameter in the theory, this result has no practical consequence for the interpretation of the phenomenological results. For economy of notation, in the rest of the paper we will use $m_a$ to denote the physical mass. 

The ALP field redefinition does not impact the parameterization of the $haa$ coupling from $\Ocah$, which appears in Eq.~\eqref{eq.redefined_aah}. It is interesting to compare it to the interaction induced by the $a^2\phi^\dag \phi$ operator: in principle the two vertices have different Lorentz structures. However, in the $h\to aa$ decay, where all three participating scalars are on-shell, there is no kinematic freedom: the scattering amplitude is
\begin{align}
\mathcal{M}_{h\to aa} &= -2iv\left[ \frac{\cgoh }{\La^2} p_{a1}\cdot p_{a2} +\lambda_{ah} \left(1-\cgoh\frac{v^2}{\La^2}\right) 
\right]
\\
&
=
-2iv\left[ \frac{\cgoh }{\La^2} \left(\frac{m_h^2}{2} -m_{a,{\rm phys}}^2\right)+\lambda_{ah} \left(1-\cgoh\frac{v^2}{\La^2}\right) 
\right]\,.
\label{eq.lambda_ah_haa}
\end{align}
This result shows that, in the calculation for $h\to aa$, the dependence on $\lambda_{ah}$ can be fully reabsorbed into a redefinition of $\cgoh$. Thus, we will not account explicitly for the presence of $\lambda_{ah}$ in the remainder of this paper. If desired, it can be easily restored via Eq.~\eqref{eq.lambda_ah_haa}.

\subsection{Current constraints on the ALP couplings}\label{sec:currentbounds}

We conclude this section with an overview of existing constraints on the parameters of interest for this work, namely $\cgamma,\cgo,\cgt$ and $\cgoh$. For this purpose, we will restrict to bounds that hold for the ALP mass ranges indicated in Eqs.~\eqref{eq:masswindowres},~\eqref{eq:masswindownonres}.

\paragraph{Constraints on $\cgamma$.}
For masses $m_a\geq\unit[3]{GeV}$, the ALP coupling to photons is constrained by collider measurements. A shared feature of these constraints is that, in general, the associated ALP signal depends on other Wilson coefficients as well, so the extraction of upper limits on $\cgamma$ typically relies on  assumptions about the remaining parameters. For this reason, in the rest of this work we will only consider the numerical values provided in the following paragraphs as indicative of the current experimental sensitivity, and we will refrain from treating these constraints as robust exclusion limits. A proper comparison among these constraints would require a global analysis, which is beyond the scope of this work.

Constraints on $\cgamma$ have been extracted from measurements of $e^+ e^- \to 3\gamma$  and $e^+e^-\to 2\gamma$ at LEP and Belle~II. An ALP signal would be produced via $e^+ e^- \to Z^{(*)}/\gamma^* \to a \gamma \to 3\gamma$, which, depending on the kinematics, can be searched for  as a $Z\to a\gamma$ decay~\cite{Jaeckel:2015jla,Mimasu:2014nea} or as a non-resonant $a+\gamma$ production process, with the ALP subsequently decaying to a photon pair~\cite{OPAL:2002vhf}. In addition, if the ALP is boosted enough that the two photons from its decay are collimated, the final state can be identified as a $2\gamma$ signature and constraints can also be set from inclusive $e^+e^-\to \gamma\gamma$ measurements~\cite{Knapen:2016moh,Belle-II:2020jti}. 
Bounds from $Z\to a\gamma$ apply for $m_a\leq m_Z$, bounds from non-resonant $e^+e^-\to a\gamma$ were provided in Ref.~\cite{OPAL:2002vhf}  for $\unit[20]{GeV}\leq m_a\leq \unit[200]{GeV}$, while bounds from the inclusive measurements apply to low ALP masses: a measurement by the OPAL experiment constrains $\cgamma$  for $\unit[50]{MeV}\leq m_a\leq \unit[8]{GeV}$~\cite{OPAL:2002vhf,Knapen:2016moh}, while a measurement by Belle~II sets bounds for $\unit[0.2]{GeV}\leq m_a\leq \unit[9.7]{GeV}$~\cite{Belle-II:2020jti}. 
The interpretation of these searches in terms of $\cgamma$ is usually performed under the assumption that $\Br(a\to\gamma\gamma)=1$, which is spoiled \eg\ for $\cgo \neq 0$ and, as will be discussed in Sec.~\ref{sec:decay}, is actually forbidden by gauge invariance for $m_a> 2m_W$. Moreover, the signal rate depends on at least two parameters simultaneously: $C_{W}$ and $C_B$, or equivalently $\cgamma$ and $C_{Z\gamma}$.  In fact, bounds on the Wilson coefficients are typically reported for the two benchmarks $C_W=0$ and $C_{Z\gamma}=0$. Investigating in detail the interplay of these measurements is beyond the scope of this work; however, we note that the absolute size of the bounds, as well as their relative impact, can depend significantly on the chosen benchmark. For instance, the constraints from $Z\to a \gamma$ decays vanish entirely in the limit $C_{Z\gamma}=0$.
With this caveat in mind, the quoted numerical values for $e^+ e^-$ bounds typically give $|\cgamma|/\La\leq\mathcal{O}(\unit[1]{TeV^{-1}})$, see \eg~\cite{AxionLimits,Mimasu:2014nea,Biekotter:2025fll}.\footnote{Note that ALP parameters are defined with different conventions in different references. The values reported in these paragraphs account for the appropriate conversion factors. For instance, Ref.~\cite{Biekotter:2025fll} reports constraints on the absolute value of a quantity $g_{a\gamma\gamma}$, which is related to $\cgamma$ defined in this work by $\cgamma = - g_{a\gamma\gamma}/e^2 \simeq -10.2\, g_{a\gamma\gamma}$.}
The constraints on $\cgamma$ extracted from $e^+e^-$ colliders are generally dominant in the mass region $\unit[3]{GeV}\lesssim m_a\lesssim \unit[5]{GeV}$, which is hard to access at hadronic colliders.

For $\unit[5]{GeV}\lesssim m_a\lesssim \unit[100]{GeV}$ the strongest constraints quoted in the literature are extracted from light-by-light (LbL) scattering measurements in ultraperipheral Pb-Pb collisions (UPC) at the LHC~\cite{Knapen:2016moh}. These measurements search for the resonant process $\gamma\gamma\to a\to \gamma\gamma$ and are sensitive to ALP masses approximately in the range $\unit[5-100]{GeV}$.  The upper limits on $\cgamma$ are of the order or $|\cgamma|/\La\leq\mathcal{O}(\unit[0.1]{TeV^{-1}})$~\cite{CMS:2018erd,ATLAS:2020hii,CMS:2024bnt}. 
At higher ALP masses, similar constraints can be extracted by measurements of central exclusive production (CEP) of photons, namely the LbL process $pp \to p (\gamma\gamma\to \gamma\gamma)p^{(*)}$ in which two colliding protons are left (nearly) intact~\cite{Baldenegro:2018hng,TOTEM:2023ewz,ATLAS:2023zfc}. As protons are collided at higher center-of-mass energies than Pb nuclei, these measurements can probe di-photon resonances with masses up to about \unit[2]{TeV}. Upper bounds on the resonant $\gamma\gamma\to a \to \gamma\gamma$ cross section are converted into limits on $|\cgamma|/\La\leq \mathcal{O}(\unit[0.1]{TeV^{-1}})$ under assumptions analogous to those adopted for UPC constraints, namely that $\Br(a\to\gamma\gamma)=1$ and neglecting potential contributions from other ALP couplings, such as $\cgo$, that could in principle enter via $gg\to a\to\gamma\gamma$~\cite{Goncalves:2020bqi}. 

Limits on $\cgamma$ can also be extracted by from di-photon resonance searches at the LHC, at both high ($\unit[200]{GeV}-\unit[5]{TeV}$)~\cite{ATLAS:2017ayi,CMS:2018dqv,ATLAS:2021uiz,CMS:2024nht} and low ($\unit[5 - 110]{GeV}$)~\cite{Mariotti:2017vtv,CMS:2024yhz,ATLAS:2024bjr,LHCb:2025gbn,CMS:2026zsp} masses, see also~\cite{dEnterria:2021ljz}. These limits are the dominant ones for $m_a\gtrsim\unit[200]{GeV}$ and they are typically quoted as of the order of $|\cgamma|/\La\leq \mathcal{O}(\unit[0.1]{TeV^{-1}})$. Let us stress that the conversion of upper bounds on $\sigma(pp\to a \to \gamma\gamma)$ into upper bounds on $\cgamma$ at large $m_a$ is actually particularly sensitive to assumptions on the remaining ALP interactions, for two reasons: on one hand, as discussed in Sec.~\ref{sec:overview}, ALP production at the LHC tends to be dominated by the $gg\to a$ channel, which is fully controlled by $\cgo$. This implies that an upper bound can only be placed on the product $(\cgo\cgamma)$.
On the other, despite being a convenient benchmark for reinterpretation, the assumption $\Br(a\to\g\g)=1$ is rather unnatural for $m_a$ of the order of a few hundreds of GeV. As mentioned above and discussed in Sec.~\ref{sec:decay}, this condition is even forbidden for $m_a\geq 2m_W$. As a consequence, the interpretation of these measurements can hardly adopt the same benchmarks chosen for the measurements listed above, and the final result is also very sensitive to the alternative benchmark choice.

Complementary information on $\cgamma$ can be extracted from ALP searches in EW processes such as (non-resonant) diboson production~\cite{Gavela:2019cmq,Carra:2021ycg,CMS:2021xor,Biswas:2023ksj}  and vector boson scattering (VBS)~\cite{Bonilla:2022pxu}, see also~\cite{Alonso-Alvarez:2018irt,Craig:2018kne,Cheung:2024qge,Sun:2025hep} for studies focusing on ALP couplings to EW gauge bosons, which could be related to $\cgamma$ via gauge invariance. 
Several non-resonant searches in multi-boson final states were recently combined into a global analysis~\cite{Esser:2025nmd}, which interpreted them within the three dimensional parameter space ($\cgo,C_W,C_B$). The corresponding bounds on the combination corresponding to $\cgamma$ are of the order of $\unit[0.1-1]{TeV^{-1}}$, depending on the assumptions on $\cgo$. As the analysis was performed assuming a non-resonant ALP signal,  the results hold approximately for any value of the ALP mass below the resonant production threshold.

Finally, indirect constraints on $\cgamma$ can be extracted from precision measurements, in processes where the ALP could enter at one loop. For instance, Ref.~\cite{Biekotter:2023mpd} 
exploited the fact that ALP interactions contribute to the renormalization group running of dimension-6 SMEFT operators~\cite{Galda:2021hbr}, by recasting a global analysis of a large number of measurements in the Higgs, top and EW sector (including EW precision observables) in terms of ALP interactions. The results are nearly independent of $m_a$ and yield constraints on the combination of Wilson coefficients corresponding to $\cgamma$ of the order of $\unit[1]{TeV^{-1}}$.

\paragraph{Constraints on $\cgo$ and $\cgt$.}

To the best of our knowledge, no experimental bound has been derived to date on the dimension-six coefficient $\cgt$.
On the other hand, the dimension-5 coupling $\cgo$ has been studied in various contexts.

Some of the processes considered in the literature can be mediated by $\cgo$ alone: the most notable examples include resonant  dijet and multi-jet production via $gg\to a$~\cite{Arganda:2018cuz}, $gg\to ag$~\cite{Ghebretinsaea:2022djg} and  $gg\to a gg$~\cite{Haghighat:2020nuh}, followed by the ALP decaying to hadrons via $a\to gg$.\footnote{
The mono-jet signature $gg\to a g$ (or multi-jet + $\slashed{E}_T$) obtained when the ALP is stable or  long-lived can also be considered in principle. However, those bounds typically do not apply for $m_a\geq\unit[3]{GeV}$, as for these masses  and $\cgo\neq 0$ the decay width to hadrons is typically large enough to make the ALP decay promptly~\cite{Mimasu:2014nea,ATLAS:2021kxv}.
} 
These searches can be sensitive to heavy ALPs: Ref.~\cite{Arganda:2018cuz} reports constraints $\cgo/\La\leq\mathcal{O}(\unit[0.01]{TeV^{-1}})$ for $m_a\in [45,300]~\unit{GeV}$, while Ref.~\cite{Ghebretinsaea:2022djg} indicates $\cgo/\La\leq\mathcal{O}(\unit[1]{TeV}^{-1})$, for $m_a$ up to 2.3~TeV. 
In both cases, the limit on the Wilson coefficients is extracted under the assumption that all other ALP interactions vanish: this can be a realistic approximation for the ALP production cross section, as the contribution from $ q \bar q\to a$ is expected to be suppressed by $m_q$. However, on the decay side, it is equivalent to assuming $\Br(a\to gg)=1$. Therefore, in the presence of other ALP decays, the bound could worsen significantly.

The ALP-gluon interaction can also be constrained via searches of the resonant processes $gg\to a\to X$, where X is some SM final state: in this case, an upper bound can only be placed on the product $(\cgo C_X)$.  
Notable channels include $X=\g\g$~\cite{ATLAS:2017ayi,CMS:2018dqv,ATLAS:2021uiz,CMS:2024nht,Mariotti:2017vtv,CMS:2024yhz,ATLAS:2024bjr,LHCb:2025gbn,CMS:2026zsp}, $VV'$ with $V^{(\prime)}$ EW gauge bosons~\cite{Biswas:2023ksj}, and $t\bar t$~\cite{Anuar:2024qsz}. 
For lighter ALPs ($m_a<m_X$), the same processes can occur in a non-resonant fashion, \ie\ $gg\to a^*\to X$~\cite{Gavela:2019cmq,Carra:2021ycg,Esser:2023fdo}. Dedicated searches for non-resonant $ZZ$~\cite{CMS:2021xor} and $jj$~\cite{CMS:2026ecv} production by the CMS Collaboration report respectively the upper bounds $|\cgo C_{ZZ}|/\La^2\leq\mathcal{O}(\unit[0.1]{TeV^{-2}})$ for $m_a\lesssim\unit[100]{GeV}$ and $\cgo/\La\leq \mathcal{O}(\unit[1]{TeV^{-1}})$ for $m_a\ll\unit[3]{TeV}$. An important advantage of non-resonant searches is that they are independent of the ALP mass and decay width, which makes these results quite robust against alternate modeling assumptions.
The global analysis of non-resonant processes presented in Ref.~\cite{Esser:2025nmd} as a function of the three parameters $(\cgo,C_W,C_B)$ confirms constraints of the order $\cgo/\La\leq \mathcal{O}(\unit[0.1]{TeV^{-1}})$ after profiling on the EW interactions, which hold for ALP masses up to few tens of GeV. 

Further collider processes with a potential sensitivity to $\cgo$ include the production of $a j \gamma$, $tja$, $tWa$ and $ t \bar t a$~\cite{Ebadi:2019gij,Rygaard:2023dlx,Hosseini:2024kuh}, with the ALP decaying to hadrons or other SM particles.
Among these, only  $tWa$ and $\bar tt a$ production with a stable (or long-lived) ALP have been confronted with LHC data so far~\cite{Hosseini:2024kuh}. The resulting constraints are of the order $\cgo/\La\leq\mathcal{O}(\unit[1]{TeV^{-1}})$, when extracted by floating simultaneously $\cgo$ and the ALP coupling to top quarks. Their applicability to masses $m_a\geq\unit[3]{GeV}$ is unclear as, in this case, the ALP prefers to be unstable. 

Finally, indirect information on $\cgo$ can be inferred through its loop contributions to other processes. 
Ref.~\cite{Butterworth:2025szb} performed a global analysis of a number of LHC measurements (including $t\bar t$, which was studied in detail in Refs.~\cite{Phan:2023dqw,Blasi:2023hvb}) in which ALP contributions only appear at one loop. Floating simultaneously the two ALP couplings to gluons and to top quarks (while assuming that the remaining ones are vanishing), the authors found a constraint of order $\cgo/\La\leq\mathcal{O}(\unit[0.1]{TeV^{-1}})$, which holds for ALPs with $m_a\lesssim\mathcal{O}(\unit[100]{GeV})$, provided --however -- that they decay invisibly or remain stable on collider scales.
Analogous constraints were derived via the global analysis of SMEFT observables in Ref.~\cite{Biekotter:2023mpd}, which reports $\cgo/\La\lesssim\mathcal{O}(\unit[0.1]{TeV^{-1}})$, holding for a wide range of ALP masses.
An interesting quantity in this context is the chromomagnetic dipole moment of the top quark~\cite{Ebadi:2019gij,Bauer:2021mvw}: when other ALP couplings are neglected, reported bounds are of the order
$\cgo/\La\lesssim\mathcal{O}(\unit[1]{TeV^{-1}})$, and hold up to ALP masses of $\mathcal{O}(\unit{TeV})$.

\paragraph{Constraints on $\cgoh$.}
For $m_a<m_h/2$, the Higgs coupling to an ALP pair is constrained by upper bounds on the branching ratio of the Higgs boson to BSM particles. 
The latter can be extracted from combined analyses by the ATLAS and CMS Collaborations in which the Higgs branching ratios to invisible and undetected particles (respectively $\Br_{\rm inv}$ and $\Br_{\rm undet}$) are introduced as free parameters alongside coupling modifiers for SM Higgs interactions, defined in the so-called $\kappa$-framework~\cite{ATLAS:2016neq,CMS:2018uag,ATLAS:2019nkf,ATLAS:2022vkf,CMS:2022dwd,CMS:2026nce}.
The parameterization adopted in these analyses is such that $\Br_{\rm inv}$ accounts for Higgs decays to invisible final states, which are targeted by dedicated searches (see \eg~\cite{CMS:2020ulv,CMS:2022qva,ATLAS:2023tkt}) included in the combination, while $\Br_{\rm undet}$ accounts inclusively for any other potential decay mode that has remained undetected or was not explicitly searched for. The most recent combined analysis by the CMS Collaboration~\cite{CMS:2026nce} reports
\begin{align}
    \Br_{\rm inv}&\leq 13\%\,,
    &
    \Br_{\rm undet}&\leq 21\%\,,
    &&\text{at 95\% CL}.
    \label{eq.BRH_bounds}
\end{align}
The results from previous analyses lie in a very similar ballpark.

As the limits in Eq.~\eqref{eq.BRH_bounds} are significantly correlated, a 
conservative bound on $\Br(h\to aa)$ can be inferred from a statistical combination of the two. Here we refrain from performing a precise estimation 
and simply choose two benchmark values for the 95\%-CL upper bound on $h\to aa$ decays, namely 
\begin{align}
    \Br(h\to aa) &\leq 10\%\,,
    &&\text{and}&
    \Br(h\to aa) &\leq 30\%\,.
\label{eq.BRhaa_bounds}
\end{align}
The latter is a conservative estimate, roughly of the order of the current bounds on $\Br_{\rm inv}+\Br_{\rm undet}$, while the former represents a more optimistic benchmark, potentially attainable within the HL-LHC lifetime.

The main advantage of extracting upper bounds on BSM Higgs decays from global analyses is that the results do not rely on any assumption about the nature of such decays. In particular, the constraints on $\Br_{\rm undet}$ can always be applied directly onto $\Br(h\to aa)$ and subsequently translated into bounds on $\cgoh$ via Eq.~\eqref{eq:htoaa}, independently of the value of the ALP mass, decay width and couplings to other SM particles. 
This situation is qualitatively different from the extraction of bounds on $\cgamma$ and $\cgo$ reviewed above, which rely on assumptions about other ALP couplings, and it makes the constraints on $\cgoh$ quite robust.

Other Higgs measurements could be also used, in principle, to constrain $\cgoh$. However, their interpretation in term of this single parameter requires assumptions on the remaining Wilson coefficients.
For instance, direct searches for $h\to aa$, have been performed in various final states, including $4\ell$~\cite{ATLAS:2021ldb,CMS:2021pcy}, $4b$~\cite{ATLAS:2025rfm,CMS:2024zfv}, $bb\tau\tau$~\cite{ATLAS:2024vpj}, $bb\mu\mu$~\cite{ATLAS:2021hbr}, $4\tau$~\cite{ATLAS:2025qyn}, 
$\g\g\tau\tau$~\cite{ATLAS:2024nnm}, $\g\g jj$~\cite{ATLAS:2018jnf}.
However, the signal predictions for these processes depend on the ALP branching ratios to final state particles.

Searches targeting invisible Higgs decays could also be employed. In this case, the results are subject to the assumption that the observed missing energy is caused by an invisibly decaying Higgs boson, and to
the condition that the ALP is sufficiently long-lived to escape detection at the LHC or that it decays to other invisible final states. In the latter case, the interpretation in terms of $\cgoh$ depends on the ALP branching fraction to invisible particles.\footnote{In the absence of other light, neutral, long-lived particles, the ALP can only decay invisibly by going to final states with neutrinos. At tree level, only the $a\to 4\nu$ channel is achievable, which is suppressed by phase space factors and, depending on the ALP mass, by the presence of one or two off-shell $Z$ boson mediators. The $a\to\bar\nu\nu$ process is loop suppressed~\cite{Bonilla:2023dtf}. } 
In the former case, a bound on $\cgoh$ can only be extracted for very small values of the $d=5$ ALP couplings, see Sec.~\ref{sec:FSDE} for quantitative estimates.

The Higgs-ALP coupling $\cgoh$ could also be constrained by measurements of the total decay width of the Higgs extracted from off-shell Higgs production, see~\cite{ATLAS:2024jry,CMS:2024eka,ATLAS:2025okx,CMS:2026igg} for recent measurements. These constraints will not be considered in this work for two reasons: they currently yield weaker bounds on $\Br(h\to aa)$ compared to the constraints on both $\Br_{\rm inv}$ and $\Br_{\rm  undet}$, and they rely on a modeling  assumption that is potentially violated in the presence of BSM states, namely that corrections to SM Higgs interactions only act as coupling modifiers that are equal on- and off-shell.

Finally, constraints on $\cgoh$ can be extracted by examining the impact of the $haa$ interaction onto low energy processes, such as meson decays and atomic spectroscopy, or even from cosmological constraints, such as the Dark Matter relic abundance, see Ref.~\cite{Bauer:2022rwf} for a comprehensive study. These results will not be considered in this work, because they also rely on specific assumptions on the remaining ALP interactions: low energy processes generally depend simultaneously on $\cgoh$ and on the (dimension-5) ALP couplings to fermions, while cosmological bounds assume that ALPs constitute Dark Matter, which requires the coupling to light SM particles to be small enough to ensure its stability on cosmological scales.

\section{Modeling of the double-ALP signal}\label{sec:modeling}

This section describes in full detail the modeling of $gg\to aa$ production and of the ALP decay into photons. We adopt the NWA for the ALP, such that the two processes fully factorize:
\begin{equation}
 \sigma(p p \to a a \to \g\g\g\g)= \sigma(p p \to a a)\times \left(\Br(a \to \g \g)\right)^2 \,. \label{eq:signalNWA}
 \end{equation}
The event simulations are performed with {\aNLO}~\cite{Alwall:2014hca, Frederix:2018nkq} using the PDF set {\tt NNPDF40\_nnlo\_as\_01180\_1000}, which adopts $\alpha_s(m_Z)=0.1180024$ as a central value~\cite{NNPDF:2021njg}. The input parameters adopted for the EW sector are those from the $\alpha(m_Z)$-scheme,
\begin{align}
 m_Z &= \unit[91.1876]{GeV}\,, 
 &
 m_W &= \unit[80.379]{GeV}\,,
 &
 \alpha_{\rm EW}&= 1/127.9\,,
\label{eq:inputs}
\end{align}
and the Higgs mass is fixed to $m_h = \unit[125]{GeV}$. All numerical results produced in this work adopt the same conventions.

\subsection{ALP production} \label{sec:prod}
As anticipated in Sec.~\ref{sec:overview}, we consider two modes for the production of ALP pairs at the LHC: the non-resonant $gg\to aa$ channel and the Higgs-resonant $gg\to h\to aa$ channel, respectively discussed in Secs.~\ref{sec:nonres} and \ref{sec:res}. 

\subsubsection{Non-resonant channel}\label{sec:nonres}

The relevant Feynman diagrams for the tree-level calculation of $gg\to aa$ are those represented in Figs.~\ref{fig:t_channel} (including both $t$- and $u$-channels) and~\ref{fig:contact}. Correspondingly, the amplitude has the form 
\begin{equation}
    \mathcal{M}_{gg\to aa} = 
    \cgo^2\, \mathcal{M}_{1} + \cgt\,\mathcal{M}_{2}\,,
\end{equation}
the squared amplitude can be parameterized as
\begin{equation}
|\mathcal{M}_{gg\to aa}|^2=
\cgo^4\,|\mathcal{M}_{1}|^2
+ \cgt^2\,|\mathcal{M}_{2}|^2
+ \cgo^2\cgt\,2\Re( \mathcal{M}_{1} \mathcal{M}_{2}^\dag)\,,
\end{equation}
and the hadronic cross section, obtained after convolution with the gluon PDFs, as 
\begin{equation}\label{eq:gg_production_crossx}
\sigma(pp \to aa) = \cgo^4  \,\sA + \cgt^{ 2}\, \sB + \cgo^2 \cgt\, \sC \,.
\end{equation}
Analytic expressions for the squared amplitude components can be easily derived:
\begin{align}
|\mathcal{M}_{1}|^2 &= \frac{g_s^8}{2} \,\frac{(t+u)^2}{\La^4} F(t,u,m_a)\,,
\label{eq:A_Feynman}
\\
|\mathcal{M}_{2}|^2 &= \frac{g_s^4}{2} \, \frac{s^2}{\La^4}\,, \label{eq:B_Feynman}
\\
2\Re( \mathcal{M}_{1} \mathcal{M}_{2}^\dag) &= 
2 g_s^6 \, \frac{s^2}{\La^4}\,,\label{eq:C_Feynman}
\end{align}
where
\begin{align}
F(t,u,m_a) &=   5 - \frac{16m_a^2}{t+u} -\frac{2m_a^4}{tu}\frac{t^2+u^2-6tu}{(t+u)^2} + \frac{m_a^8}{t^2 u^2}\,,
\end{align}
and $s,t,$ and $u$ are the Mandelstam variables for the partonic process, satisfying $s+t+u=2m_a^2$.
Thus, $|\mathcal{M}_{1}|^2$ has an angular dependence while $|\mathcal{M}_{2}|^2$ and $2\Re( \mathcal{M}_{1} \mathcal{M}_{2}^\dag)$ depend only on $s$. Moreover, the ratio $2\Re(\mathcal{M}_{1} \mathcal{M}_{2}^\dag)/|\mathcal{M}_2|^2 = 4g_s^2 = 16\pi\alpha_s\simeq 5.9$ is constant across all phase space, up to the scale dependence carried by $\alpha_s$. In the limit $m_a=0$ we have $F(t,u,0) = 5$ and $(t+u)^2=s^2$, so 
the angular dependence in $|\mathcal{M}_1|^2$ is lost and the ratio $|\mathcal{M}_1|^2/2\Re(\mathcal{M}_{1} \mathcal{M}_{2}^\dag) = 5g_s^2/4 = 5\pi\alpha_s\simeq 1.8$ becomes constant as well.

\begin{table}[t]
\centering
\begin{tabular}{|c | c| c| c| c|}
\hline
$m_a\;[\mathrm{GeV}]$ & 
$\sigma_{11}\;[\mathrm{pb}]$ & $\sigma_{22}\;[\mathrm{pb}]$ & $\sigma_{12}\;[\mathrm{pb}]$ \\
\hline
30   & 
2039   & 192.5 & 1137 \\
60   & 
1258   & 136.9 & 762  \\
100  & 
764.9  & 92.67 & 492.0 \\
170  & 
382.0  & 52.21 & 263.5 \\
200  & 
295.33 & 41.95 & 208.48 \\
300  & 
139.65 & 21.73 & 103.92 \\
500  & 
40.70  & 7.145 & 32.56 \\
1000 & 
3.444  & 0.7083 & 3.026 \\
\hline
\end{tabular}
\caption{Values of the coefficients $\sA, \sB$ and $\sC$ defined in Eq.~\eqref{eq:gg_production_crossx}, for $pp$ collisions at 13 TeV and different values of the ALP mass. $\La$ is fixed to $\unit[1]{TeV}$. 
}\label{tab:coefficients_values}
\end{table}

These features are maintained  to a good approximation  at the level of the cross section, \ie\ after phase-space integration and PDF convolution.
Tab.~\ref{tab:coefficients_values} reports the numerical values of $\sA$, $\sB$ and $\sC$ for representative values of $m_a$. They are computed simulating $pp$ collisions at a center-of-mass energy of $\unit[13]{TeV}$, with a common dynamical factorization and renormalization scale  $\mu_R = \mu_F = 2\sqrt{m_a^2 + p^2_{T,a}}$ and with fixed ${\cgo=\cgt=1}$ and ${\La=\unit[1]{TeV}}$, and subsequently reweighting the events sample to extract the individual contributions $\sigma_{ij}$. The reweighting procedure ensures that statistical uncertainties on $\sA,\sB$ and $\sC$ remain fully correlated, such that potential cancellations among the components are reproduced precisely and $\sigma(pp\to aa)\geq 0$ is always respected. 
We find that, numerically, $\sA\gtrsim \sC\gg \sB$. The ratios between these coefficients vary with $m_a$, but remain approximately close to $\sA/\sC\sim 1.5$ and $\sC/\sB\sim 5$, which are consistent with the relative scalings of the squared amplitudes. All components of the cross section decrease steeply for increasing ALP masses, due to phase space effects. 

The values of $\sA,\sB$ and $\sC$ reported in Tab.~\ref{tab:coefficients_values} are truncated according to the statistical uncertainties returned by the Monte Carlo for simulations with $10^5$ events. Since they
bear a strong dependence on $\alpha_s$ and on the gluon PDF, the largest uncertainties on $\sA,\sB$ and $\sC$ are actually the systematics associated to PDFs and scale variations. For our LO simulation, we estimate the total systematic error to be in the ballpark of $+50\%$/-$30\%$, with some variations depending on the ALP mass and the $\sigma_{ij}$ component.
As we are only interested in an order-of-magnitude estimate of the sensitivity of double-ALP searches, we choose to neglect these uncertainties in our analysis: their inclusion would require a slightly more sophisticated statistical treatment compared to the one presented in Sec.~\ref{sec:results}, but it would not change significantly the results. A reduction of the uncertainties can be achieved only  by performing the calculation at NLO in QCD, which is left to future work.

\begin{figure}[t]
    \centering
    \includegraphics[width=0.5\textwidth]{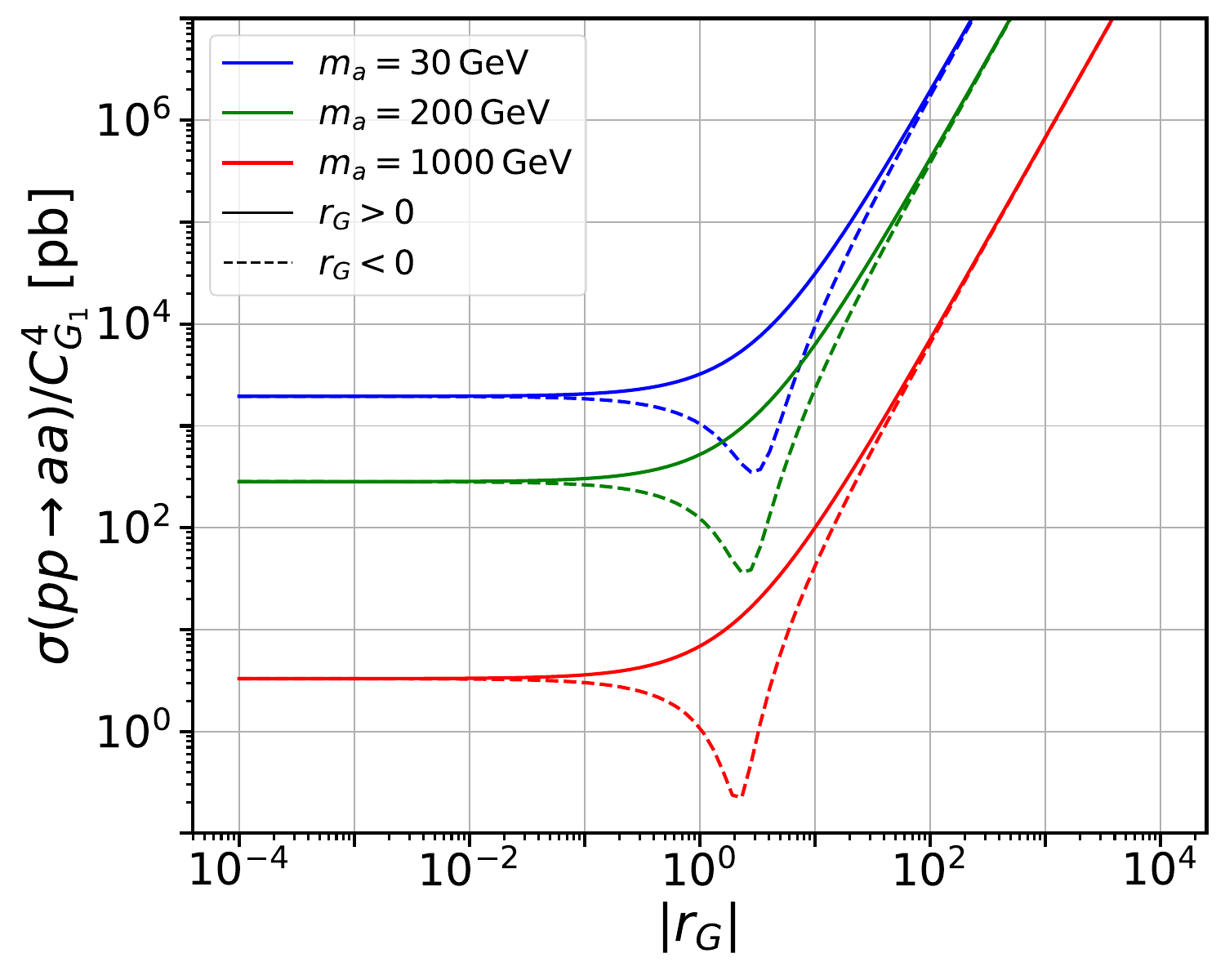}
    \caption{Non-resonant $pp\to aa$ production cross section normalized to \(\cgo^4\) 
    as a function of $|\rG| = |\cgt|/\cgo^2$  for three representative values of the ALP mass.
   }
    \label{fig:sigma_p_regimes}
\end{figure}

For the sake of generality, in our analysis we will allow the Wilson coefficients $\cgo$ and $\cgt$ to take values within a very broad range. The total production cross section can then be dominated by different combinations of $\sA,\sB$ and $\sC$, depending on the relative size of the parameters.
Fig.~\ref{fig:sigma_p_regimes} shows, for three representative values of $m_a$, the value of $\sigma(pp\to aa)$ normalized to $\cgo^4$ as a function of the ratio
\begin{equation}
\label{eq:rG}
\rG = \cgt /\cgo^2\,,
\end{equation}
that is
\begin{equation}
\label{eq:gg_production_crossx_normalized} 
\sigma(pp \to aa)/\cgo^4 = \sA  + \rG^2\,  \sB +  \rG\, \sC\,.
\end{equation}
Note that $\rG$ can take positive (solid lines) or negative (dashed lines) values, depending on the sign of $\cgt$. The $\sC$ term gives, respectively, constructive and destructive interference in the two cases. There is no sensitivity to the sign of $\cgo$, as it only enters squared in the amplitude.

Fig.~\ref{fig:sigma_p_regimes} clearly identifies three regimes:
\begin{itemize}
\item at low $|r_G|$, 
non-resonant $gg\to aa$ production is dominated by $\cgo$, and $\sigma(pp\to aa)/\cgo^4 \simeq \sA$ remains approximately constant as a function of $r_G$.\\
For $m_a=\unit[30]{GeV}$ (\unit[1]{TeV}) this occurs roughly for $|r_G|\lesssim 0.1$ $(0.06)$.\footnote{The values of $r_G$ delimiting the $\cgo$ ($\cgt)$-dominance regime were determined as those for which the dominant contribution constitutes 95\% of the cross section, \ie\ $\sA (\sB)/\sigma(pp\to aa)=0.95$.}
\item at large $|r_G|$, 
non-resonant $gg\to aa$ production is dominated by $\cgt$, and $\sigma(pp\to aa)/\cgo^4 \simeq r_G^2\,\sB$ grows quadratically with $r_G$.\\
For $m_a=\unit[30]{GeV}$ (\unit[1]{TeV}) this occurs roughly for $|r_G|\gtrsim 115$ $(85)$.
\item in the intermediate region, all contributions are relevant and have a non-trivial interplay. Since $\sB\ll\sC\lesssim\sA$, the interference term can lead to large cancellations for $r_G<0$.
\end{itemize}

\begin{figure}[t]
    \centering
    \includegraphics[width=1\textwidth]{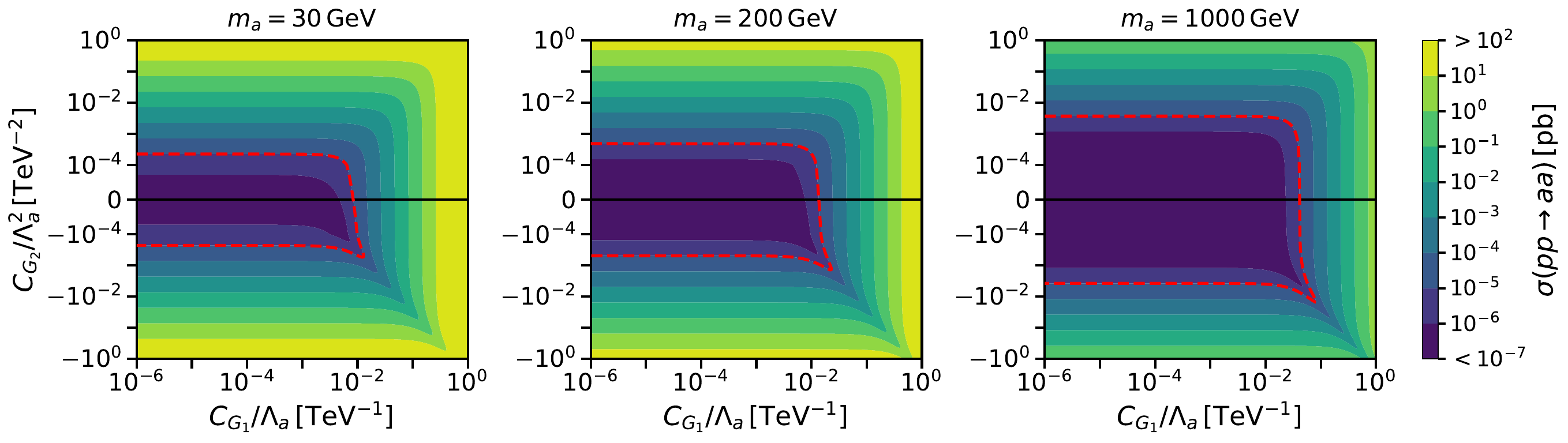}
    \caption{
Contours of constant $\sigma(pp \to aa)$ for different values of $m_a$ as a function of $(\cgo,\cgt)$.
The cross section is sensitive to the sign of $\cgt$, which is shown on a symmetric logarithmic scale. 
The red dashed lines lie at $\unit[10^{-5}]{pb}$, which corresponds to 3 events for an integrated luminosity of $\unit[300]{fb^{-1}}$.
}
\label{fig:sigma_p_contours}
\end{figure}

These features can be observed in Fig.~\ref{fig:sigma_p_contours} as well, which shows contour plots of $\sigma(pp \to aa)$ in the $(\cgo,\cgt)$ plane for the same three ALP masses. In the regions where the contours are vertical (horizontal) the cross section is dominated by $\cgo$ ($\cgt$), while those around the ``corners'' correspond to the intermediate regime of Fig.~\ref{fig:sigma_p_regimes}. For $\cgt>0$ they have rounded shapes, while for $\cgt<0$ they are elongated, due to the large negative interference.

Depending on the size of the branching ratio to photons, the sensitivity region of the double-ALP search will fall roughly within the values of $(\cgo,\cgt)$ crossed by the contours shown. 
For reference, the red dashed line marks the contour at $\unit[10^{-5}]{pb}$, corresponding to 3 events for an integrated luminosity of $\unit[300]{fb^{-1}}$ and $\Br(a\to\gamma\gamma)=1$. The decrease in cross section at larger $m_a$ is visible in these plots, which show how the analysis sensitivity to $(\cgo,\cgt)$ is correspondingly expected to worsen.

\subsubsection{Higgs-resonant channel}
\label{sec:res}

The Higgs-resonant double-ALP production $gg\to h\to aa$, proceeding through  the diagram in Fig.~\ref{fig:gghaa}, was first studied in Ref.~\cite{Bauer:2017ris}, and searches based on data collected at the LHC Run-2 have been presented by the ATLAS and CMS Collaborations in Refs.~\cite{ATLAS:2023ian} and~\cite{CMS:2022xxa,CMS:2022fyt,CMS:2026knm} respectively. For this work, we are mainly interested in understanding the sensitivity of this channel within the multi-dimensional ALP parameter space spanned by $m_a,\cgamma,\cgo$ and $\cgoh$, which,
to our knowledge, has not yet been explored in detail. 

Remarkably, the dependence
of the final bounds presented in Refs.~\cite{CMS:2022xxa,CMS:2022fyt,ATLAS:2023ian,CMS:2026knm}
on the ALP couplings is such that they can be simply re-interpreted via a rescaling \emph{a posteriori}, without repeating the simulation chain. As this procedure has the clear advantage of yielding realistic results, we will take the upper bounds on $\cgamma$ derived in Ref.~\cite{ATLAS:2023ian} under the assumptions $\Br(a\to\g\g)=1$ and $\cgoh/\La^2=\unit[1]{TeV^{-2}}$, and we will recast them for $\cgo\neq0$ and alternative values of $\cgoh$.

Our full reinterpretation procedure will be described in Sec.~\ref{sec:resresults}. In terms of modeling of $gg\to h\to aa$, following Refs.~\cite{Bauer:2017ris,ATLAS:2023ian}, it assumes a narrow-width approximation for the Higgs boson:
\begin{equation}
\label{eq:prod_resonant}
\sigma(pp \to h \to aa)=\sigma(pp \to h) \times \Br(h\to aa)\,,
\end{equation}
where
\begin{eqnarray}\label{Eq:Br_h_aa}
 \Br(h\to aa)&=&\frac{\Gamma_{h \to aa}}{\Gamma_h^{\rm tot}}\,,
 \\
 \Gamma_{h \to aa}&=&\frac{v^2 m_h^3}{32\pi} \frac{\cgoh^2}{\La^4}\left(1-\frac{2 m_a^2}{m_h^2}\right)^2\sqrt{1-\frac{4 m_a^2}{m_h^2}}\,, \label{eq:htoaa}
 \\
 \Gamma_h^{\rm tot}&=& \Gamma_h^{\text{SM}}+\Gamma_{h\to aa}\,,
\end{eqnarray}
with $\Gamma_h^{\text{SM}}=\unit[4.07\times 10^{-3}]{GeV}$.

\begin{figure}[t]
\centering
\includegraphics[width=0.45\linewidth]{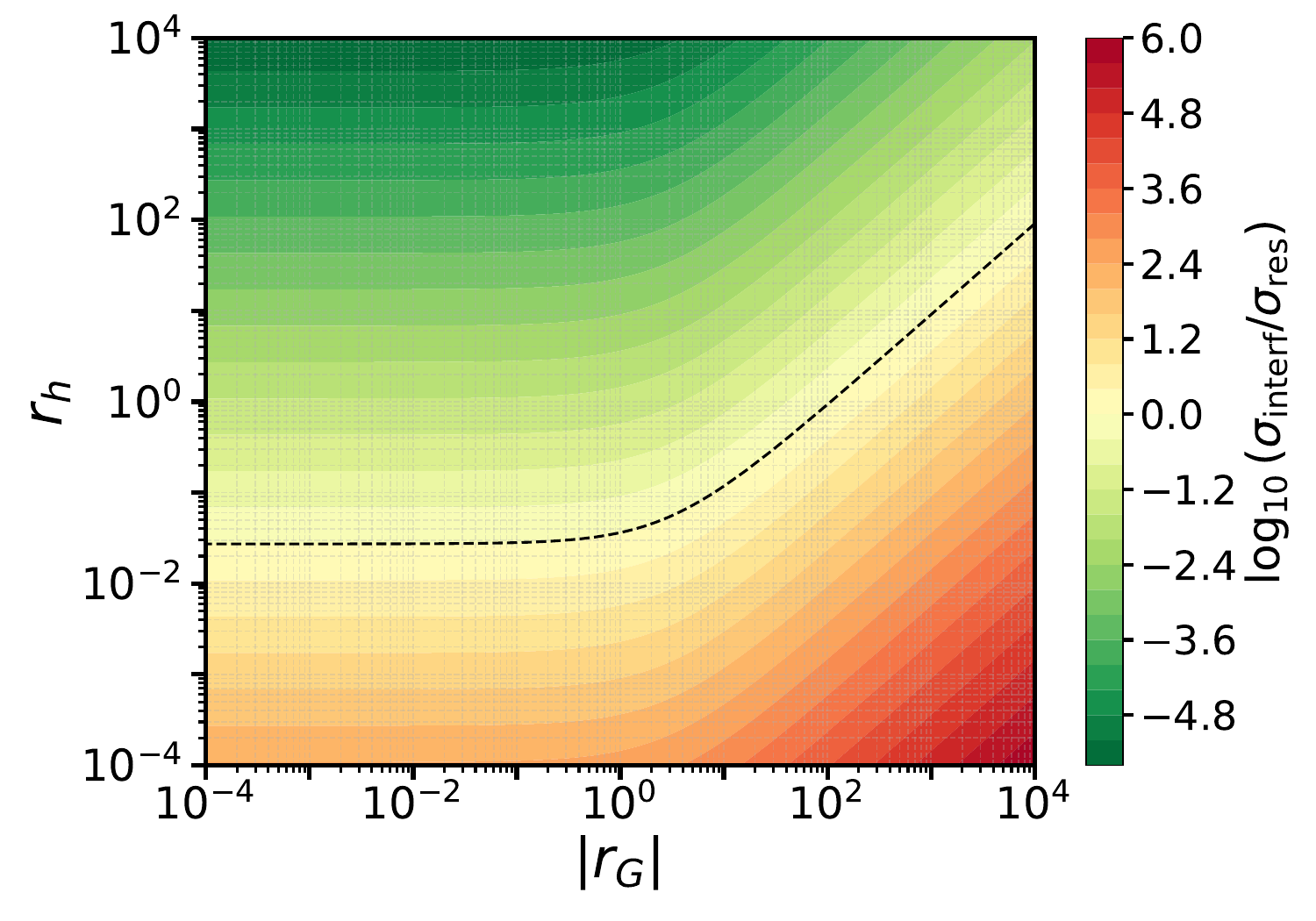}
\hfill
\includegraphics[width=0.45\linewidth]{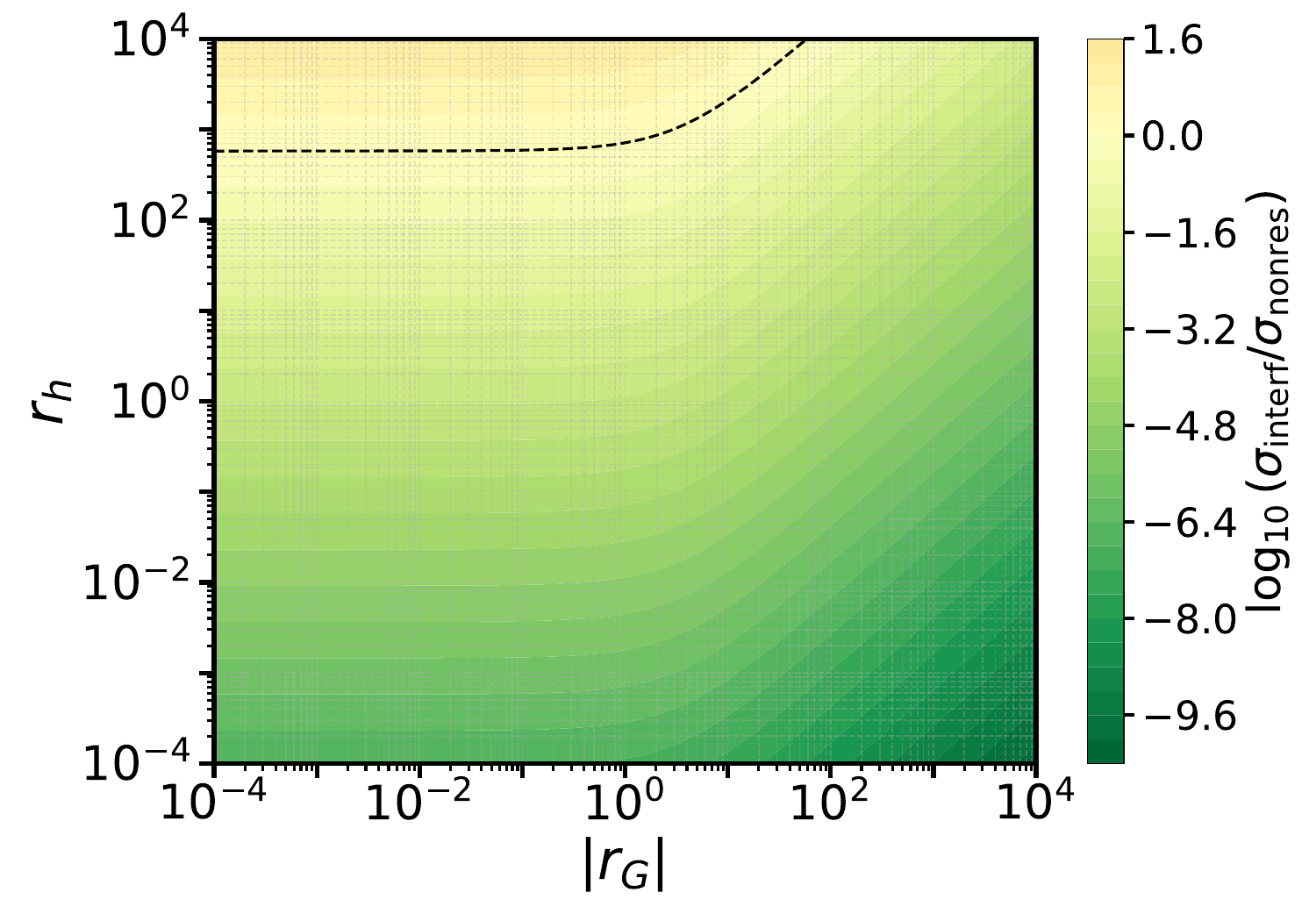}
\caption{
Impact of the interference between Higgs-resonant and non-resonant double ALP production: contour plots of $\log_{10}(\sigma_{\rm interf}/\sigma_{\rm res})$ (left) and $\log_{10}(\sigma_{\rm interf}/\sigma_{\rm nonres})$ (right) as a function of $r_G=\cgt/\cgo^2$ and $r_h=\cgoh/\cgo^2$, for $m_a = 30~\mathrm{GeV}$. Within the green regions, the interference gives a negligible contribution to the event yield, while within the red regions it is dominant. The black dashed lines mark the contours at $\sigma_{\rm interf}/\sigma_{\rm res \,(nonres)}=1$.
}
\label{fig:interference_test}
\end{figure}

In principle, the diagram in Fig.~\ref{fig:gghaa} interferes with those from non-resonant production, Figs.~\ref{fig:t_channel} and \ref{fig:contact}, leading to
\begin{align}
    \sigma_{\rm interf}(pp\to aa) =  \cgoh \cgo^2\, \sigma_{h1} + \cgoh\cgt\, \sigma_{h2}\,.
\end{align}
It can be verified analytically that $\sigma_{h1}/\sigma_{h2} = 2g_s^2 = (1/2) \sC/\sB$ independently of $m_a$. 
We are interested in estimating the impact of the interference term onto the Higgs-resonant and non-resonant signal yields. To this end, we examine the ratios
\begin{align}
\label{eq:interf_over_res}
 \frac{\sigma_{\rm interf}}{\sigma(pp\to h\to aa)} &= \frac{\sigma_{h1}+\sigma_{h2}\, r_G}{\sigma_h\, r_h}\,,
 \\
 \frac{\sigma_{\rm interf}}{\sigma(pp\to aa)} &= 
 \frac{(\sigma_{h1}+\sigma_{h2}\, r_G)r_h}{\sA+\sB\, r_G^2+\sC\, r_G}\,,
 \label{eq:interf_over_nonres}
\end{align}
where $r_G$ has been defined in Eq.~\eqref{eq:rG} while
\begin{equation} 
r_h = \cgoh/\cgo^2
\end{equation} 
and $\sigma_h \equiv \sigma(pp\to h\to aa)/\cgoh^2$. For simplicity we neglect the dependence of $\sigma_h, \sigma_{h1}$ and $\sigma_{h2}$ on $\cgoh$, by assuming that $\Gamma_h^{\rm tot}\simeq \Gamma_h^{\rm SM}$.\footnote{We also neglect the potential dependence of $\sA,\sB$ and $\sC$ on $\cgo$ and $\cgamma$ that will be discussed in Sec.~\ref{sec:FSDE}.} 
This is a good approximation as long as $\cgoh$ is not too large. For reference: for $m_a=\unit[30]{GeV}$ and $\cgoh/\La^2=\unit[1]{TeV^{-2}}$, one has $\Gamma_{h\to aa}/\Gamma_h^{\rm SM} \simeq 20\%$. This fraction remains approximately constant for smaller masses, and decreases significantly for $m_a$ close to the $m_h/2$ threshold. 
Outside of the region of validity of this approximation, the results presented below would need to be re-derived accounting for the modified $\Gamma_h$ in the Higgs propagator, which would introduce a dependence on the absolute value of $\cgoh$. However, in practice, the current bounds on BSM Higgs decays given in Eq.~\eqref{eq.BRhaa_bounds} ensure that $\Gamma_h^{\rm tot}\leq 1.4\,  \Gamma_h^{\rm SM}$. Imposing this constraint allows the contours shown in Fig.~\ref{fig:interference_test} to shift only by factors of order 1, which do not affect the main conclusions.

Fig.~\ref{fig:interference_test} shows numerical values for the ratios in Eqs.~\eqref{eq:interf_over_res} and \eqref{eq:interf_over_nonres} as a function of $r_G$ and $r_h$, for $m_a=\unit[30]{GeV}$.
We have verified that varying $m_a$ in the range $1-\unit[60]{GeV}$ the results do not change significantly. For even higher masses the Higgs-resonant signal is absent, and the interference is also negligible as the Higgs propagator is pushed significantly off-shell.

The results shown in Fig.~\ref{fig:interference_test} were obtained via dedicated event simulations of the full $pp\to 4\gamma$ process. The values of $\sA,\sB$ and $\sC$ were estimated as indicated in the previous subsection, with the treatment of $a\to\g\g$ decays and photon selection cuts detailed in Secs.~\ref{sec:decay} and \ref{sec:cuts}. The values of $\sigma_h$, $\sigma_{h1}$ and $\sigma_{h2}$ were estimated with independent event simulations, in which the $gg\to h\to aa$ matrix element is modeled in the infinite top-mass limit, treating the top loop as an effective vertex induced by 
\begin{align}
   \Lag_{ggh} &= \frac{g_s^2}{48\pi^2}G_{\mu\nu}^b G^{b\mu\nu}\frac{h}{v} + \cO(g_s^4)\,,
\end{align}
and retaining the full Higgs propagator with Higgs width fixed to $\Gamma_h^{\rm SM}$. The implementation of ALP decays and selection cuts is the same as in the non-resonant simulation, except for the additional requirement $\unit[120]{GeV}\leq m_{4\gamma}\leq \unit[130]{GeV}$ that is applied in the computation of $\sigma_{\rm interf}/\sigma(pp\to h\to aa)$ but not of $\sigma_{\rm interf}/\sigma(pp\to aa)$. The $m_{4\gamma}$ window was chosen to match as closely as possible the selections in Ref.~\cite{ATLAS:2023ian}.
Overall $K$-factors are applied to $\sigma_h$ and $\sigma_{hi}$ to account for higher-order QCD corrections. The two contributions are  multiplied by $K$ and $\sqrt{K}$ respectively, where $K\simeq 5$ is the $K$-factor required for our simulation of $pp\to h$ production to match the $\rm N^3LO$ prediction $\sigma(pp\to h)\simeq\unit[48.6]{pb}$~\cite{Anastasiou:2016cez}. 

Fig.~\ref{fig:interference_test} (right) shows that 
for 
\begin{align}
r_h&\lesssim 100+33 r_G\,,
&&\text{\ie}& 
\cgoh&\lesssim 100\cgo^2+33\cgt\,,
\label{eq.negligibile_interf_nonres}
\end{align}
the interference contribution is always negligible compared to the event yield of the non-resonant signal. The interference can also be neglected in the study of the Higgs-resonant channel (left panel) provided that 
\begin{align}
r_h&\gtrsim 0.1+0.03r_G\,,
&&\text{\ie}
&
\cgt&\lesssim 33\cgoh -3 \cgo^2\,.
\label{eq.negligible_interf_res}
\end{align}
In the region where the interference is sizable compared to the Higgs-resonant signal, it is not possible to recast the results of Ref.~\cite{ATLAS:2023ian} with the simple rescaling performed in Sec.~\ref{sec:resresults}, because both contributions need to be accounted for in the signal parameterization. 

In the rest of this work we will treat the non-resonant and Higgs-resonant channels as independent, neglecting the interference among them, and we will perform independent studies of the two processes. 
Because $\cgt$ does not contribute to pure Higgs-resonant production and, vice versa, $\cgoh$ does not contribute to the non-resonant one, the conditions above do not formally restrict the parameter space in which our analyses apply.
However, in a potential statistical combination of the two processes, it should be understood that our results only hold within the regions defined by Eqs.~\eqref{eq.negligibile_interf_nonres} and~\eqref{eq.negligible_interf_res}.

\subsection{ALP decay to a photon pair}
\label{sec:decay}
As anticipated in Sec.~\ref{sec:overview}, the relevant physical quantity for the characterization of the ALP decay to photons in our analysis is the branching ratio $\Br(a\to\g\g)$ defined in Eq.~\eqref{eq.BR_a_gamgam_def}, which, through $\Gamma_a^{\rm tot}$ at the denominator, depends in principle on all the ALP couplings to SM particles. For simplicity, in this work we will only consider tree-level, two-body decays of the ALP, assuming that other contributions are negligible.
Among the Wilson coefficients considered thus far, the dimension-6 terms $\cgt$ and $\cgoh$ can only give 1-loop contributions to ALP decays, which will be neglected. On the other hand, $\cgo$ induces decays to hadrons via $a\to gg$. For ALP masses above a few GeV,  hadronization effects are subleading and therefore  we can consider the inclusive decay rate. At leading order it is given by~\cite{Bauer:2017ris}
\begin{align}
    &\Gamma_{a\to gg}=32\pi\alpha^2_s(m_a)\,\dfrac{ m_a^3}{\La^2}\,\cgo^2\,,\label{eq:Gammaagg}
\end{align}
where we have indicated explicitly that $\alpha_s$ should be evaluated at $m_a$.
The decay width into a photon pair has an analogous expression
\begin{align}
    &\Gamma_{a\to \gamma\gamma}=4\pi\alpha^2_{\text{EW}}\dfrac{m_a^3 }{\La^2}\,\cgamma^2\,. \label{eq:Gammaaphph}
\end{align}
Note that, although in principle $\alpha_{\text{EW}}$ should also be evaluated at $m_a$, we will neglect running effects and adopt a fixed numerical value $\alpha_{\text{EW}}(m_Z)$.

As we are interested in ALP masses of order 10 -- 1000~GeV, we need to properly account for EW gauge invariance by recalling that $\cgamma$ is a linear combination of the coefficients $C_W$ and $C_B$, see Eq.~\eqref{eq:pheno_coupling}. Those two parameters control \emph{four} EW decays of the ALP, namely $a\to\g\g$, $a\to Z\g$, $a\to W^+W^-$ and $a\to ZZ$.
The corresponding decay rates can be written in a compact form as\footnote{As we only consider 2-body decays of the ALPs, we will assume that the weak bosons are produced on-shell. }

\begin{equation}
\Gamma_{a \to V_i V_j} = 
4\pi\alpha^2_{\text{EW}}
\,\dfrac{m_a^3}{\La^2}
\,\kappa_{ij}
\, C_{V_i V_j}^2
\lambda_{ij}^{3/2} 
\label{eq:general_a-VV_width}
\,.
\end{equation}
where
$\lambda_{ij}= (1 - x_i - x_j)^2 - 4x_ix_j$ is the K\"all\'en function with $x_i = m_{V_i}^2/m_a^2$. The coefficients $C_{V_iV_j}$ have been defined in Eq.~\eqref{eq:pheno_coupling} and the factors $\kappa_{ij}$ are given by
\begin{align}
\kappa_{\gamma\gamma}&=1\,,
&
\kappa_{Z \gamma}&=2/(s_wc_w)^2\,, 
&
\kappa_{WW}&=2/s_w^{4}\,,
&
\kappa_{ZZ}&=1/(s_wc_w)^{4}\,.
\end{align}
Depending on the value of the ALP mass, only a subset of these channels may be open. 

In general, the total decay width of the ALP is therefore given by\footnote{In full generality, ALP decays to fermions are also present. They are controlled by the parameters $\mathbf{C}_\psi$, which are neglected in this work and do not enter in the production process. The scenario $\mathbf{C}_\psi=0$ considered here is trivially equivalent to minimizing  $\Gamma_{a\to {\rm fermions}}$. }
\begin{align}
\Gamma_a^{\rm tot} &= 
\Gamma_{a\to gg} + \Gamma_{a\to \g\g} + \Gamma_{a\to {\rm EW}}\,,
\end{align}
where $\Gamma_{a\to {\rm EW}}$ 
represents the sum over all the kinematically open decay channels involving  $W^\pm$ and $Z$ bosons. 
It is convenient to recast the dependence of $\Gamma_{a\to {\rm EW}}$ on $C_B$ and $C_W$ into a dependence on $\cgamma$ and $C_{WW}$, via a rotation of the Wilson coefficients that yields
\begin{align}
    C_{Z \g}&=C_{WW}-s_w^2 \cgamma\,,
    &
    C_{ZZ}&=c_{2w}C_{WW}+s_w^4\cgamma\,. \label{eq:relation_pheno_coeff}
\end{align}
In this way, $\Gamma_{a}^{\rm tot}$ and $\Br(a\to \g\g)$ become functions of $\cgo,\cgamma$ and $C_{WW}$: 
the first two are our parameters of interest, for which we want to retain the full dependence, while the latter only appears in $\Gamma_a^{\rm tot}$, playing no other role in double-ALP production. In practice, retaining $C_{WW}$ 
as free parameter is equivalent to floating $\Gamma_a^{\rm tot}$ independently of $\cgo$ and $\cgamma$, which creates degeneracies in the parameterization of the double-ALP signal that can result in spurious unconstrained directions. To avoid this issue, we choose a best-case scenario, fixing $C_{WW}$ 
to the value $\bar C_{WW}$ 
that \emph{minimizes} $\Gamma_{a\to{\rm EW}}$ 
at each value of the ALP mass. This choice correspondingly maximizes $\Br(a\to\g\g)$.

The minimization of $\Gamma_{a\to {\rm EW}}$ needs to be considered carefully, taking into account the dependence on the ALP mass: 
for $m_a<m_Z$, the ALP can only decay into photons, so $\Gamma_{a\to{\rm EW}}\equiv0$ and we can simply take $\bar C_{WW}=0$. 
For $m_Z\leq m_a< 2m_W$, the ALP can decay into $Z\g$ and we still have $\Gamma^{\min}_{a\to{\rm EW}} = \Gamma^{\min}_{a\to Z\gamma}=0$, which is achieved fixing $C_{Z\gamma}=0$, \ie\ $\bar C_{WW}=s_w^2\cgamma$. For $m_a\geq 2m_W$, there are 2 or 3 (if $m_a\geq 2m_Z$) EW decay channels available, which cannot be all set to zero simultaneously, so $\Gamma_{a\to {\rm EW}}^{\min}\neq 0$ and the minimization yields more complicated expressions that we report in Appendix~\ref{sec:decayanalytic}.
The resulting expressions for $\Gamma_{a\to{\rm EW}}^{\min}$ are always proportional to $\cgamma^2$, such that
\begin{equation}\label{eq:pheno_minimum_ew_decay_width}
    F(m_a) =
    \frac{\Gamma^{\min}_{a\to{\rm EW}}(m_a,\cgamma)}{\Gamma_{a\to\g\g}(m_a,\cgamma)}\geq 0\,,
\end{equation}
is a continuous function that depends on $m_a$ but not on $\cgamma$. 
\begin{figure}[t!]
    \centering
    \includegraphics[width=0.5\textwidth]{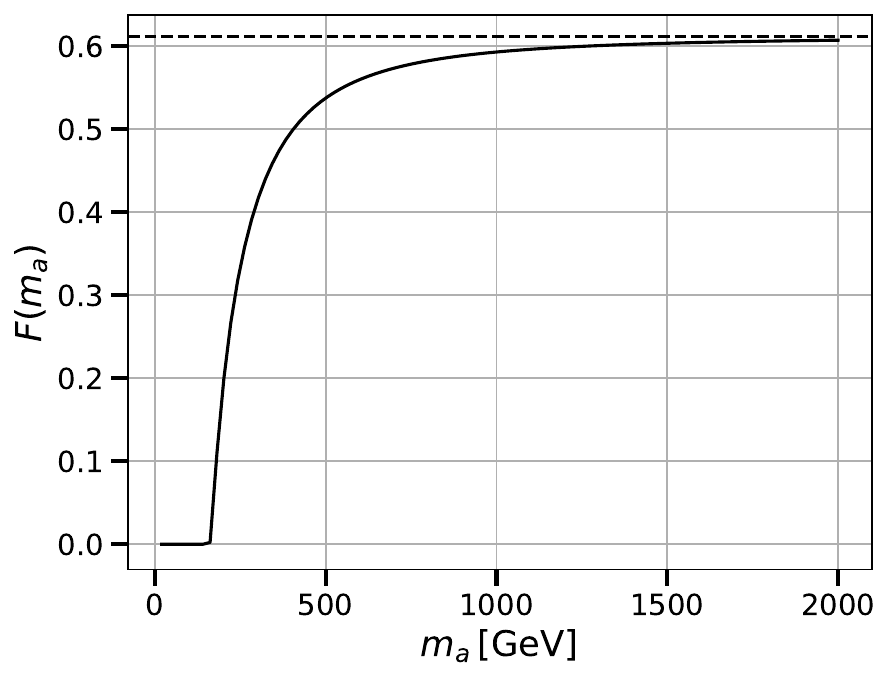}
    \caption{Numerical value of \(F(m_a)\) defined in Eq.~\eqref{eq:pheno_minimum_ew_decay_width}, as a function of the ALP mass \(m_a\).
    }
    \label{fig:F_vs_ma}
\end{figure}
In this way, we can parametrize the total decay width of the ALP in the chosen ``best-case scenario'' as
\begin{align}
\Gamma^{\rm tot,best}_a
&=
\Gamma_{a\to gg}   +
\Gamma_{a\to\g\g} +\Gamma^{\min}_{a\to{\rm EW}} 
\\
&=\Gamma_{a\to gg}   +
\Gamma_{a\to\g\g} \left(1+F(m_a)\right)\,,
\label{eq:total_decay_width}
\end{align}
and the corresponding branching ratio to photons as
\begin{equation}\label{eq:Br_a_phph_ratio}
    \Br^{\rm best}(a\to\gamma\gamma)=\dfrac{1}{1+F(m_a)+k^2(m_a)\,r_{\gamma}^2}\,
\end{equation}
where
\begin{align}
\label{eq:kdef}
k^2(m_a) &= 
\frac{\Gamma_{a\to gg}/\cgo^2}{\Gamma_{a\to\g\g}/\cgamma^2} =
\dfrac{8\alpha_s^2(m_a)}{\alpha_{\text{EW}}^2}\,, 
&
r_{\gamma}&=\dfrac{\cgo}{\cgamma}\,.
\end{align}
Thus, in our setup, the branching ratio of the ALP into photons is only a function of the ALP mass and of the ratio $r_\gamma$.

Equation \eqref{eq:Br_a_phph_ratio} indicates that necessarily 
\begin{equation}
\label{eq.maxBr}
 \Br(a\to\g\g)\leq \Br^{\rm best}(a\to\g\g)\leq \frac{1}{1+F(m_a)}\,,  
\end{equation}
with the maximum reached for $r_\gamma=0$, \ie\ in the absence of hadronic decays.
As shown in Fig.~\ref{fig:F_vs_ma}, $F(m_a)>0$ for $m_a\geq 2m_W$, reflecting the presence of irremovable EW decays above this threshold.
In the limit $m_a\gg m_W,m_Z$, we see that $F(m_a)$ tends to an asymptotic value that only depends on the weak mixing angle: 
\begin{equation}\label{eq:F_asym}
    F(m_a\to\infty)
=\frac{2 s_w^2 (1+2 c_w^2)}{c_{2w}^2+2c_w^2}\approx 0.6\,.
\end{equation}
Correspondingly, the maximum value of the branching ratio tends to $(1+0.6)^{-1}\approx 0.6$.

\begin{figure}[t]\centering
\includegraphics[width=0.5\textwidth]{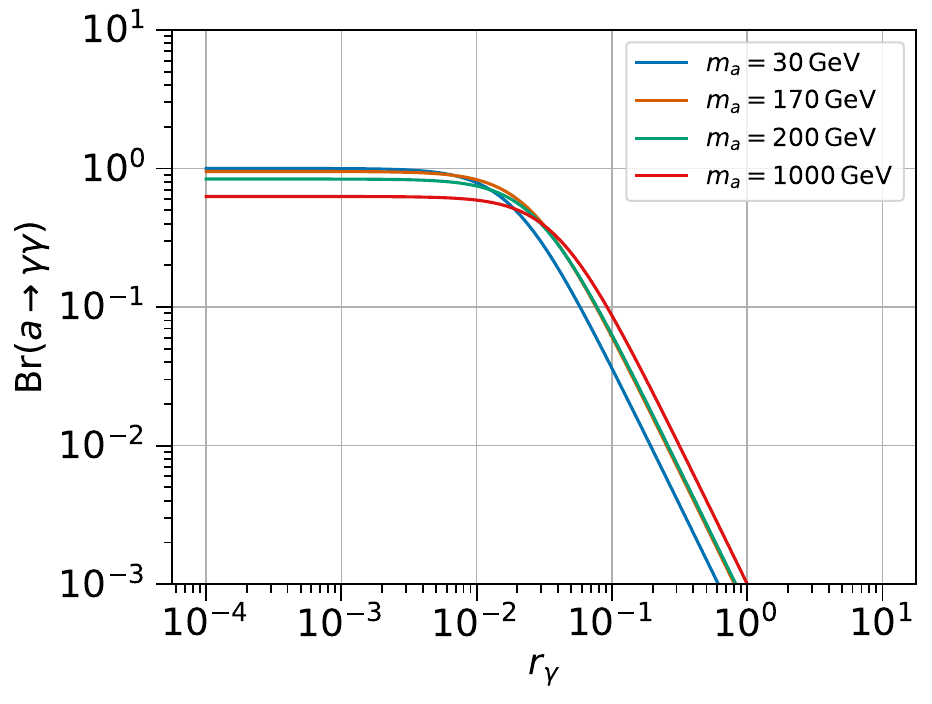}
    \caption{
$\Br(a\to\g\g)$ evaluated in the best-case scenario that minimizes the decay rates of the ALP into 
weak bosons (Eq.~\eqref{eq:Br_a_phph_ratio}),   as a function of the ratio $r_\gamma=\cgo/\cgamma$ and for representative values of~$m_a$. 
}
\label{fig:Br_all_vs_r}
\end{figure}

The behavior of $\Br^{\rm best}(a\to\g\g)$ as a function of $r_\gamma$ is shown in
Fig.~\ref{fig:Br_all_vs_r} for representative values of $m_a$. In this figure, as in the rest of the paper, we drop the ``best'' superscript and implicitly assume that the ALP branching ratios and decay width are always computed in the best-case scenario defined here.
Similar to the non-resonant production case, we can identify three regimes:\footnote{The numerical values of $r_\gamma$ delimiting the three regimes were identified as those for which $\Br(a\to\g\g)\sim r_\gamma^{-0.05}$ and $\Br(a\to\g\g)\sim r_\gamma^{-1.95}$ respectively.}

\begin{itemize}
\item for $r_\gamma\lesssim 0.004$ (\ie\ $\cgamma\gtrsim 250 \, \cgo$) 
$\Br(a\to\g\g)$ is approximately constant and equal to its maximum allowed value. In this region $\Br(a\to\g\g)$ is independent of $r_\gamma$ and therefore of both $\cgo$ and $\cgamma$. Note that this behavior is different from the production cross-section case, as $\sigma(pp\to aa)$ scales with $\cgo^4$ for small $r_G$ and with $\cgt^2$ for large $r_G$, and therefore it is always sensitive to one of the two Wilson coefficients.
\item for $r_\gamma\gtrsim 0.15$ (\ie\ $\cgo\gtrsim 7\,\cgamma$), $\Br(a\to\g\g)$ drops to zero as $r_\gamma^{-2}$. Therefore it is sensitive to both $\cgamma$ and $\cgo$, with a fixed power-law dependence.
\item in the intermediate region, $\Br(a\to\g\g)$ is sensitive to both $\cgamma$ and $\cgo$ variations, with a scaling that becomes gradually harder going from smaller to larger $r_\gamma$.
\end{itemize}
The fact that the various curves in Fig.~\ref{fig:Br_all_vs_r} cross each other is due to the $m_a$-dependence carried by the $k^2(m_a)$ term in Eq.~\eqref{eq:Br_a_phph_ratio}, which in turn reflects the running of $\alpha_s(m_a)$.

\subsection{Selection cuts and reconstruction efficiency}
\label{sec:cuts}

The last steps in our simulation of the double-ALP signal are the estimation of the signal reconstruction efficiency and the application of selection cuts on the four final-state photons. These steps are only applied to the non-resonant production channel, as for the Higgs-resonant analysis we will simply rescale the limits from Ref.~\cite{ATLAS:2023ian} without repeating the simulation chain. Nevertheless, our selection criteria reproduce as closely as possible those applied in Ref.~\cite{ATLAS:2023ian} (without the $m(4\g)=m_h$ requirement), which we take as a benchmark for a realistic measurement.

\paragraph{Selection cuts.} We require four isolated photons, using Frixione isolation \cite{Frixione:1998jh} with separation radius
 \begin{equation}
 \label{eq.deltaR}
    \Delta R = \sqrt{(\Delta \eta)^2 + (\Delta \phi)^2}>0.4\,,
 \end{equation}
where \(\Delta\eta\) is the difference in pseudorapidity and \(\Delta\phi\) the difference in azimuthal angle in the detector coordinate system. We also require
\begin{align}
   |\eta_\gamma| &< 2.4\,,
   &
    p_{T,\gamma} &>15 ~\rm GeV\,,
    \label{eq:etapt_cuts}
   \\ 
   p_{T,\g_1} &> 30\,\text{GeV}, 
   &
   p_{T,\g_2} &> 18\,\text{GeV}\,,
   \label{eq:ptgamma12cuts}
\end{align}
where $\g_1$ and $\g_2$ are respectively the first and second $p_T$-leading photons. The selections in Eq.~\eqref{eq:etapt_cuts} are applied to all photons, ensuring that they are central and within the barrel part of the detector.

\begin{figure}[t]
    \centering
\includegraphics[width=0.5\textwidth]{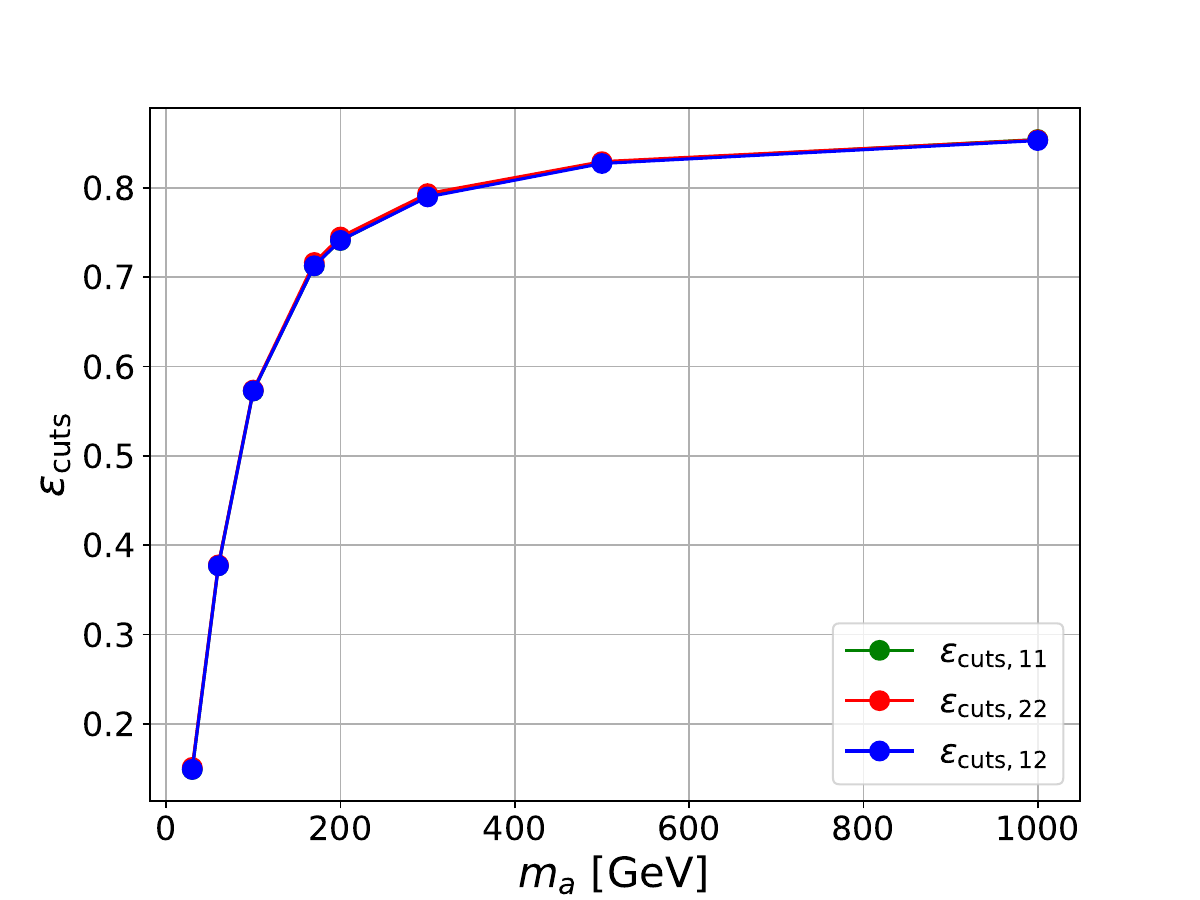}
    \caption{Selection efficiency $\epsilon_{\rm cuts}$ as a function of $m_a$, evaluated for individual components of the non-resonant production cross section $\sA,\sB$ and $\sC$. The three lines completely overlap.
    }
    \label{fig:ratio_coeff_cuts}
\end{figure}
Working in NWA, we can absorb the impact of the selection cuts
into an effective production cross section. In our parameterization of the double-ALP signal, this is done individually for the three terms in the $\cgo$ and $\cgt$ polynomial~in Eq.~\eqref{eq:gg_production_crossx}:
\begin{equation}
 \sigma^{\text{cuts}}_{ij}(m_a)\equiv \sigma_{ij}(m_a)  \times \epsilon_{\text{cuts},ij} \,,
 \hspace*{2cm}
 ij=\{11,22,12\}\,,
\end{equation}
where $\epsilon_{\text{cuts},ij}$ is the selection efficiency of the cuts in Eqs.~\eqref{eq.deltaR}--\eqref{eq:ptgamma12cuts}. In practice, $\sigma^{\rm cuts}_{ij}$ are extracted directly by simulating $pp\to aa\to 4\gamma$ events in the same way described in Sec.~\ref{sec:nonres}, using NWA for the two ALPs and sampling only the phase space allowed by the selection cuts.  One event generation is performed with $\cgamma=\cgo=\cgt=1$ and $\La=\unit[1]{TeV}$, which is then reweighted to extract the $\sA^{\rm cuts},\sB^{\rm cuts}$ and $\sC^{\rm cuts}$ components.  The total cross section after selection cuts is
\begin{align}
    \sigma^{\rm cuts}(pp\to aa) = 
    \cgo^4\, \sA^{\rm cuts} + \cgt^2\, \sB^{\rm cuts} + \cgo^2\cgt\, \sC^{\rm cuts}\,.
\end{align}

Fig.~\ref{fig:ratio_coeff_cuts} shows the value of $\epsilon_{{\rm cuts},ij}$ for each component, extracted by comparing the simulations with and without generation cuts. We see that, numerically, the selection efficiency is essentially identical for all three terms. This can be understood by observing that the differential distributions $d\sigma_{ij}/dp_{T,a}$ are very close in shape for all the masses considered, see Appendix~\ref{app.pT_dist}.
Fig.~\ref{fig:ratio_coeff_cuts} also shows that $\epsilon_{\rm cuts}$ is quite large for heavy ALPs ($\epsilon_{\rm cuts}\simeq 85\%$ for $m_a=\unit[1]{TeV}$), but it falls steeply for $m_a\lesssim\unit[100]{GeV}$. 
At 30 GeV, which is the lowest mass shown in the plot, $\epsilon_{\text{cuts}}\simeq0.15$. Below this point, it drops even faster, reaching $\epsilon_{\text{cuts}}\simeq0.002$ at 10 GeV. At this mass, in addition, $\epsilon_{\rm cuts}$ becomes quite sensitive to the specific thresholds chosen for the cuts in Eqs.~\eqref{eq:etapt_cuts} and  \eqref{eq:ptgamma12cuts}. Moving to even lower masses, this effect is further amplified. To avoid large uncertainties associated with this instability, we keep 30~GeV as the lower bound of the $m_a$ range explored in this work.

\paragraph{Reconstruction efficiency.}
As our final state is quite simple, we do not perform a full detector simulation, but we  account for a flat photon identification efficiency of 95\%~\cite{ATLAS:2019qmc}, 
resulting in a total reconstruction efficiency 
\begin{equation}
\label{eq.eps_reco}
\epsilon_{\text{reco}}=(0.95)^4 \approx 81\%\,.
\end{equation}
We adopt this as a flat correction factor for the non-resonant double-ALP signal, across all the parameter space considered.

\paragraph{Backgrounds.}
We conclude this section by discussing the expected size of the SM background.
The dominant background source is the prompt production of four photons in the SM, whose 
cross section is about $\unit[5.5\times10^{-6}]{pb}$ after applying the selection cuts listed above. This corresponds to $\sim 1.65$ events for an integrated luminosity of $\unit[300]{fb^{-1}}$.
Another possible background source is the production of $n_j$ hadronic jets plus $(4-n_j)$ photons, with the jets being misidentified as photons. 
The misidentification rate for a single jet is about $10^{-3}$ for $p_{T,j}\simeq \unit[20]{GeV}$ and it drops below $10^{-4}$ for $p_{T,j}>\unit[100]{GeV}$, see \eg\ Ref.~\cite{ATLAS:1999uwa}.
We have verified with dedicated simulations that these suppression factors are strong enough to compensate for the increase in cross-section obtained when replacing a photon with a jet in the final state. As a result, this background is completely negligible.
 
Even though it is already quite small, the continuum SM background can be reduced even further by requiring the simultaneous presence of two photon pairs with invariant mass close to~$m_a$, \ie, satisfying
\begin{align}
    |m_{\g\g}/m_a - 1| &< 0.1\,.
\end{align}
By definition, this additional selection cut has no impact on the signal simulation performed in NWA, but it reduces the background cross section to $\unit[6\times 10^{-7}]{pb}$ or less, depending on the value of $m_a$ assumed for the cut. This corresponds to 0.18 events for an integrated luminosity of $\unit[300]{fb^{-1}}$ and 1.8 events with $\unit[3000]{fb^{-1}}$, indicating that the background would remain negligible even at the HL-LHC.

\subsection{Finite-size detector effects (FSDE)}
\label{sec:FSDE}
In order to produce a four-photon signature, the two ALPs must decay before escaping the detector volume.
This condition, however, is not always satisfied across the ALP parameter space: for low $m_a$ and tiny Wilson coefficients, the average distance traveled by the ALP before decaying can become commensurate with the size of the detector. In this case, the expected number of four-photon signal events decreases, and the measurement correspondingly loses sensitivity. Intuitively, in the limit of a very long-lived ALP, it becomes impossible to extract meaningful bounds from four-photon searches, and signatures with large missing energy become relevant instead.

This subsection describes how this phenomenon, which we will refer to as finite-size detector effects (FSDE), is modeled in our analysis for the two production modes considered. In the Higgs-resonant case we follow closely the treatment proposed in Ref.~\cite{Bauer:2017ris} and adopted in Ref.~\cite{ATLAS:2023ian}.

The proper lifetime of the ALP is given by
\begin{equation}\label{eq:alp_lifetime}
    \tau_a=\dfrac{1}{\Gamma_a^{\rm tot}}\,,
\end{equation}
where \(\Gamma_a^{\rm tot}\) is the total decay width of the ALP, which we compute via Eq.~(\ref{eq:total_decay_width}), adopting the best-case scenario described in Sec.~\ref{sec:decay}. 
Assuming a cylindrical 
detector, the relevant decay length is in the direction
perpendicular to the beam axis, and it is given by
\begin{equation}\label{eq:L_aX_Perp}
L_a^{\perp}\equiv L_a\sin{\theta}=\gamma_a\beta_a \tau_a \sin{\theta}\, ,
\end{equation}
where $\theta$ is the angle between the ALP momentum and 
the beam axis in the laboratory frame, 
while $\gamma_a$
is the relativistic boost factor of the ALP 
\begin{equation}\label{eq:alp_boost_factor}
    \gamma_a=\dfrac{E_a}{m_a}\,,
\end{equation}
 and $\beta_a$ its velocity
 \begin{equation}\label{eq:beta_factor}
    \beta_a=\dfrac{|\vec p_a|}{E_a}\,.
\end{equation}
Here $E_a$ and $\vec p_a$ are respectively the energy and three-momentum of the ALP in the laboratory frame. 
Using these relations we can rewrite Eq.~\eqref{eq:L_aX_Perp} as 
 \begin{equation}\label{eq:L_aX_Perp_pT}
L_a^{\perp}=\dfrac{p_{T,a}}{m_a}\dfrac{1}{\Gamma^{\rm tot}_a}\,,
\end{equation}
indicating that $L_a^{\perp}$ is directly proportional to the transverse momentum of the ALP, $p_{T,a}$, which can be reconstructed via the momenta of the detected photons.

If no additional recoil is present, as in a LO calculation,\footnote{Even taking into account additional gluon radiation, in either a purely parton-shower approach or at NLO QCD accuracy, the main contribution would originate from radiation that is collinear to the beam axis, thus not inducing additional transverse momentum.}  $p_{T,a_1}= p_{T,a_2}$ and therefore $L_{a_1}^\perp= L_{a_2}^\perp$. Then the probability that both ALPs decay inside the detector volume
is 
\begin{equation}\label{eq:decay_probability}
    P_{aa}=\left(1-\exp(-L_{\text{det}}/L_a^{\perp})\right)^2\,,
\end{equation}
where \(L_{\text{det}}\) is the detector radius. In our simulations we take  \(L_{\text{det}}=\unit[1.08]{m}\). This value ensures excellent numerical agreement with the results of Ref.~\cite{ATLAS:2023ian} (see Sec.~\ref{sec:resresults}) and it accounts for the fact that photon reconstruction requires the ALPs to decay well before reaching the electromagnetic calorimeter.\footnote{
In the barrel, the ATLAS electromagnetic calorimeter extends between \unit[1.5]{m} and \unit[2.0]{m} from the beam axis~\cite{ATLAS:2008xda}, 
while the CMS electromagnetic calorimeter extends approximately between \unit[1.3]{m} and \unit[1.75]{m}~\cite{CMS:2008xjf}. Photon reconstruction requires energy deposits in the very first layer of the electromagnetic calorimeter. As a consequence, the detection efficiency for ALPs decaying within the calorimeter or very close to its inner radius is essentially zero. We thank K.~Schmieden for private communications on this point. }
Assuming $p_{T,a}/m_a\sim\cO(1)$, which is a good approximation for the processes considered here, FSDE will be relevant for $\Gamma_a^{\rm tot}\lesssim \unit[10^{-16}]{GeV}$: for values of $m_a,\cgo$ and $\cgamma$ within that region, $P_{aa}<1$ and the number of predicted signal events is reduced, while away from it $P_{aa}\simeq 1$ and FSDE can be safely neglected.

In the absence of additional recoil, it is also possible to parametrize  $L_a^\perp$ as
\begin{align}
\label{eq:L_a_perp_maa}
L_a^\perp &= \dfrac{\sqrt{\gamma_a^2-1}}{\Gamma_{a}^{\text{tot}}}\sin{\theta}\,,\qquad
\text{with} \qquad
\g_a = \frac{m_{aa}}{2m_a}
\end{align}
where $m_{aa}$ is the invariant mass of the ALP pair. 

In the Higgs-resonant case, the form in Eq.~\eqref{eq:L_a_perp_maa} is most convenient, as $m_{aa}= m_h$ is constant over all simulated events. Then the FSDE factorize from the production cross section, and they can be accounted for via an overall scaling factor~\cite{Bauer:2017ris}
\begin{equation}\label{eq:effective_sigma_p_resonant}
    \sigma^{\text{FSDE}}(pp\to h \to aa)\equiv\sigma(pp\to h \to aa) \times f_{aa}\,,
\end{equation}
where 
\begin{equation}
f_{aa}\equiv \int_{\Phi} d\theta \sin \theta P_{aa} \,,
\label{eq:faa}
\end{equation}
with $\Phi$ the angular volume defined by the selection cuts. Following Ref.~\cite{ATLAS:2023ian}, we approximate it as
\begin{align}
  \Phi &= ( 2\arctan e^{-2.37}\leq \theta\leq 2\arctan e^{-1.52}) \cup (2\arctan e^{-1.37} \leq \theta\leq \pi/2)\,.
  \label{eq.phi_faa}
\end{align}
Let us stress that in Ref.~\cite{ATLAS:2023ian} $f_{aa}$ is estimated via a dedicated simulation such that the integration is performed summing over the signal events that pass all the selection cuts. 
Eq.~\eqref{eq:faa} inevitably represents a numerical approximation to this estimate, as the selection cannot be reproduced exactly and the ATLAS analysis combines multiple signal regions, each satisfying different requirements.
The integration volume in Eq.~\eqref{eq.phi_faa} approximately accounts for the requirement that the final state photons have $|\eta|< 2.37$ and fall outside the transition region between barrel and endcap $1.37< |\eta|< 1.52$. We have checked that this definition of $f_{aa}$ is consistent with the function adopted in the ATLAS analysis, by verifying that we reproduce with good accuracy their bounds on $\cgamma$ by recasting the reported limits on $\Br(h\to aa\to 4\gamma)$, for both  $\cgoh=1$ and $\cgoh=0.1$.

\medskip

In the non-resonant production case, the approximation of constant $m_{aa}$ does not hold, so FSDE need to be convoluted with the kinematic distribution. To this end,  the parameterization of $L_a^\perp$ in Eq.~\eqref{eq:L_aX_Perp_pT} is more convenient. We can define effective cross sections
\begin{equation}\label{eq:effective_sigma_p}
    \sigma^{\text{FSDE,cuts}}_{ij}\equiv\displaystyle\int_{\rm cuts}  d p_{T}(a) \;\dfrac{d\sigma_{ij}}{dp_{T}(a)}P_{aa}\,,
\end{equation}
for each component $\sA,\sB$ and $\sC$,  where the integral is performed over the phase space defined by Eqs.~\eqref{eq.deltaR}--\eqref{eq:ptgamma12cuts}. In practice $\sigma^{\rm FSDE,cuts}_{ij}$ is computed by reweighting the event sample generated for the estimate of $\sigma^{\rm cuts}_{ij}$ (see Sec.~\ref{sec:cuts}), multiplying each event's weight by the corresponding $P_{aa}(p_{T,a})$. 
By analogy with the previous sections, we can define
\begin{equation}
\label{eq.sigma_FSDE_cuts}
    \sigma^{\rm FSDE,cuts}(pp\to aa) 
    =
    \cgo^4\, \sA^{\rm FSDE,cuts} + \cgt^2\, \sB^{\rm FSDE,cuts} + \cgo^2\cgt\, \sC^{\rm FSDE,cuts}\,,
\end{equation}
where it is important to note that now $\sigma^{\rm FSDE,cuts}_{ij}$ bear an implicit dependence on $\cgo$ and $\cgamma$ through the dependence on $\Gamma_a^{\rm tot}$. 

\begin{figure}[t]
\centering
\includegraphics[width=0.45\textwidth]{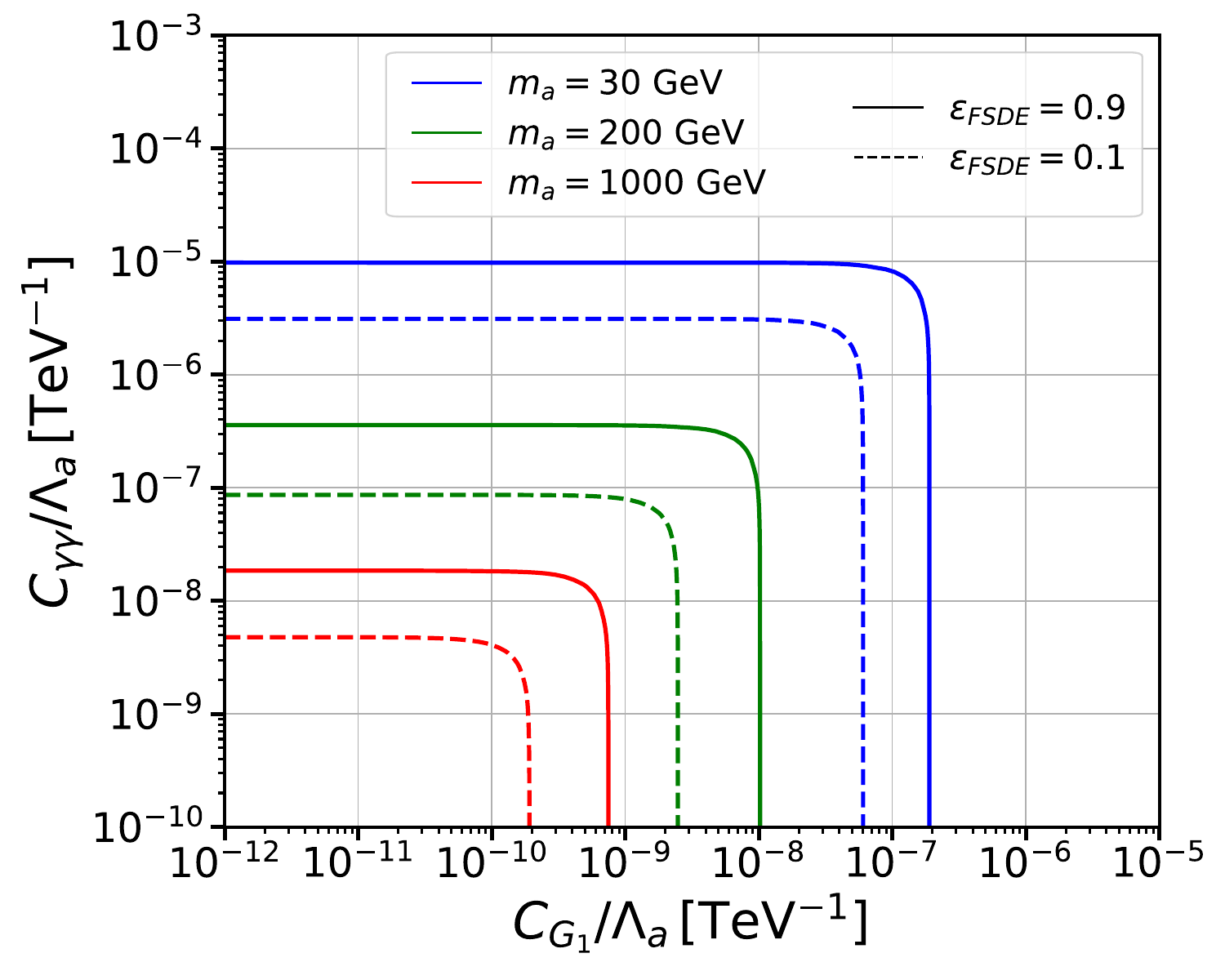}
\hfill
\includegraphics[width=.45\textwidth]{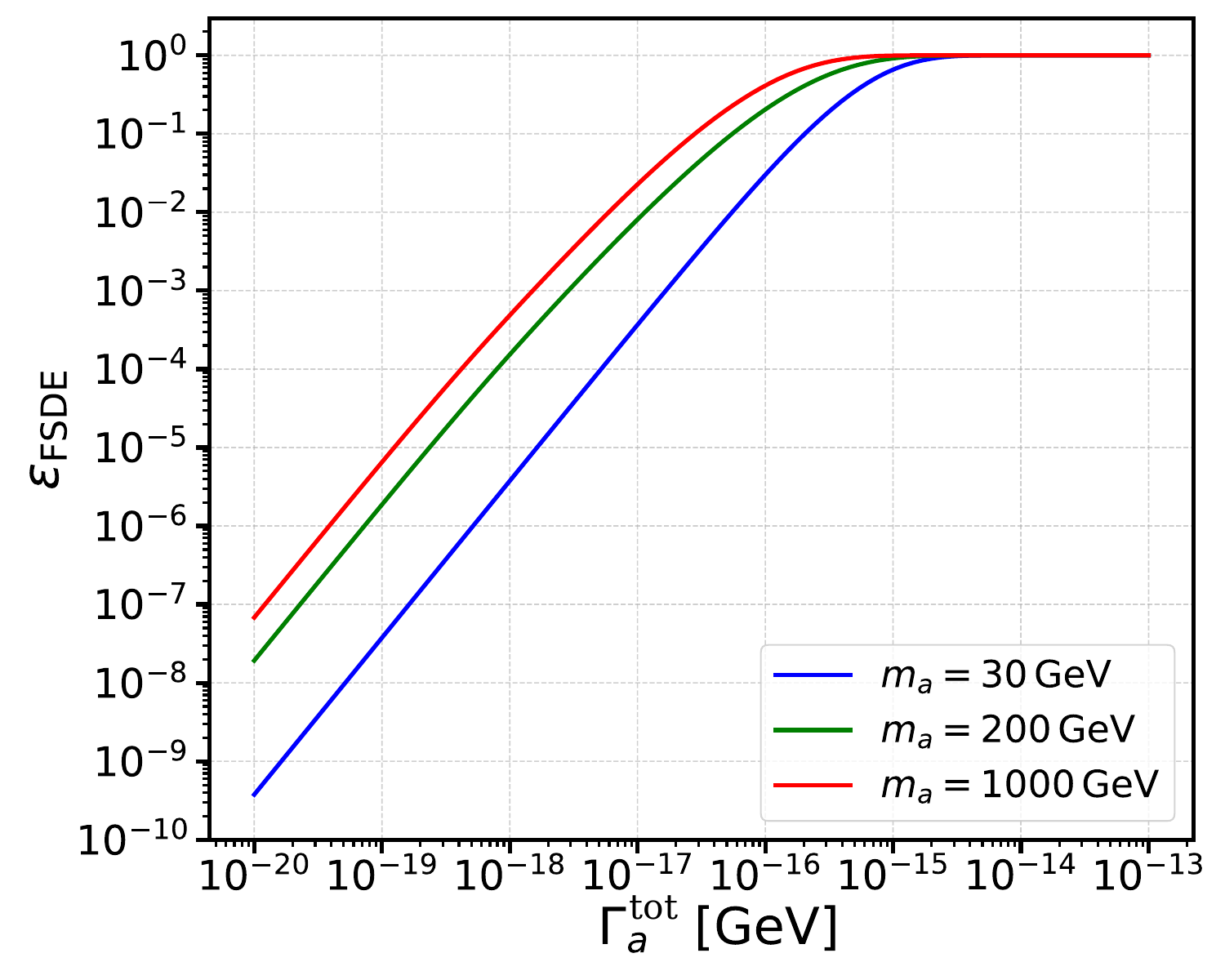}
\caption{Left: Contours of constant $\epsilon_{\rm FSDE} = \sigma^{\rm FSDE,cuts}/\sigma^{\rm cuts}$ for non-resonant $pp\to aa$ production, as a function of $(\cgo, \cgamma)$ and for representative ALP masses. The figure was produced with a fixed $\cgt=\cgo$, but the result is approximately independent of $\cgt$.
Right: $\epsilon_{\rm FSDE}$ as a function of $\Gamma_a^{\rm tot}$, for representative values of $m_a$. 
}
\label{fig:FSDE_less_90_ma_s}
\end{figure}

Fig.~\ref{fig:FSDE_less_90_ma_s} (left) shows contours (solid lines) of constant  $\sigma^{\rm FSDE,cuts}/\sigma^{\rm cuts}=0.9$ in the $(\cgo,\cgamma)$ plane, computed with fixed $\cgt=\cgo$ and for representative values of the ALP mass. 
Along these lines,
10\% of the events feature at least one ALP  decaying outside the detector.  Going towards smaller values of $\cgo$ and $\cgamma$, this fraction increases, further reducing the number of detectable events in the four-photon final state. The cases $\sigma^{\rm FSDE,cuts}/\sigma^{\rm cuts}=0.1$, where 90\% of the events feature at least one ALP  decaying outside the detector, are displayed as dashed lines. For values of $\cgo$ and $\cgamma$ that are much larger than those marked by the shown contours, FSDE become irrelevant. As expected from the scaling of $\Gamma_a^{\rm tot}\sim m_a^{3}C_i^{2}$, the portion of $(\cgo,\cgamma)$ space exposed to FSDE is much larger at lower $m_a$. 

We have verified that, for all the values of $m_a,\cgo$ and $\cgamma$ within the range shown in Fig.~\ref{fig:FSDE_less_90_ma_s} (left),  the three ratios $\sigma_{ij}^{\rm FSDE,cuts}/\sigma_{ij}^{\rm cuts}$ are actually equal at the sub-percent level. This happens because the differences in the normalized $p_{T,a}$ distributions for $\sA,\sB$ and $\sC$ are very mild (see App.~\ref{app.pT_dist}) and therefore the modulation due to $P_{aa}$ is essentially the same. As a consequence, the effective cross section $\sigma^{\rm FSDE,cuts}$ approximately factorizes into
\begin{align}
  \sigma^{\rm FSDE,cuts} &\simeq
  \sigma^{\rm cuts}
  \times \epsilon_{\rm FSDE}(m_a,\cgo,\cgamma)
  \nonumber\\
  &= \left(
  \cgo^4\, \sA^{\rm cuts} + 
  \cgt^2\, \sB^{\rm cuts} +
  \cgo^2\cgt\, \sC^{\rm cuts}\right)
  \times \epsilon_{\rm FSDE}(m_a,\cgo,\cgamma)\,,
  \label{eq:sigma_FSDE_cuts}
\end{align}
where $\epsilon_{\rm FSDE} \simeq \sigma^{\rm FSDE,cuts}_{ij}/\sigma^{\rm cuts}_{ij}$. The factorization in Eq.~\eqref{eq:sigma_FSDE_cuts} guarantees that Fig.~\ref{fig:FSDE_less_90_ma_s} (left) is  representative of the values of $\cgo$ and $\cgamma$ for which FSDE impact significantly the expected number of events, even for $\cgt\neq \cgo$. 

In the region of parameter space where FSDE are relevant, the parametric dependence of the effective cross section on $\cgo$ will deviate from the pure polynomial structure, as $\epsilon_{\rm FSDE}$ depends on $\cgo$ and $\cgamma$ through $\Gamma_a^{\rm tot}$. 
We model this dependence by repeating the $P_{aa}$ reweighting of $\sigma^{\rm cuts}$ for several values of $\Gamma_a^{\rm tot}$, and interpolating the $\epsilon_{\rm FSDE}$ obtained. The result is displayed in the right panel of Fig.~\ref{fig:FSDE_less_90_ma_s}, which shows that, for $\Gamma_a^{\rm tot}$ sufficiently small, $\epsilon_{\rm FSDE}$ scales with $(\Gamma_a^{\rm tot})^2$. This result is consistent with the fact that $P_{aa}$ scales with $(L_{\det}/L_a^{\perp})^2\sim (\Gamma_a^{\rm tot})^2 (m_a/p_{T,a})^2$ in the limit $L_{\rm det}/L_a^\perp\ll1$~\cite{Bauer:2017ris}. We also observe that, for the same $\Gamma_a^{\rm tot}$, $\epsilon_{\rm FSDE}$ is larger for heavier ALPs and that the asymptotic behavior is reached faster for lighter ALPs. This effect is due to the change in the typical $(m_a/p_{T,a})$ ratio when varying $m_a$. 

Finally, we can note that restoring non-minimal ALP decays to gauge bosons (and fermions) in the calculation of $\Gamma_a^{\rm tot}$ would increase the total decay width. If the partial width for these additional channels was very small ($\Gamma_{a\to {\rm extra}}\lesssim \unit[10^{-16}]{GeV}$), FSDE would remain relevant for smaller values of $\cgo$ and $\cgamma$ compared to those in Fig.~\ref{fig:FSDE_less_90_ma_s} (left). On the other hand, if $\Gamma_{a\to {\rm extra}}\gtrsim \unit[10^{-16}]{GeV}$, the ALP would always decay promptly, yielding $\epsilon_{\rm FSDE}\simeq 1$ over the entire ($\cgo,\,\cgamma$) parameter space.


\section{Results}\label{sec:results}

\subsection{Non-resonant production}
\label{sec:nonresresults}

This section presents the main novel results of this work, examining the expected sensitivity of a non-resonant $pp\to aa\to 4\gamma$ search at the LHC, with $pp$ collisions at $\sqrt{s}=\unit[13]{TeV}$ and an integrated luminosity $\mathcal{L}=\unit[300]{fb^{-1}}$.

As this search is essentially background-free, we assume that the number of observed events~$n_{\rm obs}$ follows a Poisson distribution with mean $\mu=N_{\rm signal}$, given by the number of predicted signal events. Then, under the hypothesis $n_{\rm obs}=0$, 95\%CL upper bounds on the ALP parameters can be derived by requiring $N_{\rm signal}\leq 3$~\cite{ParticleDataGroup:2024cfk}. We take
\begin{equation}
   N_{\rm signal} = \mathcal{L} \, \epsilon_{\rm reco} \, \sigma^{\rm FSDE, cuts}(pp\to aa) \times (\Br(a\to\g\g))^2\,,
\end{equation}
where $\sigma^{\rm FSDE,\rm cuts}$ is computed as  in Eq.~\eqref{eq:sigma_FSDE_cuts} and $\epsilon_{\rm reco}$ was defined in Eq.~\eqref{eq.eps_reco}. The branching ratio of the ALP to photons is modeled in the ``best-case scenario'' described in Sec.~\ref{sec:decay}. Over most of the parameter space, where FSDE effects are irrelevant,  $\sigma^{\rm FSDE,cuts}$ is a polynomial function of $\cgo$ and $\cgt$. For $\cgo$ and $\cgamma$ such that $\Gamma_a^{\rm tot}\lesssim\unit[10^{-16}]{GeV}$, it acquires an additional dependence on $\cgo$ and $\cgamma$ through FSDE, as indicated in Eq.~\eqref{eq:sigma_FSDE_cuts}. $\Br(a\to\g\g)$ is a function of the ratio $r_{\gamma}\equiv\cgo/\cgamma$, as given in Eq.~\eqref{eq:Br_a_phph_ratio}. $N_{\rm signal}$ is  sensitive to the sign of only $\cgt$, as $\cgo$ and $\cgamma$ always appear squared in the analytic expressions. We will restrict  the analysis to $\cgo,\cgamma\geq0$ with the understanding that the bounds are fully symmetric if  signs are flipped.

\begin{figure}[t!]
\centering
\includegraphics[width=0.48\textwidth]{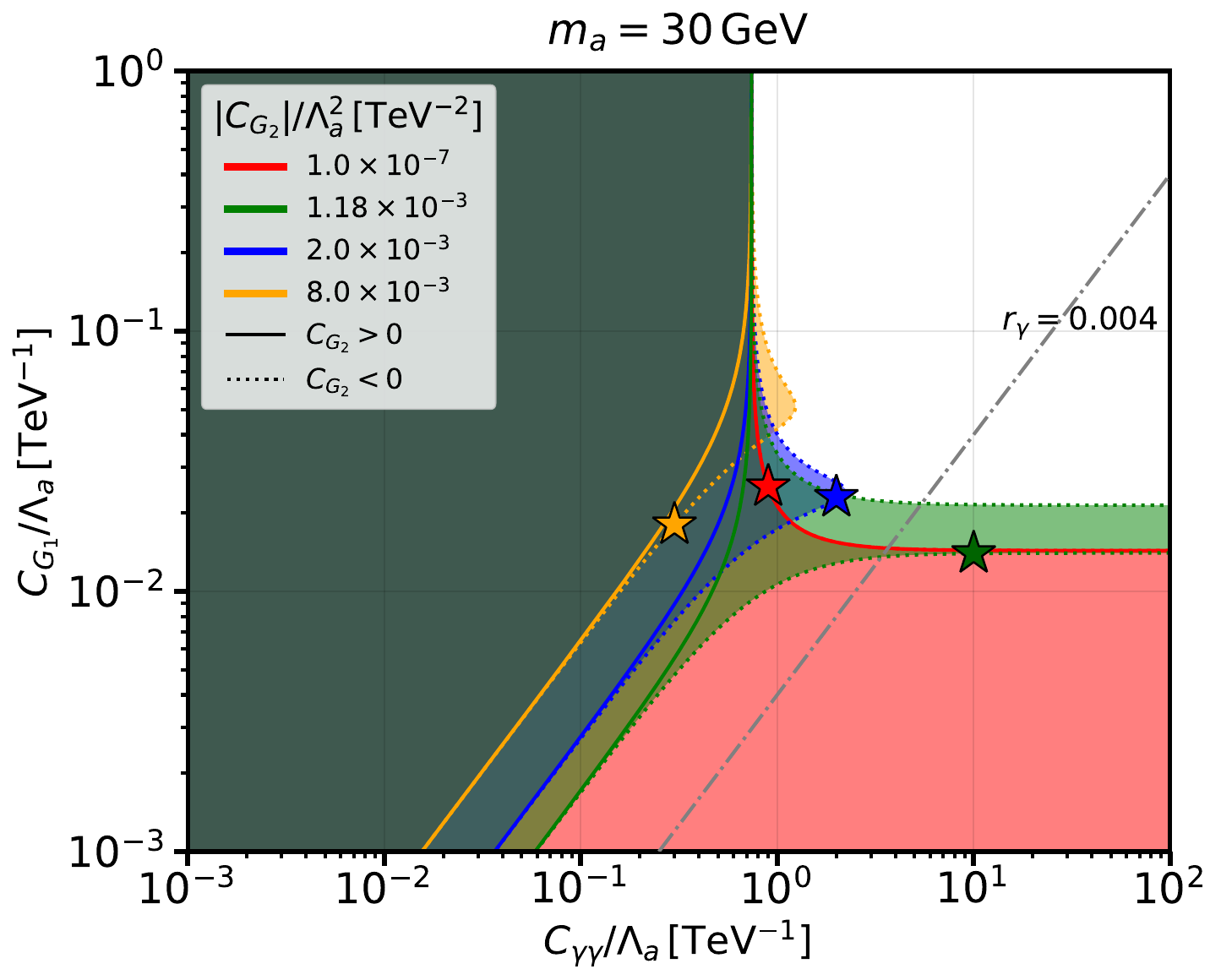}
\hfill
\includegraphics[width=0.48\textwidth]{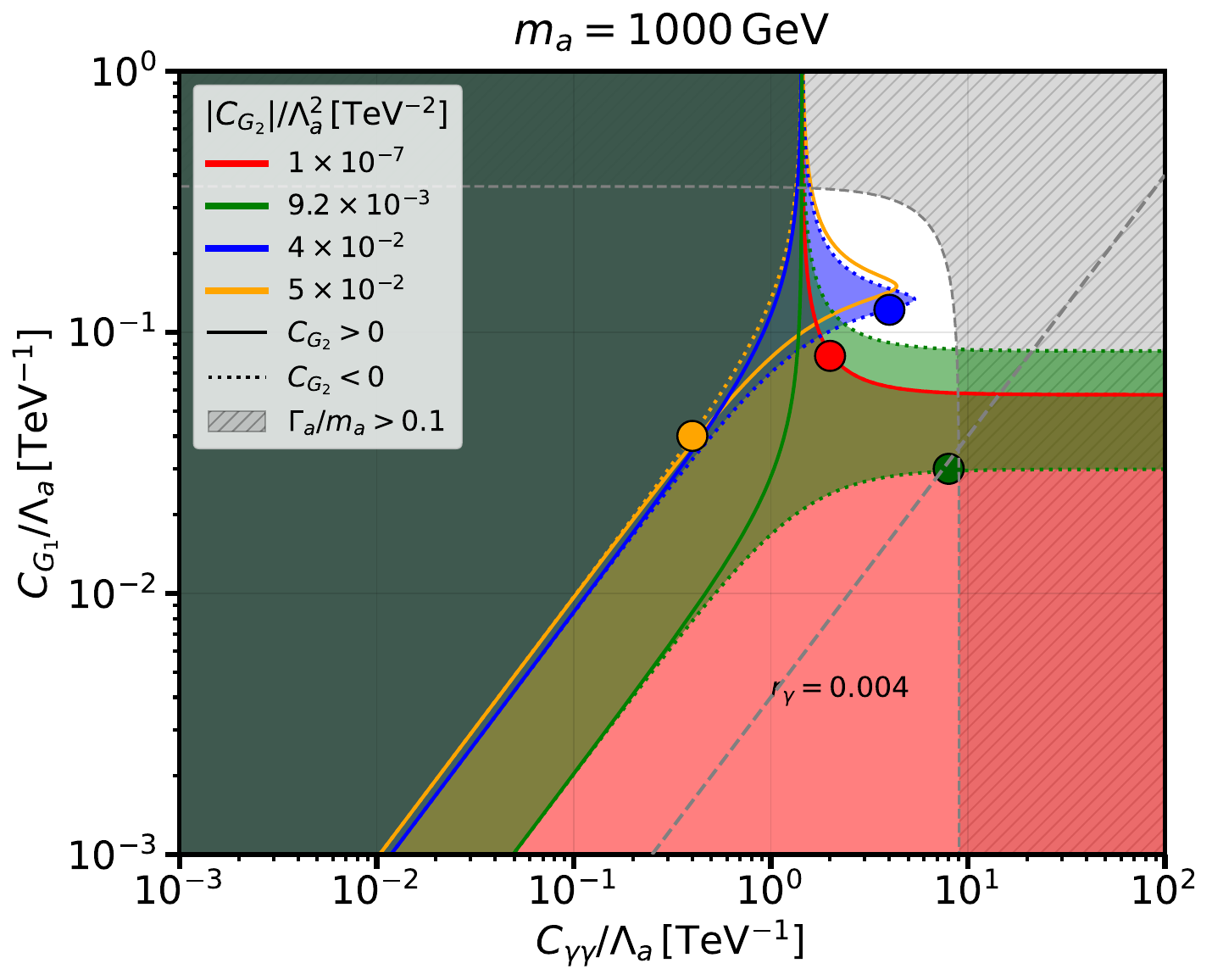}
\\[5mm]
\includegraphics[width=0.48\textwidth]{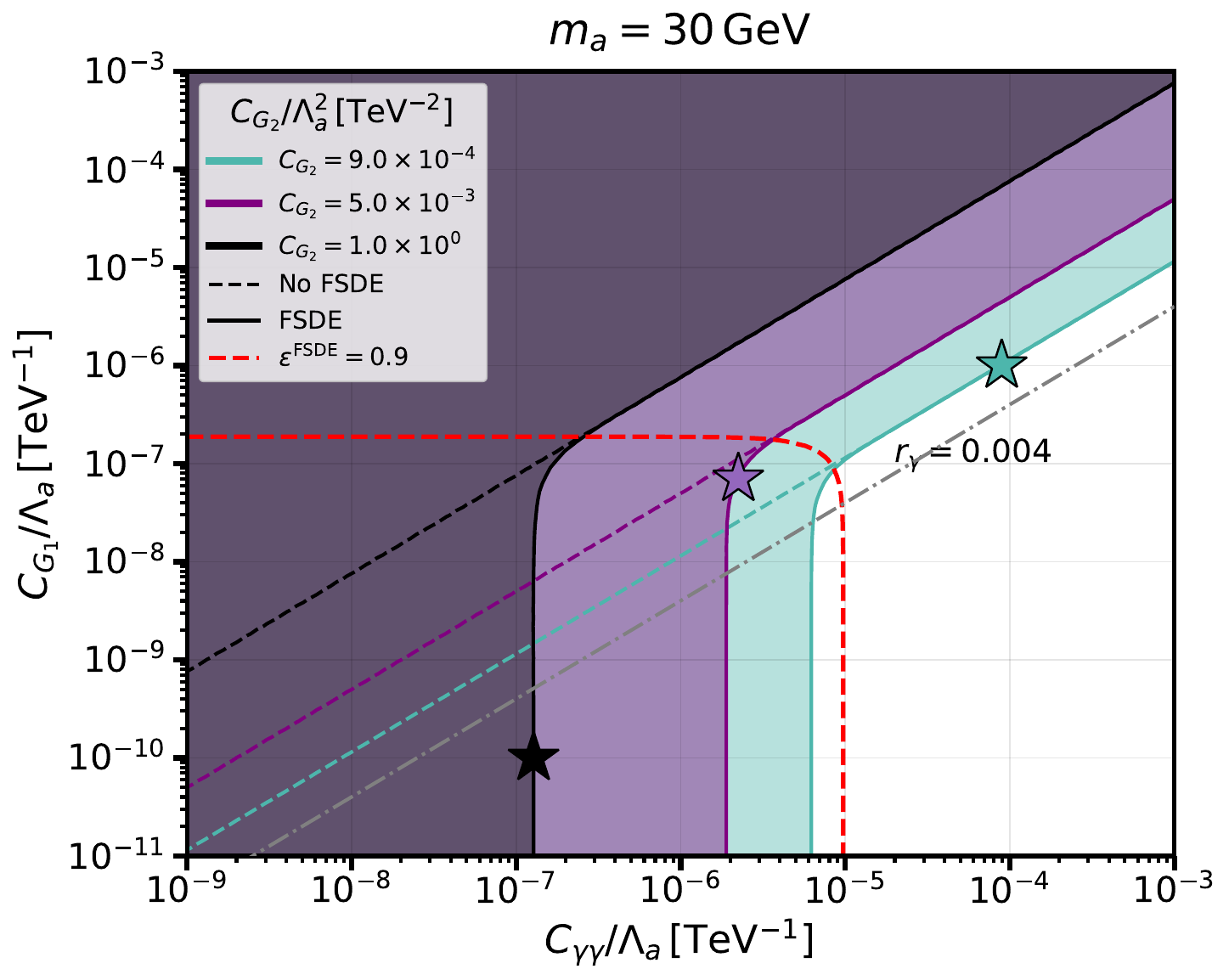}
\hfill
\includegraphics[width=0.48\textwidth]{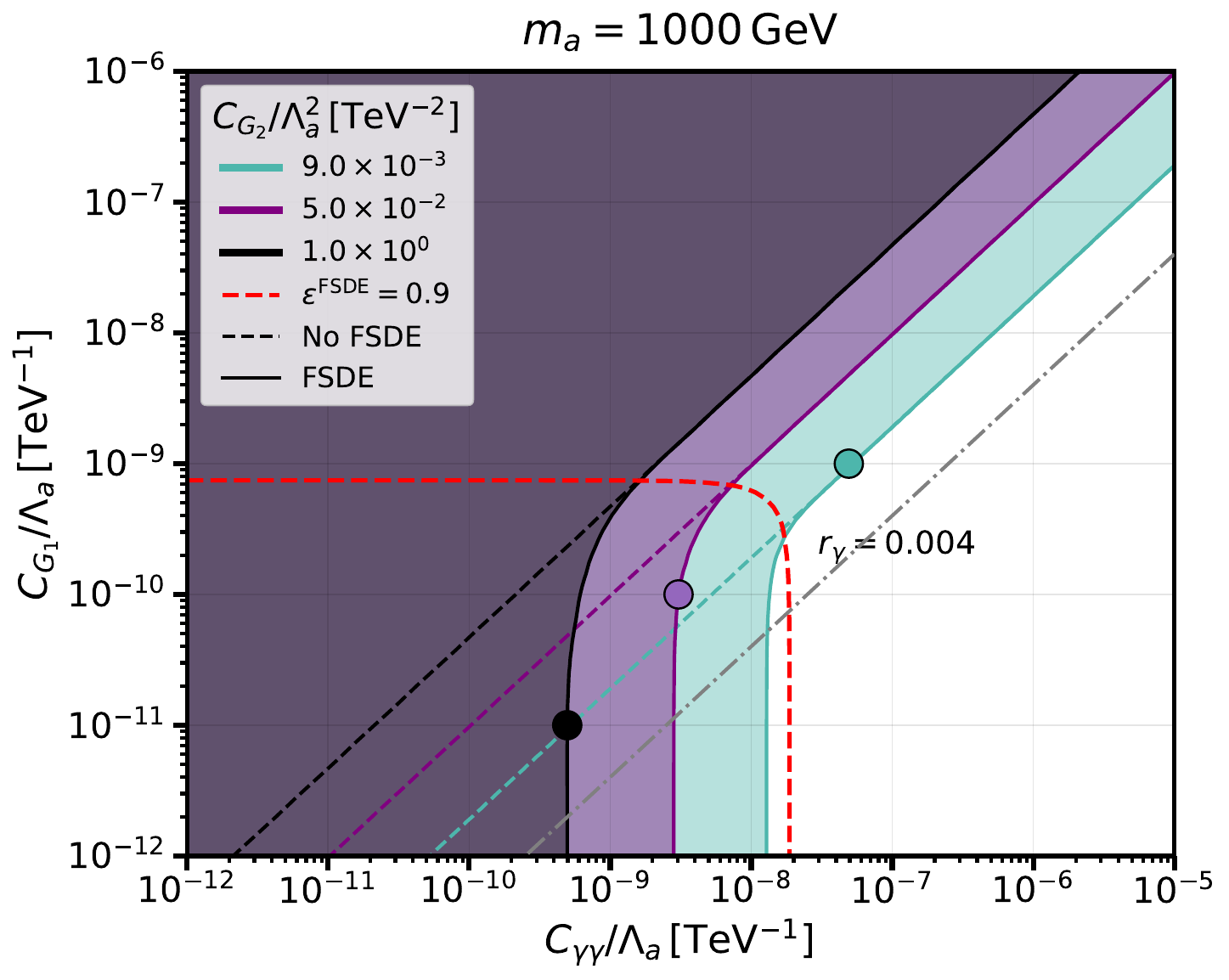}
\caption{Expected sensitivity of non-resonant $pp\to aa\to 4\gamma$ at the LHC with $\Lag=\unit[300]{fb^{-1}}$: the colored areas show the 95\%CL-allowed region in the ($\cgamma,\cgo)$ plane for fixed values of $\cgt$ and of $m_a=\unit[30]{GeV}$ (left) or $m_a=\unit[1]{TeV}$ (right).  The upper and lower panels differ in the $\cgamma$ and $\cgo$ ranges shown. In the upper panels, solid (dotted) boundaries are derived for positive (negative) values of $\cgt$. In the lower panels, solid (dashed) boundaries are derived accounting for (neglecting) FSDE. The red dashed line marks the region where FSDE become relevant ($\epsilon^{\rm FSDE}=0.9$). The gray dot-dashed line marks the transition point beyond which $\Br(a\to\g\g)$ saturates to its maximum allowed value. The star and bullet markers correspond to the benchmark points listed in Tab.~\ref{tab.stars}. Within the gray shaded area in the upper-right panel, the narrow width approximation for the ALP fails, as $\Gamma_a^{\rm tot}/m_a>10\%$.
}
\label{fig:cg1vscgamma}
\end{figure}

\begin{figure}[t!]
\centering
\includegraphics[width=0.49\textwidth]{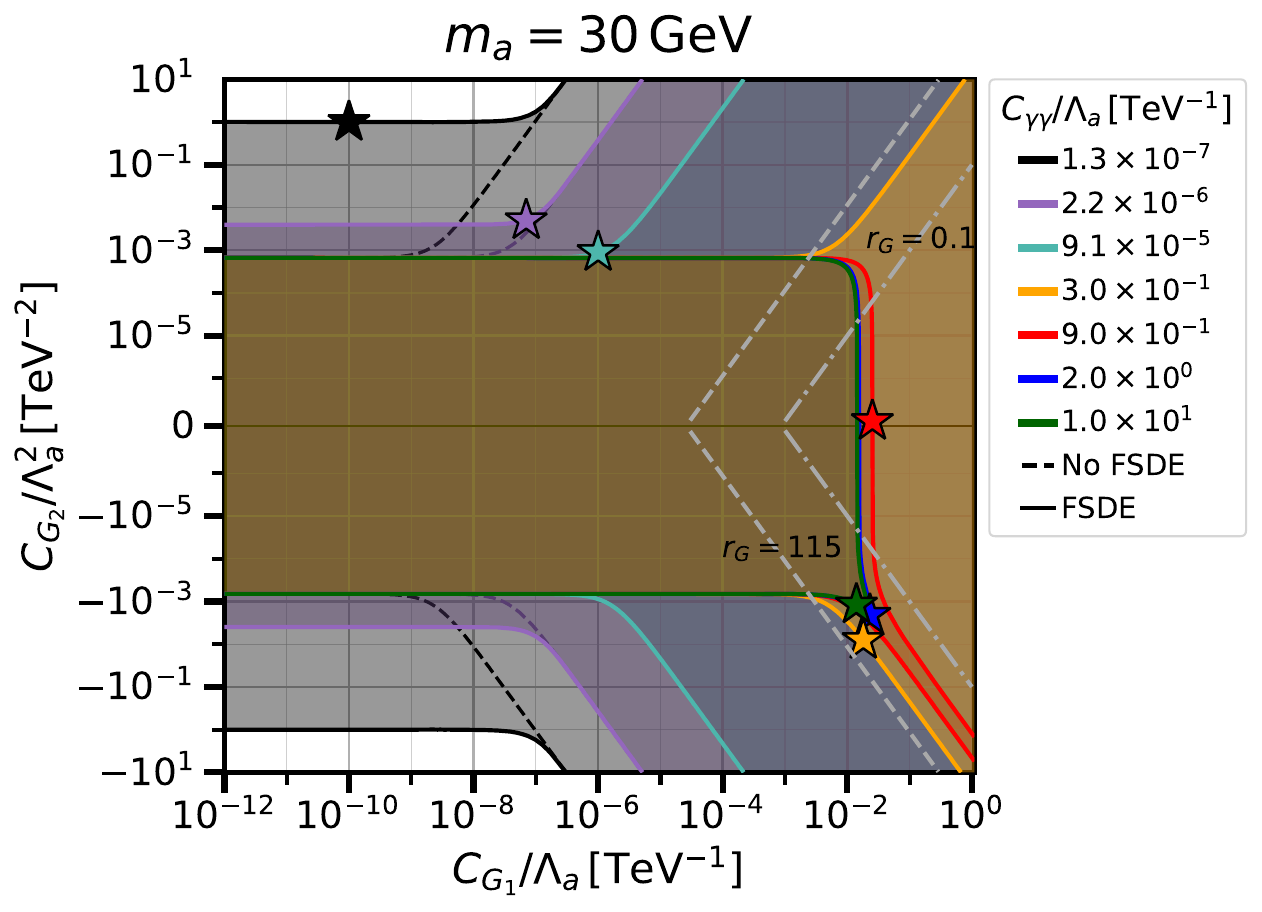}
\hfill
\includegraphics[width=0.49\textwidth]{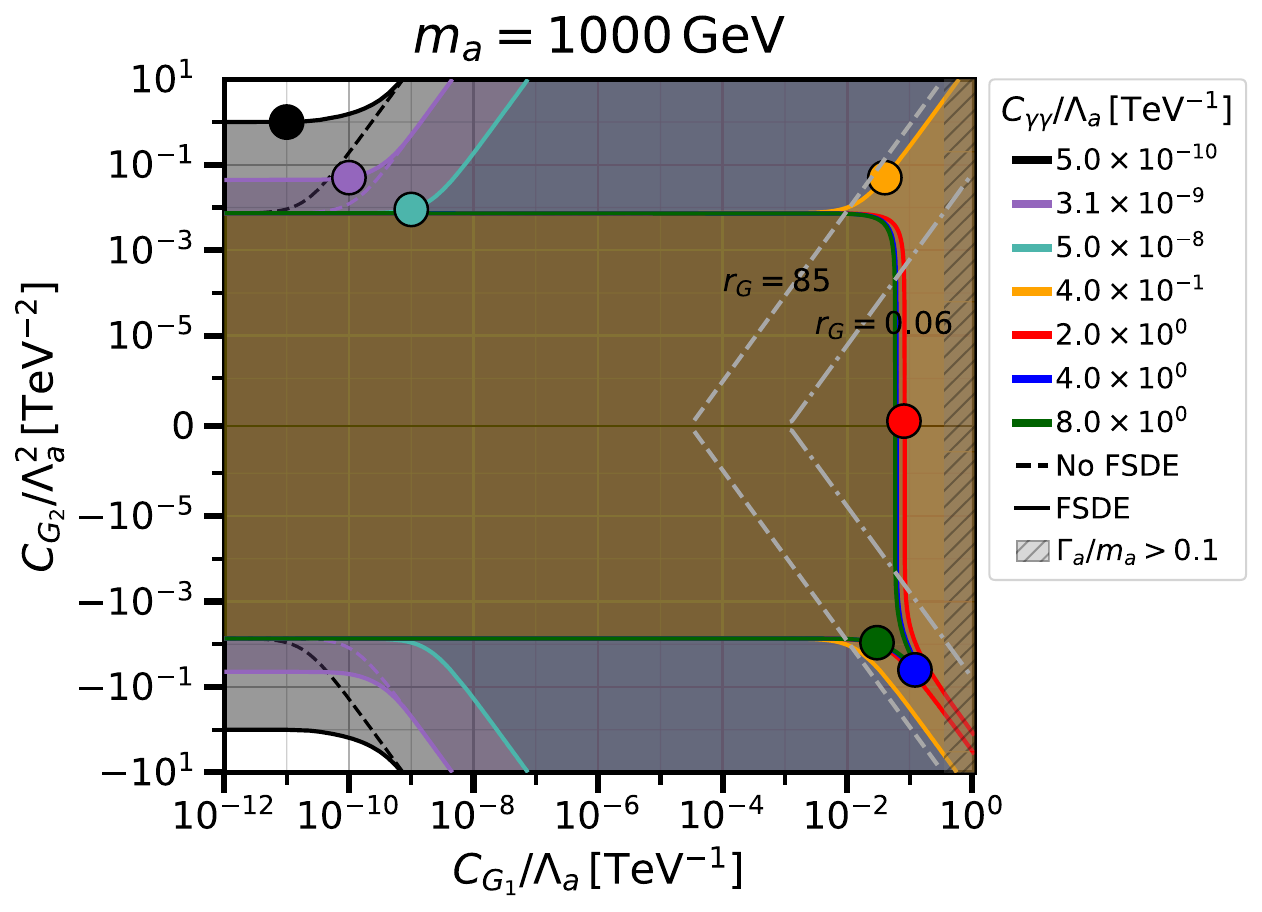}
\\
\includegraphics[width=0.49\textwidth]{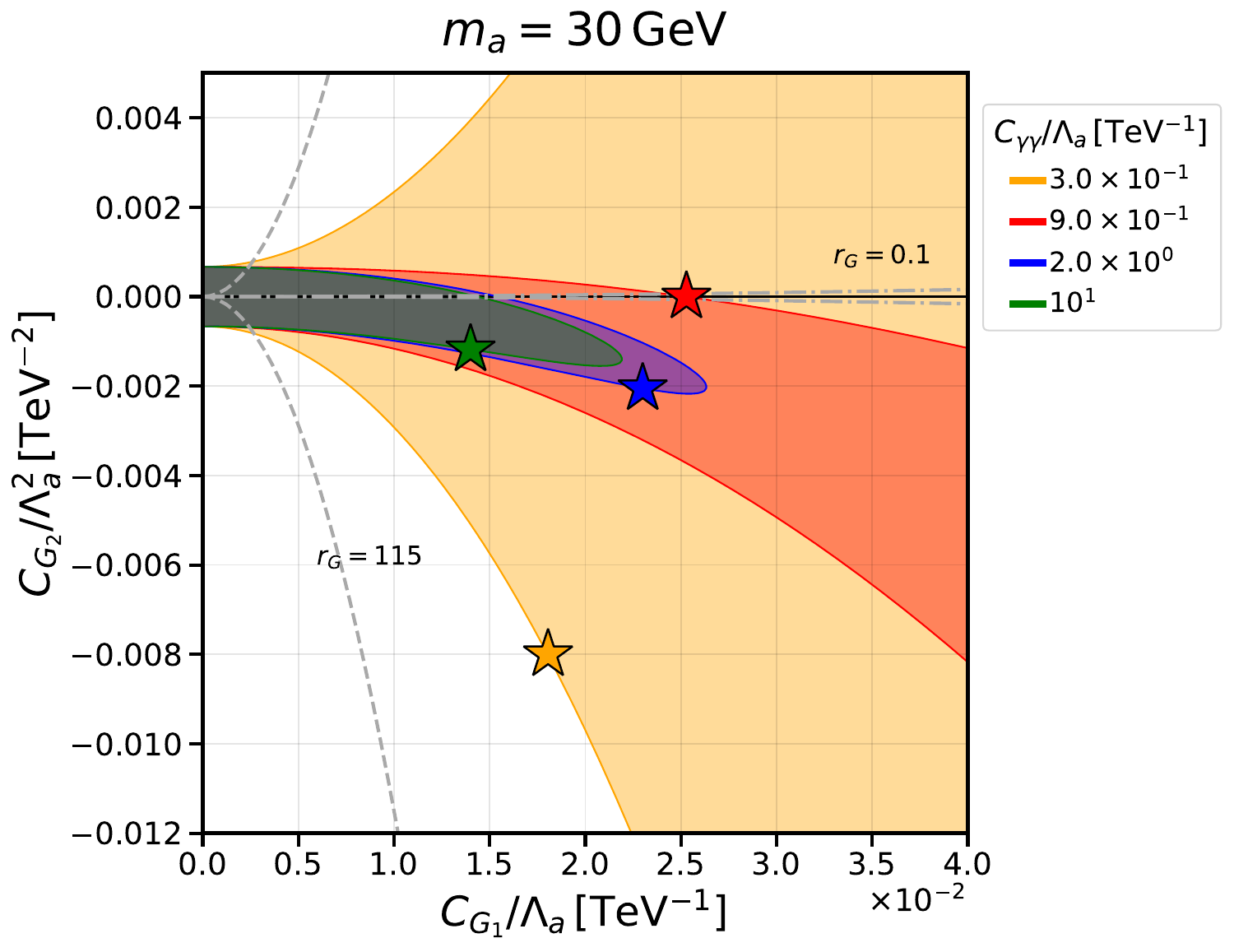}
\hfill
\includegraphics[width=0.49\textwidth]{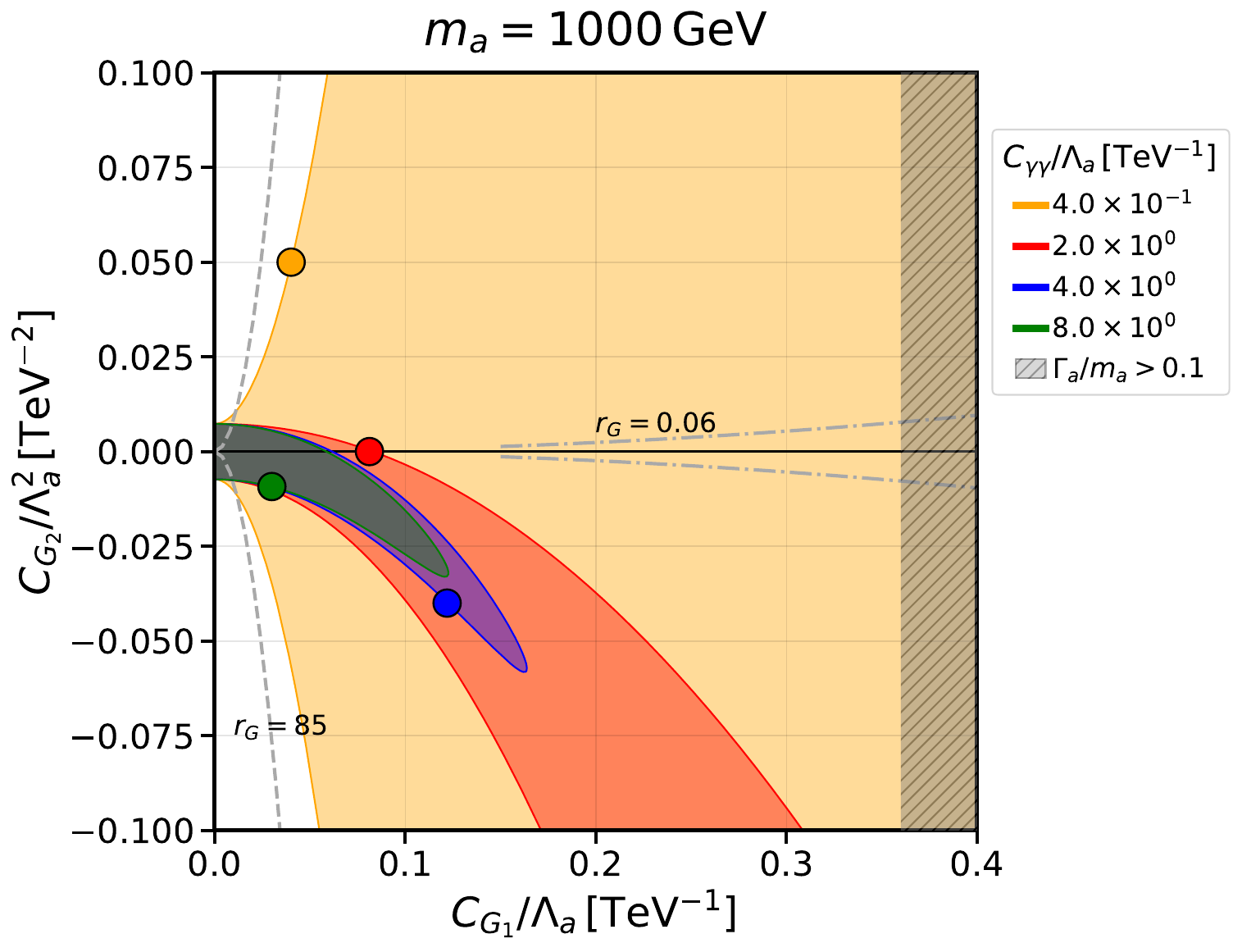}
\caption{
Expected sensitivity of non-resonant $pp\to aa\to 4\gamma$ at the LHC with $\Lag=\unit[300]{fb^{-1}}$: the colored areas show the 95\%CL-allowed region in the ($\cgo,\cgt)$ plane for fixed values of $\cgamma$ and of $m_a=\unit[30]{GeV}$ (left) or $m_a=\unit[1]{TeV}$ (right).  Upper (lower) panels show the same constraints in logarithmic (linear) scale.  Solid (dashed) boundaries are derived accounting for (neglecting) FSDE. The gray dashed and dot-dashed lines mark the transition points between $\cgo$- and $\cgt$-dominance regimes in the production cross section.  The star and bullet markers correspond to the benchmark points listed in Tab.~\ref{tab.stars}. Within the gray shaded area in the right panels, the narrow width approximation for the ALP fails, as $\Gamma_a^{\rm tot}/m_a>10\%$.
}
\label{fig:cg2vscg1}
\end{figure}

\begin{figure}[t!]
\centering
\includegraphics[width=0.48\textwidth]{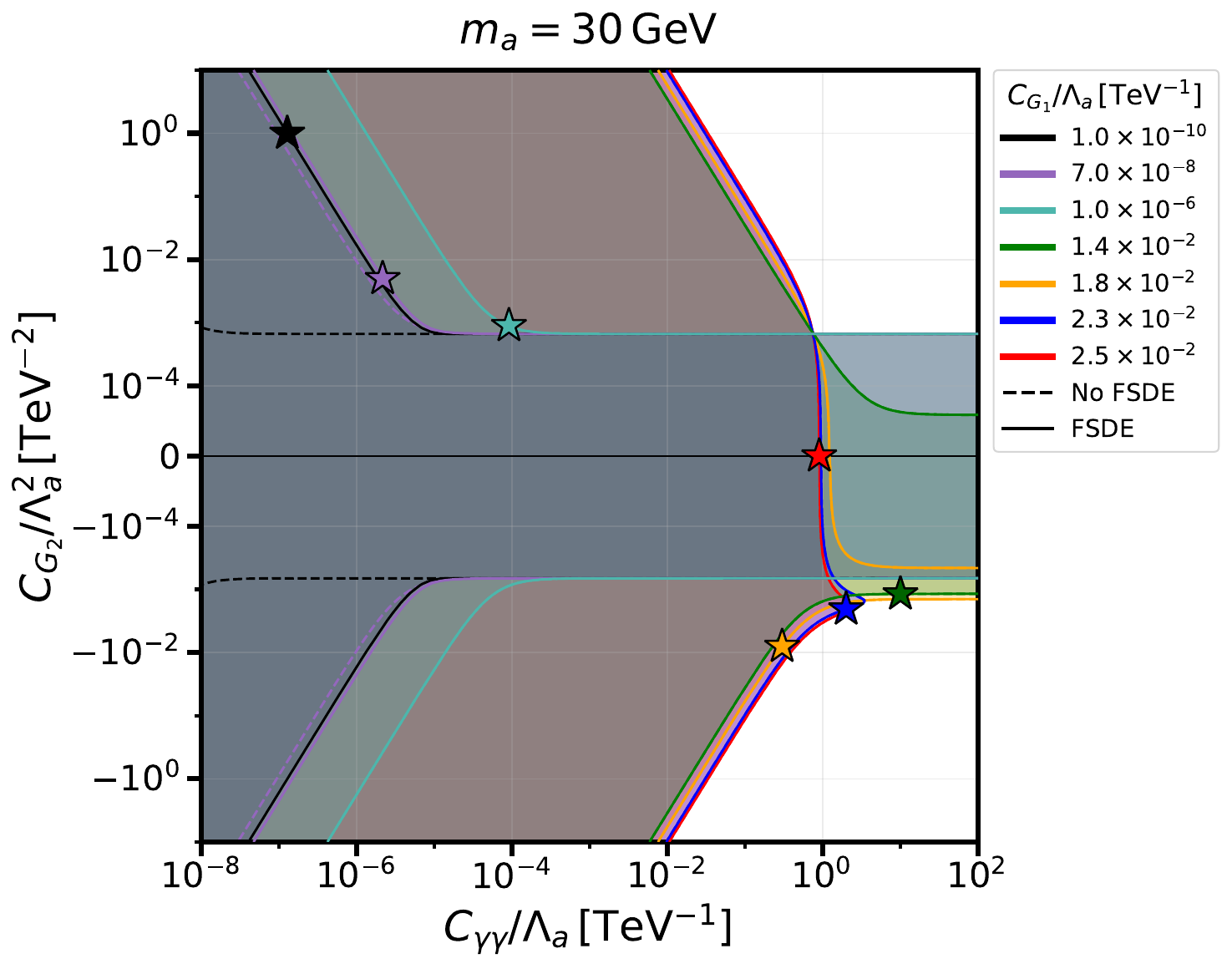}
\hspace{0.3cm}
\includegraphics[width=0.48\textwidth]{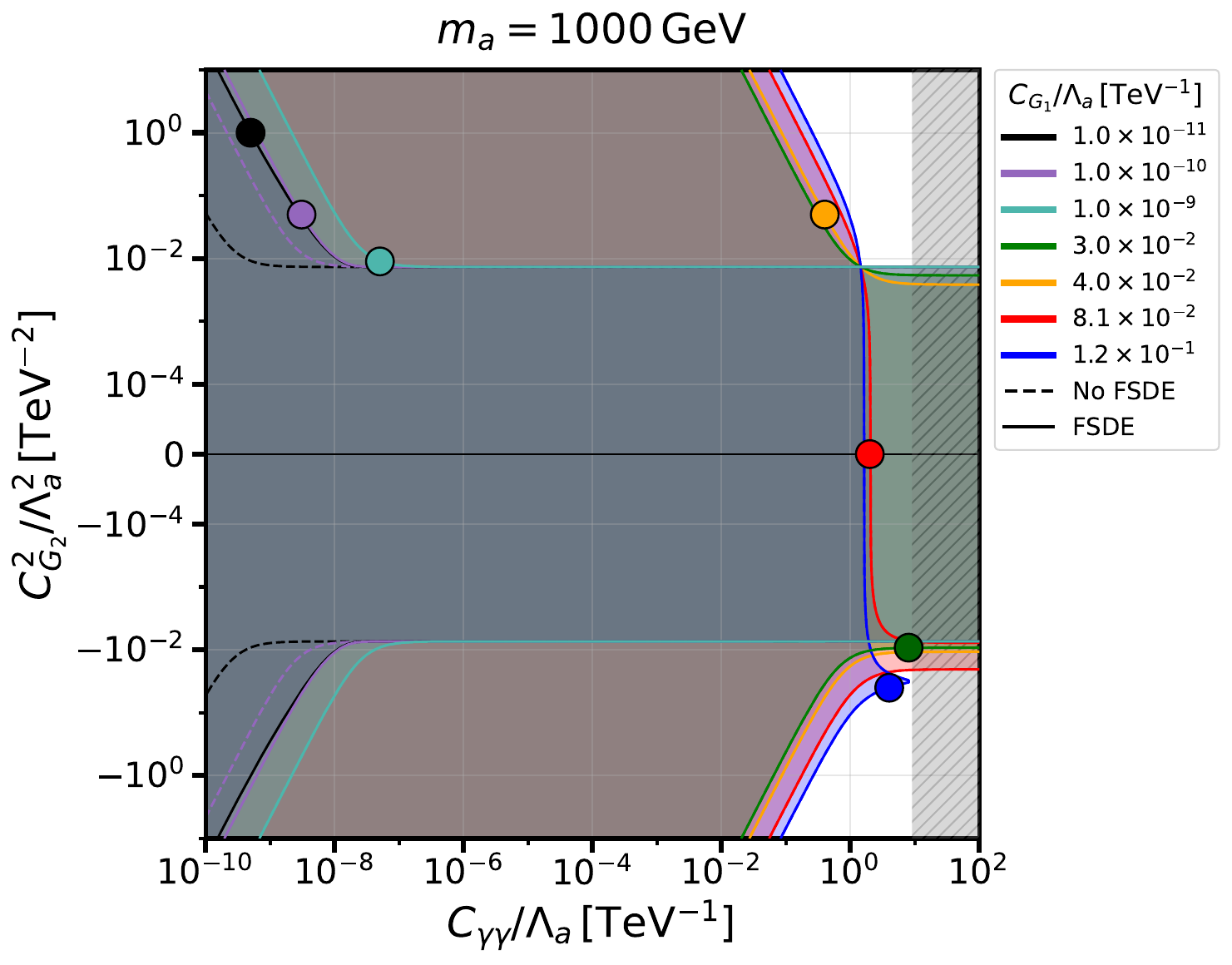}
\\
\includegraphics[width=0.48\textwidth]{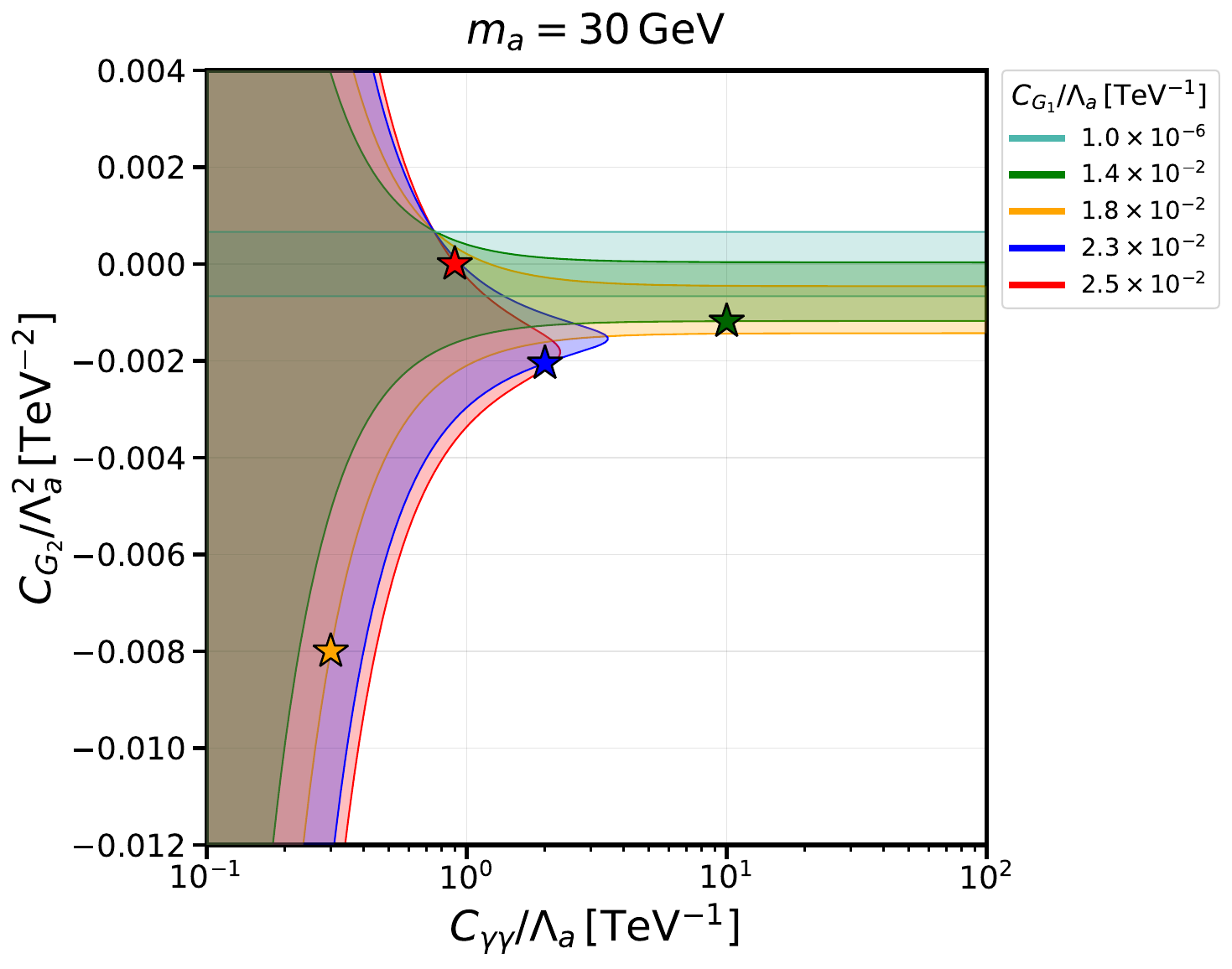}
\hfill
\includegraphics[width=0.48\textwidth]{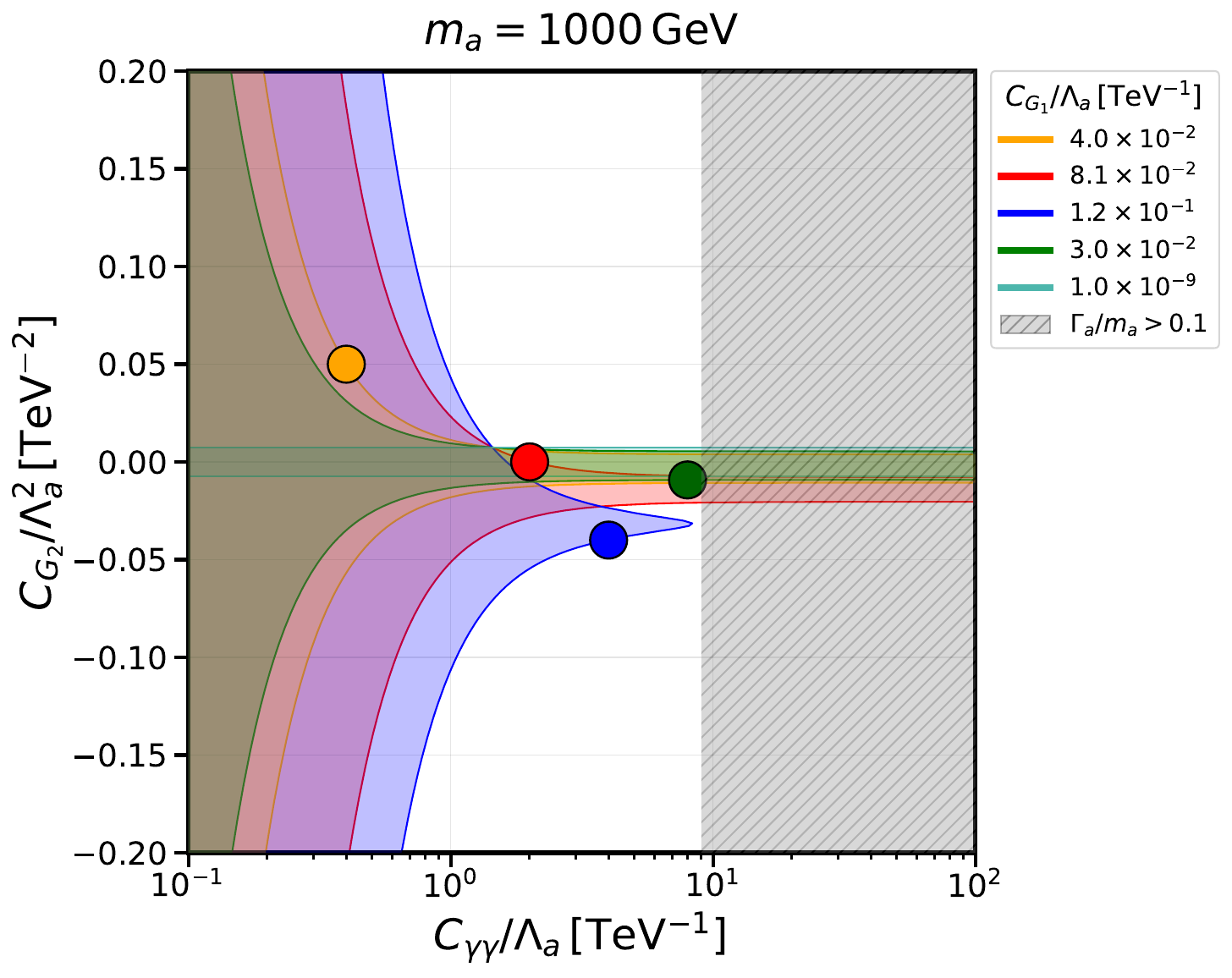}
\caption{
Expected sensitivity of non-resonant $pp\to aa\to 4\gamma$ at the LHC with $\Lag=\unit[300]{fb^{-1}}$: the colored areas show the 95\%CL-allowed region in the ($\cgamma,\cgt)$ plane for fixed values of $\cgo$ and of $m_a=\unit[30]{GeV}$ (left) or $m_a=\unit[1]{TeV}$ (right).  Upper (lower) panels show the same constraints in log-log (log-linear) scale. Solid (dashed) boundaries are derived accounting for (neglecting) FSDE.  The star and bullet markers correspond to the benchmark points listed in Tab.~\ref{tab.stars}. Within the gray shaded area in the right panels, the narrow width approximation for the ALP fails, as $\Gamma_a^{\rm tot}/m_a>10\%$.
}
\label{fig:cg2vscgamma}
\end{figure}

Figs.~\ref{fig:cg1vscgamma}, \ref{fig:cg2vscg1} and \ref{fig:cg2vscgamma} show two-dimensional slices of the 95\%CL-allowed parameter space, showing the bounds on a pair of coefficients among $\cgo,\cgt$ and $\cgamma$, obtained for fixed values of the third. Specifically, Fig.~\ref{fig:cg1vscgamma} shows 95\%CL-allowed regions in the $(\cgamma,\cgo)$ plane for fixed values of $\cgt$, while Figs.~\ref{fig:cg2vscg1} and \ref{fig:cg2vscgamma} show, respectively, allowed regions in the $(\cgo,\cgt)$ plane for fixed values of $\cgamma$ and allowed regions in the $(\cgamma,\cgt)$ plane for fixed values of $\cgo$. The results are shown for two representative values of the ALP mass, $m_a=\unit[30]{GeV}$ and $m_a=\unit[1]{TeV}$, which lie at the two extremes of the spectrum examined in this work.  We have verified that, for intermediate masses, the results are qualitatively similar, with the allowed ranges interpolating smoothly between the two cases shown.

The allowed regions display various non-trivial shapes, which are due to the interplay between a number of physical effects occurring in different regions of parameter space.  A general conclusion is that a double-ALP measurement alone is not sufficient to close the allowed region: arbitrarily large values of $\cgo,\cgt$ and $\cgamma$ remain allowed along specific directions. 
The main features and relevant phenomena at play are described in the next paragraphs, while general conclusions about the sensitivity of non-resonant double-ALP searches are deferred to Sec.~\ref{sec:discussion}.

To guide the reader's eye, we fix a number of benchmark points lying on the $N_{\rm signal}=3$ hypersurface, whose coordinates are listed in Table~\ref{tab.stars}. In Figs.~\ref{fig:cg1vscgamma}--\ref{fig:cg2vscgamma}, these points are indicated by star ($m_a=\unit[30]{GeV}$) and bullet ($m_a=\unit[1]{TeV}$) markers, and the contours of the corresponding colors represent the allowed region obtained keeping one of the benchmark coordinates fixed while varying the remaining two. We also show a gray hatched region corresponding to $\Gamma_a^{\rm tot}/m_a \geq 10\%$, to highlight where the NWA adopted for the ALP may no longer be reliable. 

\begin{table}[t]
\centering
\renewcommand{\arraystretch}{1.3}
\footnotesize
\begin{tabular}{|c|*3{>{$}r<{$}}|*3{>{$}r<{$}}|}
\hline
&
\multicolumn{3}{c|}{$m_a=\unit[30]{GeV}$}&
\multicolumn{3}{c|}{$m_a=\unit[1]{TeV}$}
\\\hline
\\[-4.5mm]
&
\dfrac{\cgo}{\La}\,[\unit{TeV^{-1}}]&
\dfrac{\cgt}{\La^2}\,[\unit{TeV^{-2}}]&
\dfrac{\cgamma}{\La}\,[\unit{TeV^{-1}}]&
\dfrac{\cgo}{\La}\,[\unit{TeV^{-1}}]&
\dfrac{\cgt}{\La^2}\,[\unit{TeV^{-2}}]&
\dfrac{\cgamma}{\La}\,[\unit{TeV^{-1}}]
\\[2mm]\hline

\orange &
1.81\times10^{-2} & -8.0\times10^{-3} & 3.0\times10^{-1} &
4.01\times10^{-2} & 5.0\times10^{-2} & 4.0\times10^{-1}
\\

\red &
2.53\times10^{-2} & 1.0\times10^{-7} & 9.0\times10^{-1} &
8.12\times10^{-2} & 1.0\times10^{-7} & 2.0
\\

\blue &
2.30\times10^{-2} & -2.05\times10^{-3} & 2.0 &
1.22\times10^{-1} & -4.0\times10^{-2} & 4.0
\\

\green &
1.40\times10^{-2} & -1.18\times10^{-3} & 10.0 &
3.0\times10^{-2} & -9.23\times10^{-3} & 8.0
\\

\cyan &
1.0\times10^{-6} & 9.0\times10^{-4} & 9.12\times10^{-5} &
1.0\times10^{-9} & 9.0\times10^{-3} & 5.04\times10^{-8}
\\

\purple &
7.0\times10^{-8} & 5.0\times10^{-3} & 2.16\times10^{-6} &
1.0\times10^{-10} & 5.0\times10^{-2} & 3.07\times10^{-9}
\\

black &
1.0\times10^{-10} & 1.0 & 1.28\times10^{-7} &
1.0\times10^{-11} & 1.0 & 4.97\times10^{-10}
\\\hline

\end{tabular}
\caption{Benchmark ALP couplings corresponding to the starred and dotted points shown in Figs.~\ref{fig:cg1vscgamma}--\ref{fig:cg2vscgamma}, for $m_a=30~\mathrm{GeV}$ and $m_a=1~\mathrm{TeV}$. }
\label{tab.stars}
\end{table}

\paragraph{Relevant physical regimes.}
Before examining the results presented in Figs.~\ref{fig:cg1vscgamma}--\ref{fig:cg2vscgamma}, it is useful to summarize the expected behavior of the main physical quantities in specific regions of the parameter space.
As established in Sec.~\ref{sec:prod}, the production cross section is dominated by the $\cgo$ contribution when
\begin{align}
|r_G| = |\cgt\/\cgo^2&\lesssim 0.1\, (0.06)
&
\text{for } m_a&=\unit[30]{GeV}\, (\unit[1]{TeV}) \,.
\label{eq.cg1_dominance}
\end{align}
This relation can be equivalently written as $\cgo\gtrsim 3\, (4)\, |\cgt|^{1/2}$ or $|\cgt|\lesssim 0.1\, (0.06)\, \cgo^2$. Within this regime, which we dub ``$\cgo$ dominance'', $\sigma^{\rm cuts}\simeq \sA^{\rm cuts}\, \cgo^4$ is approximately independent of~$\cgt$.
We have, instead, a ``$\cgt$ dominance'' regime when
\begin{align}
|r_G| = \cgt/\cgo^2&\gtrsim 115\, (85)
&
\text{for } m_a&=\unit[30]{GeV}\, (\unit[1]{TeV}) \,.
\label{eq.cg2_dominance}
\end{align}
This condition can be cast equivalently as  $\cgo\lesssim 0.09\, (0.1)\, |\cgt|^{1/2}$ or $|\cgt|\gtrsim 115\, (85)\, \cgo^2$. In this region, $\sigma^{\rm cuts}\simeq \sB^{\rm cuts}\, \cgt^2$ and the cross section is independent of $\cgo$.
\\
For values of $\cgo$ and $\cgt$ that lie in the intermediate regime, \eg \ $0.1\lesssim |r_G|\lesssim 115$ for $m_a=\unit[30]{GeV}$,  the production cross section is sensitive to both $\cgo$ and $\cgt$ and interference effects are relevant.

As discussed in  Sec.~\ref{sec:decay}, the parametric dependence of the branching ratio $\Br(a\to \gamma\gamma)$ defines two main physical regimes. In particular, if
\begin{align}
  r_\gamma = \cgo/\cgamma &\lesssim 0.004\, ,
 \label{eq.BR_saturation}  
\end{align}
which is equivalent to  $ \cgamma\gtrsim 250 \,\cgo$ or  $\cgo\lesssim 0.004\, \cgamma$, 
the branching ratio saturates to its maximum allowed value,
becoming essentially independent of both $\cgamma$ and $\cgo$. Otherwise, away from this region, it scales as $\Br(a\to\gamma\gamma)\sim r_\gamma^{-2}=\cgamma^2/\cgo^2$.

\paragraph{Bounds in the ($\cgamma,\cgo)$ plane for fixed $\cgt$.}
We are now ready to examine the main results of this section. 
We start by discussing the upper panels of Fig.~\ref{fig:cg1vscgamma}. Within the ranges shown here, FSDE are irrelevant, \ie\ $\sigma^{\rm FSDE,cuts}\simeq \sigma^{\rm cuts}$.
The allowed regions depicted in these figures have boundaries along three main directions: vertical, horizontal and diagonal, which join in transition regions where shapes are more varied.
To understand the origin of these features, it is useful to note that, in this plane, the region where the production cross section is dominated by $\cgo$ lies in the upper part of the plot, while the $\cgt$-dominated region is at the bottom; the intermediate band corresponds to the regime where interference effects are relevant. At the same time, the branching ratio $\Br(a\to\gamma\gamma)$ increases in a direction orthogonal to the diagonal lines shown in the figure, from the upper left to the lower right, and saturates (\ie\ reaches its maximum) to the right of the gray dot-dashed line.

With this in mind,  we can see that:
\begin{itemize}

\item 
in the region where $\Br(a\to\g\g)$ saturates (to the right of the gray dot-dashed line) $N_{\rm signal}$ is independent of $\cgamma$ and the $N_{\rm signal}\leq3$ constraint is equivalent to an upper bound on $\sigma^{\rm cuts}$. If $|\cgt|$ is fixed to a large value, this bound is automatically violated and the region is excluded (the \orange, \blue\ curves remain outside this region).  If on the other hand $|\cgt|$ is below a certain threshold, the bound translates into an upper limit on $\cgo$, \ie\ the allowed region has a \emph{horizontal boundary} (\red\ curve). 
This can only occur away from $\cgt$-dominance, and, as a consequence, the position of the horizontal limit has a negligible dependence on $\cgt$.

\item in the region above the gray dot-dashed line, $\Br(a\to\g\g)\sim r_\gamma^{-2}$, $N_{\rm signal}$ is sensitive to $\cgamma$ and we can observe different behaviors depending on the value of $\cgo$.

Where $\sigma^{\rm cuts}$ is dominated by $\cgt$ (lower part of the plot) the production cross section is constant, and the constraint $N_{\rm signal}\leq 3$ is equivalent to an upper bound on $\Br(a\to\g\g)$. If $|\cgt|$ is too small, the cross section is too low to produce 3 events even with maximum branching ratio, and this region is automatically allowed (\red\ curve). If $|\cgt|$ is above a certain threshold, instead, the bound translates into a lower limit on $r_\gamma$ and the allowed region has a \emph{diagonal boundary} (\orange, \blue, \green\ curves). 
The position of this boundary depends on the value of $\cgt$, because for lower $|\cgt|$, the cross section is smaller, which allows a larger branching ratio.

\item in the region above the gray dot-dashed line where $\sigma^{\rm cuts}$ is dominated by $\cgo$ (top part of the plot), the constraint $N_{\rm signal}\leq 3$ affects both $\sigma^{\rm cuts}$ and $\Br(a\to\g\g)$ simultaneously. Interestingly, in the limit of large $\cgo$,  $\sigma^{\rm cuts}\sim \cgo^4$ and $\Br^2(a\to\g\g)\sim \cgo^{-4}$, so the dependence on $\cgo$ cancels in the product, leaving no sensitivity to this parameter and a boundary along the \emph{vertical direction}. 
The position of the vertical bound is independent of the value of $\cgt$, because, in this limit, the latter gives a negligible contribution to the production cross section ($\cgo$ dominance).
  
\item in the region above the gray dot-dashed line where both $\cgo$ and $\cgt$ are relevant,  we have a transition region in which the bounds take different shapes.

For $\cgt>0$, we simply have a smooth transition curve between the diagonal and vertical boundaries. For $\cgt<0$, the boundary of the allowed region can have a cusp extending to higher values of $\cgamma$ (\orange, \blue\ dotted lines). This happens because the large negative interference lowers the production cross section, which hits a minimum in this region as shown in Fig.~\ref{fig:sigma_p_regimes}. Correspondingly, a larger $\Br(a\to\g\g)$, and therefore $\cgamma$, is allowed. 
There exists a window of negative $\cgt$ values such that, within the interference region, the cross section drops below the minimum required to reach $N_{\rm signal}=3$ with the maximum branching ratio. This effect generates a thin horizontal band that extends to arbitrarily large $\cgamma$ (\green\ dotted line).
\end{itemize}

The diagonal boundaries appearing in the upper panels of Fig.~\ref{fig:cg1vscgamma} may give the misleading impression that arbitrarily strong bounds could be set on $\cgamma$, by choosing a sufficiently small value of $\cgo$. This is not the case because, moving towards smaller $\cgo$ and $\cgamma$, one eventually reaches the region where the ALP becomes long-lived. This phenomenon is captured in the lower panels of Fig.~\ref{fig:cg1vscgamma}, which focus on the region of the ($\cgamma,\cgo$) plane where FSDE become relevant.
To highlight their effect, the figure shows the bounds obtained with and without accounting for FSDE (respectively solid and dashed lines). We also show for reference the same gray dot-dashed line as in the upper panel, marking the point beyond which $\Br(a\to\g\g)$ saturates to the maximum value, and, in red, the curve marking the values of $\cgamma$ and $\cgo$ for which FSDE reduce the event yield by 10\%, which was already shown in Fig.~\ref{fig:FSDE_less_90_ma_s}.

FSDE contribute significantly to shaping the allowed region only when $\cgo$ is tiny and the production cross section is large enough for $N_{\rm signal}$ to remain close to 3 for $\epsilon_{\rm FSDE}<1$, that is: $\cgt$ must be dominating $\sigma^{\rm cuts}$ and it must be above a certain threshold. If $\cgt$ were too small, it would be impossible to reach 3 events and all the region shown would become allowed. Then we have that the production cross section is essentially fixed to $\cgt^2\sB^{\rm cuts}$ (which is independent of the sign of $\cgt$) and the $N_{\rm signal}\leq 3$ constraint acts as an upper bound on the product $\epsilon_{\rm FSDE}\Br(a\to\g\g)^2$. As noted in Sec.~\ref{sec:FSDE}, in the limit of small $\Gamma_a^{\rm tot}$ (\ie\ small $\cgo$ and $\cgamma$ in the figure), $\epsilon_{\rm FSDE}$ scales with $(\Gamma_a^{\rm tot})^2$, so the dependence on the total width actually cancels in the product and the upper bound effectively acts on $\Gamma_{a\to\g\g}^2\sim \cgamma^4$, yielding a vertical boundary that extends to arbitrarily small $\cgo$ (\cyan, \purple\ and black curves). In practice, FSDE reduce the expected sensitivity compared to the ideal case where all ALPs decay within the detector. In our case, this limitation translates into the impossibility of probing values of $\cgamma$ below a certain threshold. 

Comparing left {\it vs.}~right panels in Fig.~\ref{fig:cg1vscgamma}, we find that the bounds obtained are weaker for heavier ALP masses. This is expected because, for the same $\cgo$ and $\cgt$, the production cross section is significantly lower at large $m_a$. However, the limits on $\cgamma$ that are controlled by FSDE are stronger for heavier ALPs. This happens because of two factors adding up in the same direction: for fixed $\cgo$ and $\cgamma$, $\Gamma_a^{\rm tot}$ is much larger for heavy ALPs, as $\Gamma_a^{\rm tot}\sim m_a^3$. At the same time, for fixed $\Gamma_{a}^{\rm tot}$, the FSDE correction is less significant at larger $m_a$ (see Fig.~\ref{fig:FSDE_less_90_ma_s}, right), due to the different $p_{T,a}$ distribution. These two factors together overcompensate for the reduced production cross section, leaving a stronger limit on $\cgamma$, for $\cgt$ relatively large and $\cgo\to 0$.

The origin of the upper bounds on $\cgamma$ in this region of parameter space is very similar to that of the bounds extracted from the Higgs-resonant channel in Refs.~\cite{Bauer:2017ris,ATLAS:2023ian}. As will be discussed in Sec.~\ref{sec:resresults}, both derivations assume a large ALP production cross section, controlled by a dimension-6 operator, and that the ALP decay width is dominated by the $\g\g$ channel. The ALP lifetime, controlled by $\cgamma$, is then required to be long enough to suppress the number of detectable signal events below the observation threshold. This strategy leads to very strong constraints on $\cgamma$, at the cost of selecting a region of parameter space in which the validity of the EFT expansion is called into question, as 
dimension-6 Wilson coefficients are orders of magnitude larger than the dimension-5 ones.
We will return to this point in Sec.~\ref{sec:discussion}.

\paragraph{Bounds in the ($\cgo,\cgt)$ plane.}
Fig.~\ref{fig:cg2vscg1} shows the allowed regions for fixed values of $\cgamma$. Our discussion will focus on the upper panels, which use a logarithmic scale. 
The lower panels show the same constraints in linear scale, retaining only the curves that can be meaningfully resolved. 

In the ($\cgo,\cgt)$ plane, we can easily visualize the behavior of the production cross section: we have $\cgt$ dominance to the left of the gray dashed line. Here $\sigma^{\rm cuts}\sim \cgt^2$ varies along the vertical direction. To the right of the gray dot-dashed line we have $\cgo$ dominance, and $\sigma^{\rm cuts}\sim \cgo^4$ varies along the horizontal direction. In the region in between, both coefficients compete and interference effects are relevant: for $\cgt>0$, curves of constant $\sigma^{\rm cuts}$ transition smoothly from horizontal to vertical, while for $\cgt<0$ they develop an elongated shape due to the negative interference (see also Fig.~\ref{fig:sigma_p_contours}). The branching ratio $\Br(a\to \g\g)$ varies along the horizontal direction: in the left-most region of the plot, it saturates to its maximum value, while moving towards larger $\cgo$ it drops towards zero as $\cgo^{-2}$. The value of $\cgo$ at which the transition between these regimes takes place depends on $\cgamma$. Finally, FSDE play a relevant role only for very small $\cgamma$ and in the corners where $\cgo$ is very small but $\cgt$ is large. Bounds obtained with and without accounting for FSDE are shown as solid and dashed lines respectively. 

The various shapes taken by the allowed region can be understood as follows:
\begin{itemize}
\item in the left-most region, where $\cgo$ is tiny, $\Br(a\to\g\g)$ is saturated, and therefore the constraint $N_{\rm signal}\leq 3$ acts as an upper bound on the product $\sigma^{\rm cuts}\epsilon_{\rm FSDE}$. At the same time, $\sigma^{\rm cuts}$ is dominated by $\cgt$. Then, if $\cgamma$ is large enough that $\epsilon_{\rm FSDE}\simeq 1$, we find an upper bound on $|\cgt|$ that is independent of both $\cgamma$ and $\cgo$ (\orange, \blue, \red, \green, \cyan\ overlapping, horizontal curves). 

If $\cgamma$ is very small, this bound weakens due to $\epsilon_{\rm FSDE}<1$. 
In the limit of tiny $\cgo$, the dependence on $\Gamma_a^{\rm tot}$ cancels in the product $\epsilon_{\rm FSDE}\Br(a\to\g\g)^2\sim \Gamma_{a\to\g\g}^2$, so the upper bound on $|\cgt|$ remains independent of $\cgo$, but its position depends on $\cgamma$ (\purple\ and black horizontal curves). 

\item moving towards larger values of $\cgo$ while staying to the left of the gray dashed line,  the approximations $\epsilon_{\rm FSDE}\simeq 1$ and $\sigma^{\rm cuts}\simeq \cgt^2\sB^{\rm cuts}$ remain valid. However, for small $\cgamma$ (\cyan, \purple\ and black lines), we enter the regime where $\Br(a\to\g\g)\sim r_\gamma^{-2}$. Then the constraint $N_{\rm signal}\leq 3$ effectively acts on $\sigma^{\rm cuts}\Br(a\to\g\g)^2\sim \cgt^2 \cgamma^4/\cgo^4$ and, for fixed $\cgamma$, it translates into an upper bound on the ratio $r_G=\cgt/\cgo^2$. The boundary of the allowed region correspondingly transitions from being horizontal to diagonal (\cyan, \purple\ and black curves. This explanation applies to the \orange\ curve between the dashed and dot-dashed gray lines as well).

\item in the region to the right of the gray dot-dashed line,  $\sigma^{\rm cuts}\simeq \cgo^4\sA^{\rm cuts}$, so the bounds are independent of $\cgt$, and $\epsilon_{\rm FSDE}=1$. 
If $\cgamma$ is very small, $\Br(a\to\g\g)$ is too small to reach $N_{\rm signal}=3$ and this whole region is allowed (\cyan, \purple, black and \orange\ curves). 

If $\cgamma$ is very large, $\Br(a\to\g\g)$ is close to its maximum, and the $N_{\rm signal}\leq 3$ constraint effectively acts only on $\sigma^{\rm cuts}$, yielding an upper bound on $\cgo$  that is independent of $\cgamma$ (\green, \blue\ vertical curves). 
For intermediate values of $\cgamma$, the constraint affects both $\sigma^{\rm cuts}$ and $\Br(a\to\g\g)\lesssim 1$. As $\cgamma$ is fixed, this results into an upper bound on $\cgo$, whose position depends on the value of $\cgamma$ (\red\ vertical curve).

There exists a precise value of $\cgamma$ (located between the \red\ and \orange\ curves) that separates the regime in which the allowed region closes at $\cgt=0$ and the regime where it remains open. This value corresponds to the position of the vertical boundary in the upper panels of Fig.~\ref{fig:cg1vscgamma} and, in the linear scale plot (lower panels of Fig.~\ref{fig:cg2vscg1}), it would produce a band whose upper boundary remains exactly horizontal for $\cgo\to\infty$. 

\item finally, in the region where the interference between $\cgo$ and $\cgt$ diagrams is relevant, we see that, for $\cgt>0$, the bounds simply exhibit a smooth transition between the $\cgo$ and $\cgt$ dominance regimes, while for $\cgt<0$ the large negative interference can produce new effects. 
In the $(\cgo,\cgt)$ plane, $\sigma^{\rm cuts}$ is minimized along a curve of constant $r_G<0$, which cuts diagonally through the parameter space in log scale.
Moving along this curve towards larger $\cgo$, the minimum value of $\sigma^{\rm cuts}$ grows with $\cgo^4$.
If $\cgamma$ is large enough that $\Br(a\to\g\g)$ remains nearly saturated, the boundary of the allowed region has a small, finite cusp (\green\ and \blue\ curves). The values of $\cgt$ spanned by the cusp are those yielding either a cusp or a thin band extending to infinitely large $\cgamma$ in Fig.~\ref{fig:cg1vscgamma}.

As $\cgamma$ becomes smaller, larger values of $\cgo$ become allowed and the cusp elongates. 
Once the cusp reaches into large enough values of $\cgo$, it enters the region where $\Br(a\to\g\g)\sim r_\gamma^{-2}$. 
Since along the curve of constant $r_G$ that minimizes $\sigma^{\rm cuts}$ we have $\sigma^{\rm cuts}\sim \cgo^4$, in this region the dependence on $\cgo$ cancels out in the product $\sigma^{\rm cuts}\Br(a\to\g\g)^2$. As a result, $N_{\rm signal}$ remains approximately constant along the curve of minimum $\sigma^{\rm cuts}$ and the allowed region forms a thin band around it that extends up to arbitrarily large $\cgo,|\cgt|$ (\red\ curve).

\end{itemize}

Overall, we observe that the allowed region in the $(\cgo,\cgt)$ plane closes only in the limit of very large $\cgamma$. Moving towards smaller and smaller $\cgamma$, the allowed region initially elongates, develops a flat direction along the curve where $\sigma^{\rm cuts}$ is minimized by destructive interference, and eventually opens up yielding \emph{lower} bounds on $\cgo$.

\paragraph{Bounds in the ($\cgamma,\cgt)$ plane.}
Fig.~\ref{fig:cg2vscgamma} shows the allowed regions for fixed values of $\cgo$. The upper and lower panels present the same bounds in logarithmic and log-linear scale respectively. For definiteness, we only discuss the representation in the upper panels. 

Focusing on the $\cgt>0$ plane, the production cross section $\sigma^{\rm cuts}$ is dominated by $\cgt$ in the upper-most region and by $\cgo$ in the lower-most region, while in the intermediate band both coefficients are relevant and interference effects are visible. The same is true in the $\cgt<0$ plane, with to the appropriate mirroring.
The branching ratio $\Br(a\to\g\g)$ tends to saturate to its maximum value at the right-most end of the plot.
$\epsilon_{\rm FSDE}$ drops below 1 for tiny $\cgo$ and $\cgamma$,  and it has a relevant impact on the bounds for large $|\cgt|$, \ie\ at the upper- and lower-left corners of the plot.

The visible features in the ($\cgamma,\cgt)$ plane are:
\begin{itemize}

\item in the regions at the top and bottom of the figure, the production cross section is dominated by $\cgt$, and the $N_{\rm signal}\leq 3$ constraint acts on the product $\sigma^{\rm cuts}\Br(a\to\g\g)^2\sim \cgt^2/\cgamma^4$, resulting in the allowed region having a diagonal boundary (all curves). This occurs for any $\cgo$, provided that $\cgamma$ is sufficiently small for $\Br(a\to\g\g)$ to be away from the maximum.

At the left-most end of those bands, where $\cgamma$ is tiny, FSDE are relevant if $\cgo$ is also very small. Their net effect is a shift of the diagonal boundaries towards the right (black and \purple\ curves): this occurs because when $\epsilon_{\rm FSDE}<1$, a larger $\Br(a\to\g\g)$ is allowed. 

\item 
for values of $\cgamma\gtrsim 250 \cgo$, $\Br(a\to\g\g)$ saturates to its maximum, and the $N_{\rm signal}\leq 3$ constraint acts on $\sigma^{\rm cuts}$. 

If $\cgo$ is very small, $\sigma^{\rm cuts}\simeq \sB \cgt^2$ and we have an upper limit on $|\cgt|$ that extends to arbitrarily large $\cgamma$, \ie\ the allowed region becomes a horizontal band (black, \purple, \cyan\ curves). The boundary of this band lies at the same value of $\cgt$ as the horizontal boundary in Fig.~\ref{fig:cg2vscg1}. This position is independent of $\cgo$ because neither $\sigma^{\rm cuts}$ nor $\Br(a\to\g\g)$ are sensitive to it in this regime. The values of $\cgo$ for which this condition occurs are those below the horizontal boundary in Fig.~\ref{fig:cg1vscgamma}.

If $\cgo$ is very large, it dominates the cross section in the central region of the plot (small $|\cgt|$). In this limit, the dependence on $\cgo$ cancels in the product $\sigma^{\rm cuts}\Br(a\to\g\g)^2$, and as a result we find an upper bound on $\cgamma$ (\ie\ a vertical boundary) that is independent of $\cgo$ (\blue, \red, \orange\ curves). This vertical bound is the same one found in Fig.~\ref{fig:cg1vscgamma}. In the transition region between $\cgt$ and $\cgo$ dominance (\ie\ the corner between diagonal and vertical boundary), interference effects are relevant and, for $\cgt<0$, they can produce cusps where a larger $\cgamma$ is allowed (\blue, \red\ curves), or even thin bands where an arbitrarily large $\cgamma$ is admitted (\orange\ curves). These features are analogous to cusps and bands in Fig.~\ref{fig:cg1vscgamma}. 

Finally, if $\cgo$ has an intermediate value (\green\ curve), the maximum value of $\Br(a\to\g\g)$ is allowed, but the cross section is not fully dominated by $\cgt$. Then we still find a horizontal allowed band, whose size depends on $\cgo$ and that is asymmetric around the $\cgt$ axis: for $\cgt<0$, $\sigma^{\rm cuts}$ is smaller and the bound on $|\cgt|$ weaker. 

\end{itemize}

It is interesting to note that the curves for all values of $\cgo$ appear to intersect at a single point in the $\cgt>0$ plane.
This point corresponds to a horizontal bound in the lower panels  of Fig.~\ref{fig:cg2vscg1}. In fact, its $\cgamma$ coordinate is given by the value of $\cgamma$ marking the transition between open and closed allowed regions in the $(\cgo,\cgt)$ plane, 
while the $\cgt$ coordinate is given by corresponding upper bound on $\cgt$, extracted in the $\cgo\to \infty$ limit.
The functional dependence of $N_{\rm signal}$ on $\cgo$ and $\cgt$ is such that, for that value of $\cgamma$, the bound on $\cgt$ remains
approximately flat for \emph{all} values of $\cgo$, \ie\ it would be a horizontal line in Fig.~\ref{fig:cg2vscg1}. 
As a consequence, all the $N_{\rm signal}=3$ curves in Fig.~\ref{fig:cg2vscgamma} pass approximately through this same point even when a small $\cgo$ is assumed. The intersection points among the various lines do not coincide exactly, but the separation between them is too small to be visible in the figure.

\paragraph{Constraints on $\cgt$ as a function of $m_a$, for fixed $\cgamma$ and $\cgo$.}

\begin{figure}[t]
\centering
\includegraphics[width=0.48\textwidth]{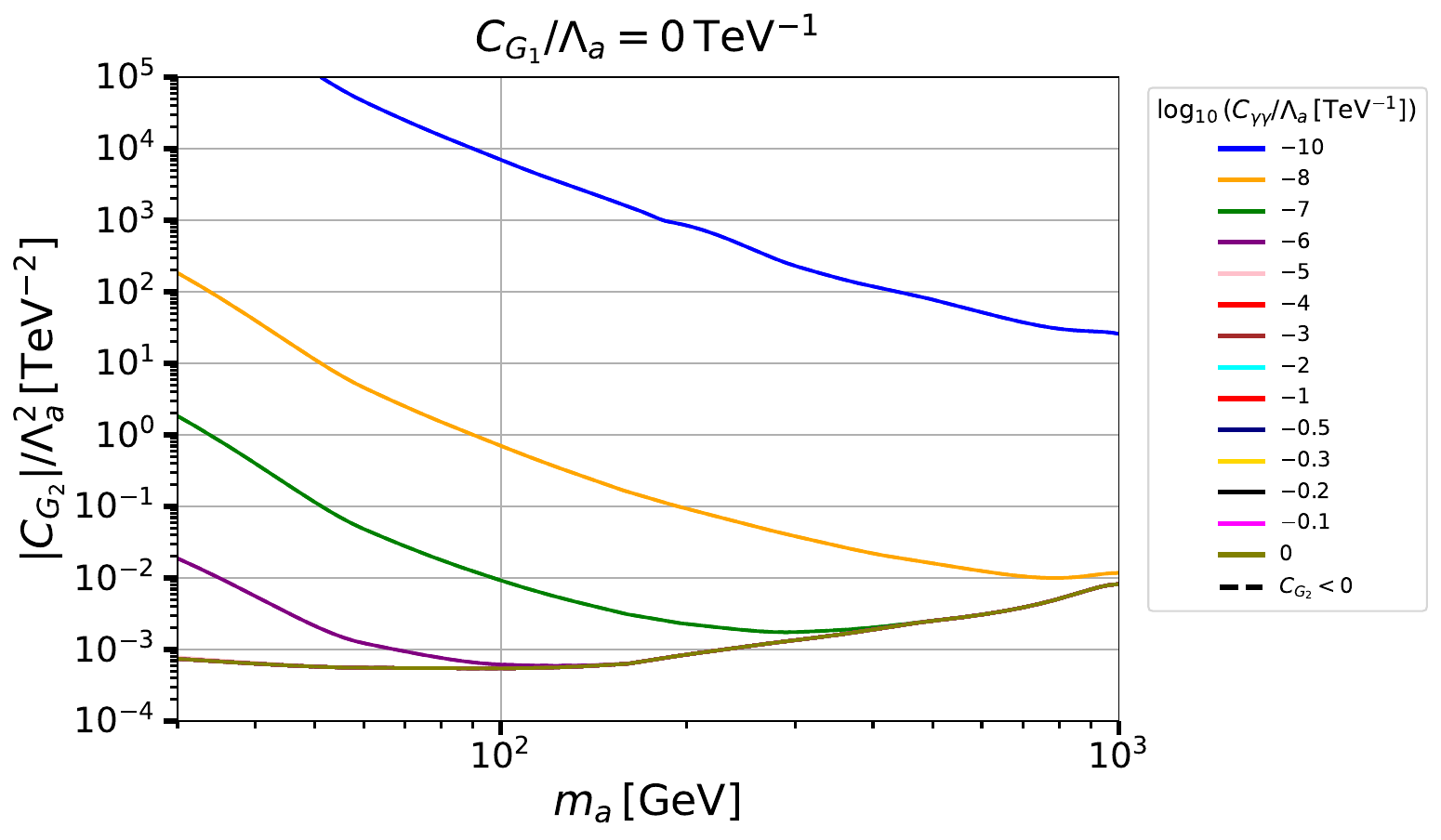}
\hfill
\includegraphics[width=0.48\textwidth]{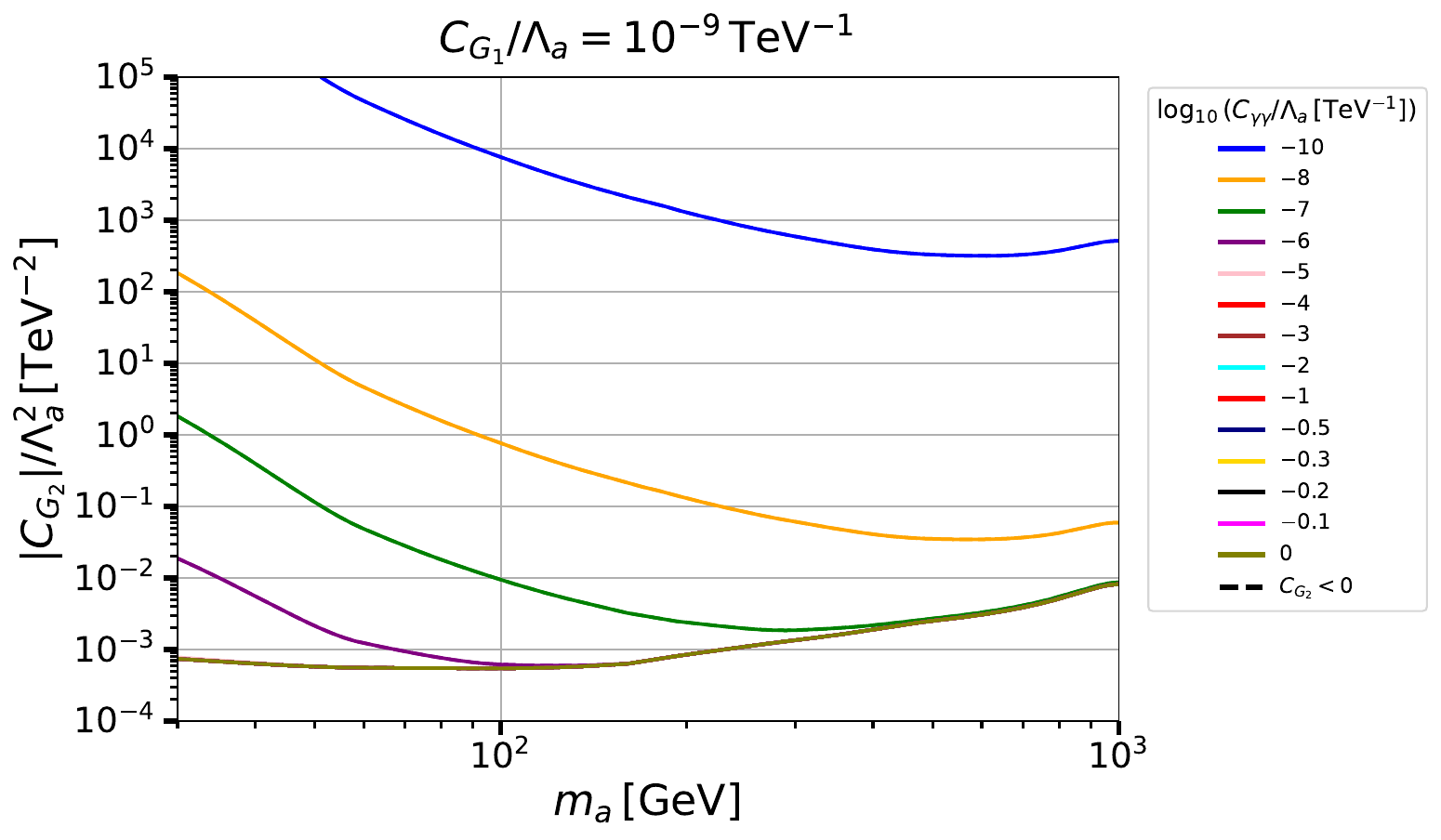}
\\[0.3cm]
\includegraphics[width=0.48\textwidth]{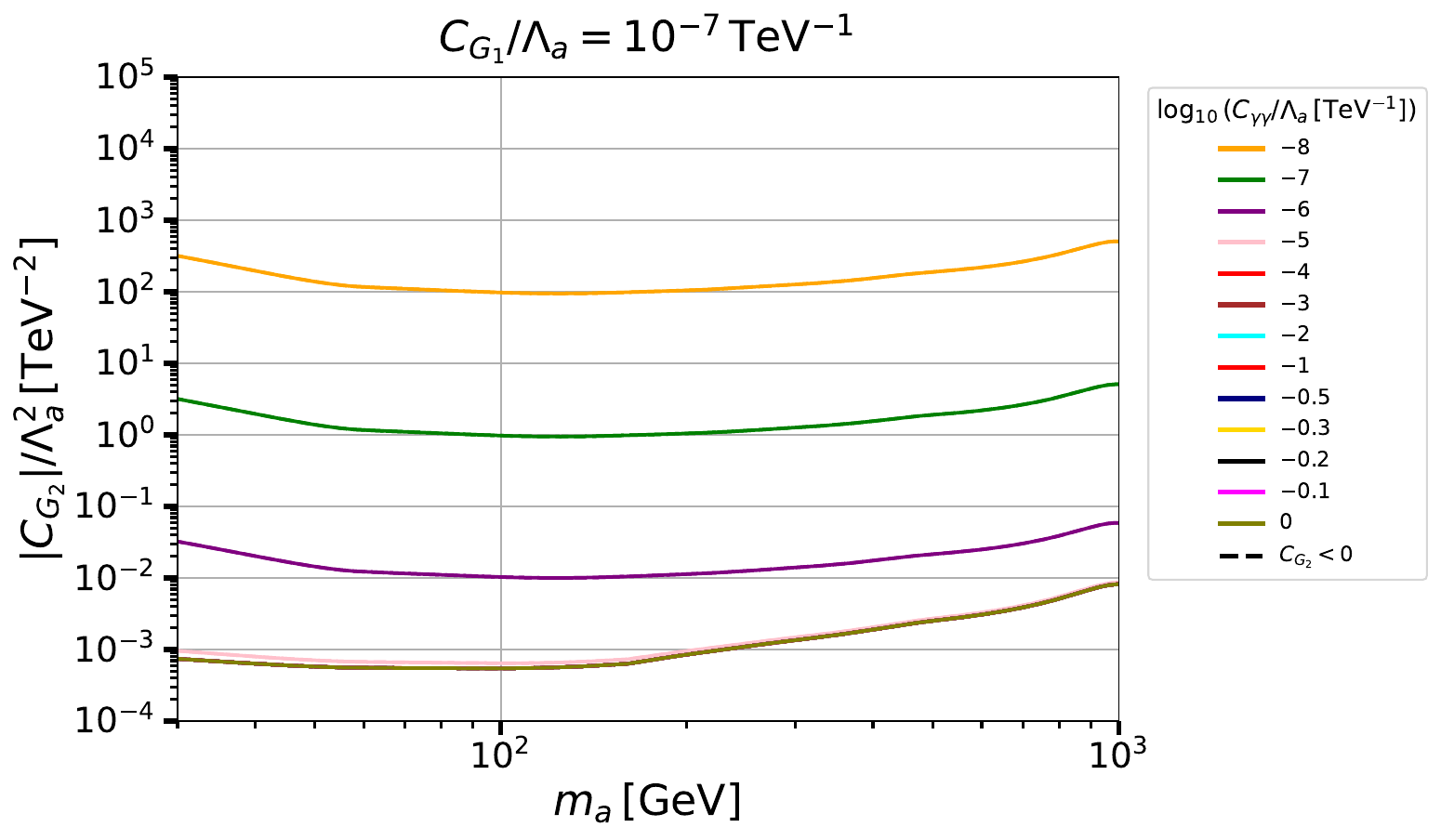}
\hfill
\includegraphics[width=0.48\textwidth]{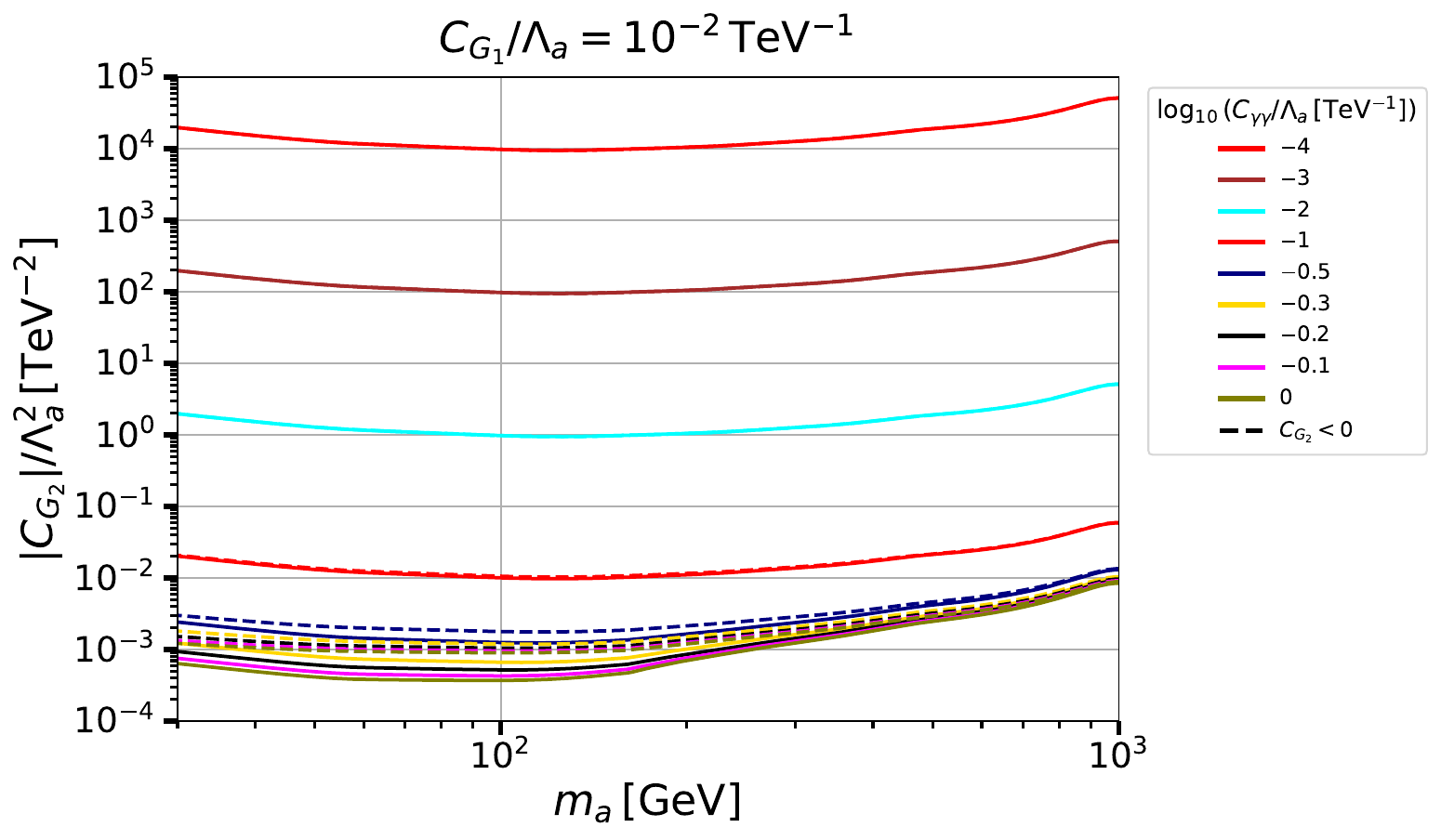}
\\[0.3cm]
\includegraphics[width=0.48\textwidth]{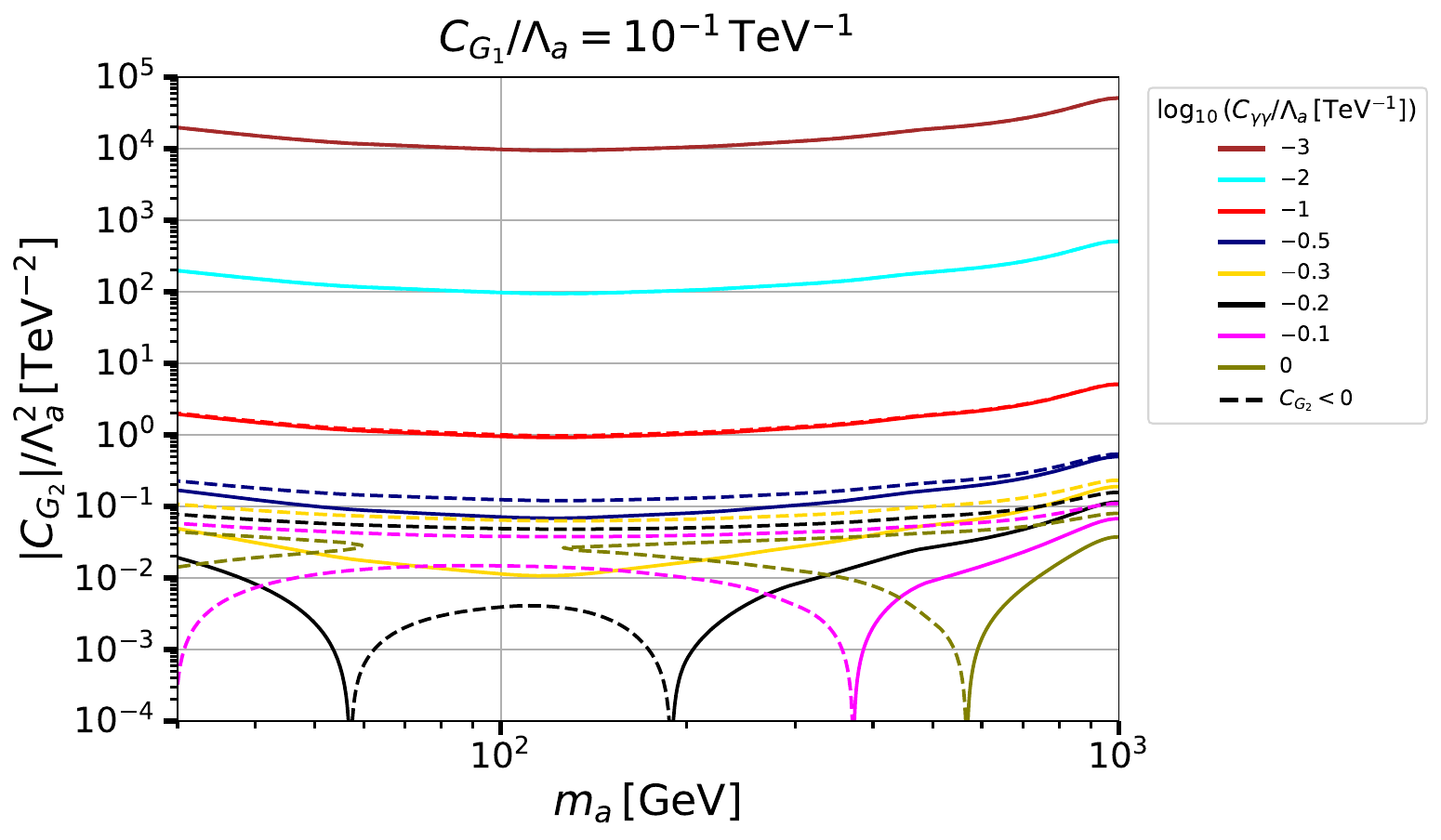}
\hfill
\includegraphics[width=0.48\textwidth]{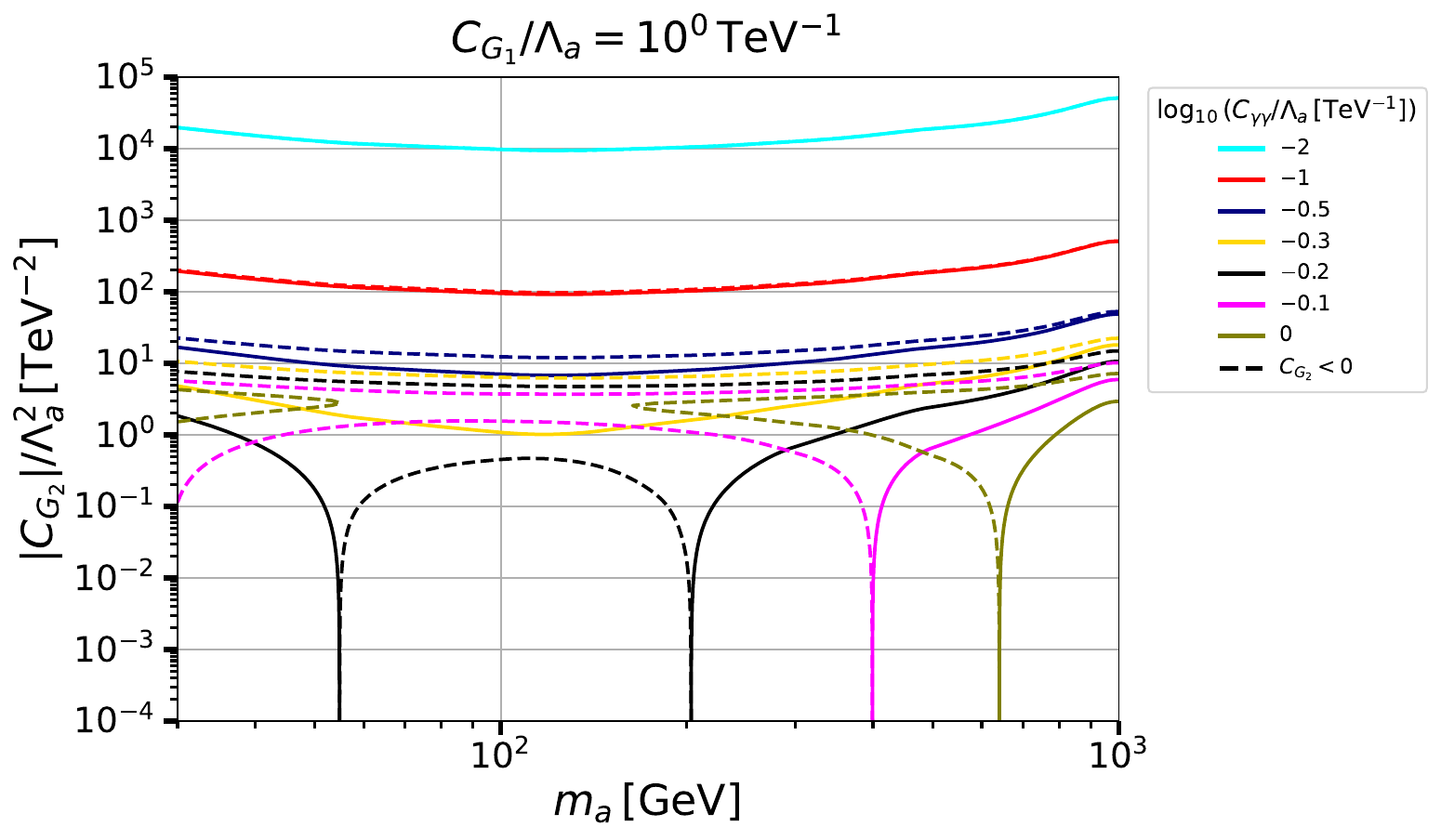}

\caption{95\%CL upper bounds on $\cgt/\La^2$  achievable with non-resonant $gg\to aa\to 4\gamma$ searches at the LHC with $\Lag=\unit[300]{fb^{-1}}$, as a function of $m_a$ and for fixed values of $\cgo/\La$ (panels) and  $\cgamma/\La$ (curves). In the first three plots, the curves for $\cgamma/\La\in [10^{-5} -1]~\unit{TeV^{-1}}$ overlap and there is no sensitivity to the sign of $\cgt$. }
\label{fig:Cg2_vs_ma_log_lin}
\end{figure}

To conclude this section, we explore the sensitivity to $\cgt$ as a function of $m_a$, for fixed values of $\cgamma$ and $\cgo$.
The rationale behind this study is the following:  non-resonant $gg\to aa$ production is the simplest process to which the dimension-6 operator $\Ocgt$ can contribute at tree level. Therefore we expect that, within a potential global analysis of the ALP EFT, $\cgamma$ and $\cgo$ will be bounded through a  set of measurements including those mentioned in Sec.~\ref{sec:currentbounds}, while $gg\to aa$ will mainly act as a constraint on $\cgt$. One can then ask how strong such a constraint would be, depending on the (best-fit) values of $\cgamma$ and $\cgo$.

Fig.~\ref{fig:Cg2_vs_ma_log_lin} shows 95\%CL upper limits on $\cgt/\La^2$ as a function of $m_a$, inferred from non-resonant $gg\to aa$ production for various values of $\cgamma$ and $\cgo$.
For $\cgo/\La\lesssim \unit[10^{-3}]{TeV^{-1}}$ (first three panels) one has $\cgt$ dominance over all the $\cgt$ range shown, while for $\cgo/\La\gtrsim\unit[10^{-2}]{TeV^{-1}}$ (last three panels) one has $\cgt$ dominance in the upper part of the plot, $\cgo$ dominance at the bottom of the plot, and an intermediate band in between them where interference effects are relevant. Only within the latter we have sensitivity to the sign of $\cgt$.

Let us start by examining the first three plots. In all three, the production cross section is dominated by $\cgt$ and the best bound attainable on this coefficient is roughly $\unit[10^{-3}]{TeV^{-2}}$. This value corresponds to the position of the horizontal boundaries in the upper panels of Figs.~\ref{fig:cg2vscg1} and~\ref{fig:cg2vscgamma}: it is achieved when $\cgamma$ is large enough that $\Br(a\to\g\g)$ is saturated to its maximum allowed value and FSDE are irrelevant. 
Within this regime, the bound is independent of both $\cgo$ and $\cgamma$: indeed all the lines for $\cgamma/\La$ between $10^{-5}$ and $\unit[1]{TeV^{-1}}$ fully overlap, and their position remains essentially unchanged across the three plots.  The bound only worsens slightly at larger $m_a$, due to the smaller production cross section at high mass: the fact that it is nearly flat at lower masses is due to the drop in the selection efficiency $\epsilon_{\rm cuts}$, which (over)compensates for the higher cross section, see Fig.~\ref{fig:ratio_coeff_cuts}.

If $\cgamma$ takes smaller values, the bound on $\cgt$ worsens due to two effects that reduce $N_{\rm signal}$: $\Br(a\to \g\g)$ drops below its maximum, and FSDE become relevant. The former effect kicks in when $\cgamma\lesssim 10\cgo$ (\ie\ $r_\gamma\gtrsim 0.1$). Because $\Br(a\to \g\g)$ depends only mildly on $m_a$, it causes the constraints to shift approximately rigidly upwards, by a factor proportional to $r_\gamma^2$. 
The latter effect enters when both $\cgo$ and $\cgamma$ are tiny, and its impact depends significantly on $m_a$, so it changes the bounds shapes, worsening them more strongly at low $m_a$.

In the limit $\cgo=0$ (first panel), $\Br(a\to \g\g)$ is always saturated to its maximum, and a dependence on $\cgamma$ enters only through FSDE. Consistently with Fig.~\ref{fig:FSDE_less_90_ma_s} (left), this effect becomes visible for $\cgamma/\La\lesssim \unit[10^{-6}]{TeV^{-1}}$ at $m_a=\unit[30]{GeV}$ and for $\cgamma/\La\lesssim \unit[10^{-8}]{TeV^{-1}}$ at $m_a=\unit[1]{TeV}$. We have verified that the results shown in this plot remain essentially unchanged for $\cgo/\La\lesssim\unit[10^{-10}]{TeV^{-1}}$.

When $\cgo/\La = \unit[10^{-9}]{TeV^{-1}}$ (second panel), $\Br(a\to\g\g)$ drops below its maximum for $\cgamma/\La\leq \unit[10^{-8}]{TeV^{-1}}$. Thus, both FSDE and $\Br(a\to\g\g)$ reduction are present at the same time.
Once $\cgo/\La\gtrsim\unit[10^{-7}]{TeV^{-1}}$ (third panel), FSDE become irrelevant across the whole mass range, and a dependence on $\cgamma$ remains only through $\Br(a\to\g\g)$. This behavior is maintained until $\cgo/\La\lesssim\unit[10^{-3}]{TeV^{-1}}$.

For $\cgo/\La\gtrsim\unit[10^{-2}]{TeV^{-1}}$ (last three panels), we observe new effects. For $\cgamma/\La\lesssim\unit[10^{-1}]{TeV^{-1}}$, the bounds on $\cgt$ still fall within the $\cgt$-dominance region, and therefore they follow the same pattern as in previous panels. This regime corresponds to values of $\cgamma$ smaller than the orange curves in Fig.~\ref{fig:cg2vscg1} (upper). For larger $\cgamma$, however, the bound on $\cgt$ reaches into the regime where the interference between $\cgo$ and $\cgt$ is relevant. As a result, the constraints for $\cgt>0$ (solid) and $\cgt<0$ (dashed) depart.  

For $\cgo/\La=\unit[10^{-2}]{TeV^{-1}}$ (fourth panel), this effect is only mild and it makes the constraints on $\cgt<0$ slightly weaker than those on $\cgt>0$. In both cases, the strongest constraint is obtained for large $\cgamma$. In fact, this value remains stable in the limit $\cgamma\to\infty$, a feature shared with all the previous panels. This regime corresponds to cutting very close to the vertical boundary in the upper panels of Fig.~\ref{fig:cg2vscg1}: the weakening of the constraints for $\cgt<0$ matches the presence of a cusp in that figure.

Moving past that vertical boundary, \eg\ for $\cgo/\La\gtrsim\unit[10^{-1}]{TeV^{-1}}$ (fifth panel) we have various regimes. For $\cgamma$ not too large (dark blue and yellow curves) we recover the same features as in the previous panel: it is possible to set an upper bound on $|\cgt|$ which is worse for negative $\cgt$. This regime corresponds to values of $\cgamma$ similar to the orange curve in Fig.~\ref{fig:cg2vscg1}. Moving to larger $\cgamma$ (black curve), we find the same result only for some values of $m_a$. For other values of $m_a$, all values of $\cgt>0$ are excluded, and $\cgt<0$ is only allowed within a finite window, contained between two dashed lines. This latter regime corresponds to the case shown with a red curve in Fig.~\ref{fig:cg2vscg1}.
For even larger $\cgamma$, we find values of $m_a$ for which any value (positive or negative) of $\cgt$ is excluded. In the limit $\cgamma\to \infty$, the entire space shown is ruled out. This behavior corresponds to choosing a value of $\cgo$ to the right of the closed region found for large $\cgamma$ in Fig.~\ref{fig:cg2vscg1}.

Finally, moving to even larger $\cgo$ (sixth panel), the bounds simply move upwards. We have verified that, in the limit $\cgo\to \infty$, the entire region shown becomes allowed if $\cgamma/\La\lesssim\unit[0.1]{TeV^{-1}}$. On the other hand, if $\cgamma/\La > \unit[0.1]{TeV^{-1}}$, we remain with vertical bounds that should be interpreted as lower limits on $m_a$:
as $\cgo$ cancels out between production and decay, $\cgt$ is subdominant and $\cgamma$ is fixed, $N_{\rm signal}$ depends only on $m_a$.

\subsection{Reinterpretation of Higgs-resonant search}

\label{sec:resresults}
In this section we focus on the Higgs-resonant production channel. As anticipated in Sections~\ref{sec:overview} and~\ref{sec:res}, rather than repeating an independent sensitivity study for this process, we re-interpret the results presented in Ref.~\cite{ATLAS:2023ian} by the ATLAS collaboration. Our main goal is to perform a consistent comparison between the sensitivities of the Higgs-resonant and non-resonant production modes, and to examine the features emerging when interpreting these measurements within a multi-dimensional parameter space.

The relevant physical quantity measured in the ATLAS search is the effective branching fraction of the Higgs boson, which can be parameterized as
\begin{align}
\Br(h\to aa\to 4\g)_{\rm eff} &= \Br(h\to aa) \Br(a\to\g\g)^2 f_{aa}\,.
\label{eq.BRh_eff}
\end{align}
$\Br(h\to aa)$ was given in Eq.~\eqref{Eq:Br_h_aa} and it is a function of $\cgoh^2$. Considering only ALP decays to photons and gluons,
$\Br(a\to\g\g)$ can be computed as in Eq.~\eqref{eq:Br_a_phph_ratio} and is a function of $r_\gamma=\cgo/\cgamma$.\footnote{As we restrict to $m_a<m_h/2$, none of the 2-body electroweak decay channels of the ALP is kinematically open, so $F(m_a)\equiv 0$.} Finally, $f_{aa}$ is the fraction of double-ALP events decaying within the detector volume, which we estimate as explained around Eq.~\eqref{eq:faa}. The quantity $f_{aa}$ depends on $\cgamma$ and $\cgo$ through the total ALP decay width, and it departs from unity only for very small ALP couplings. All the quantities in Eq.~\eqref{eq.BRh_eff} depend on the ALP mass. The total number of expected signal events is directly proportional to $\Br(h\to aa\to4\gamma)_{\rm eff}$ as:
\begin{align}
  N_{\rm signal} = \Lag \, \epsilon^{\rm cuts} \,\sigma(pp\to h)\,  \Br(h\to aa\to 4\gamma)_{\rm eff}\,,
\end{align}
where $\Lag=\unit[140]{fb^{-1}}$ is the integrated luminosity  and $\epsilon^{\rm cuts}$ is an overall suppression factor that accounts for the efficiency of the event reconstruction and acceptance of the selection cuts.

Ref.~\cite{ATLAS:2023ian} presents upper bounds on $\Br(h\to aa\to 4\gamma)_{\rm eff}$ over a broad range of ALP masses ($\unit[0.1-62.5]{GeV}$) for four benchmark values of the ALP lifetime 
(Fig.~6 in Ref.~\cite{ATLAS:2023ian}). 
These bounds are then translated into an upper limit on $\cgamma/\La$ (Fig.~9 in Ref.~\cite{ATLAS:2023ian}), under the assumptions $\cgo=0$ and $\cgoh/\La^2=\unit[1]{TeV^{-2}}$, or alternatively $\cgoh/\La^2=\unit[0.1]{TeV^{-2}}$. Within this setup $\Br(h\to aa)$ is only a function of $m_a$ and $\Br(a\to \gamma\gamma)=1$, so $\Br(h\to aa\to 4\g)_{\rm eff}$ depends on $\cgamma$ only through~$f_{aa}$. Physically, this amounts to assuming a large  cross section for the process $pp\to h\to aa\to 4\gamma$, and then requiring the ALP lifetime to be long enough to deplete the expected number of \emph{visible} signal events, such that it falls below the observation threshold.

For the study presented in this section, we only consider ALP masses $m_a\geq \unit[3]{GeV}$ because, below this threshold, it is not possible to model correctly hadronic ALP decays as a function of $\cgo$ using the formalism of Section~\ref{sec:theo}. 
Within this mass window, the upper bounds on $\Br(h\to aa\to4\gamma)_{\rm eff}$ reported in Ref.~\cite{ATLAS:2023ian} for the three benchmarks labeled as $\cgamma=5\times 10^{-4},\,0.01,\,1$ are approximately equal.
We have verified that the parameterization 
in Eq.~\eqref{eq.BRh_eff} with $f_{aa}$ defined as in Sec.~\ref{sec:FSDE} allows us to reproduce with very good accuracy the translation performed in Ref.~\cite{ATLAS:2023ian} of any of those upper limits on the Higgs branching ratio into limits on $\cgamma/\La$. 
For the purpose of the numerical studies presented in this section, we extract all the relevant limits presented in Ref.~\cite{ATLAS:2023ian} from the associated 
HEPData database~\cite{hepdata.144534}.  

\begin{figure}[t]
 \includegraphics[width=0.48\textwidth]{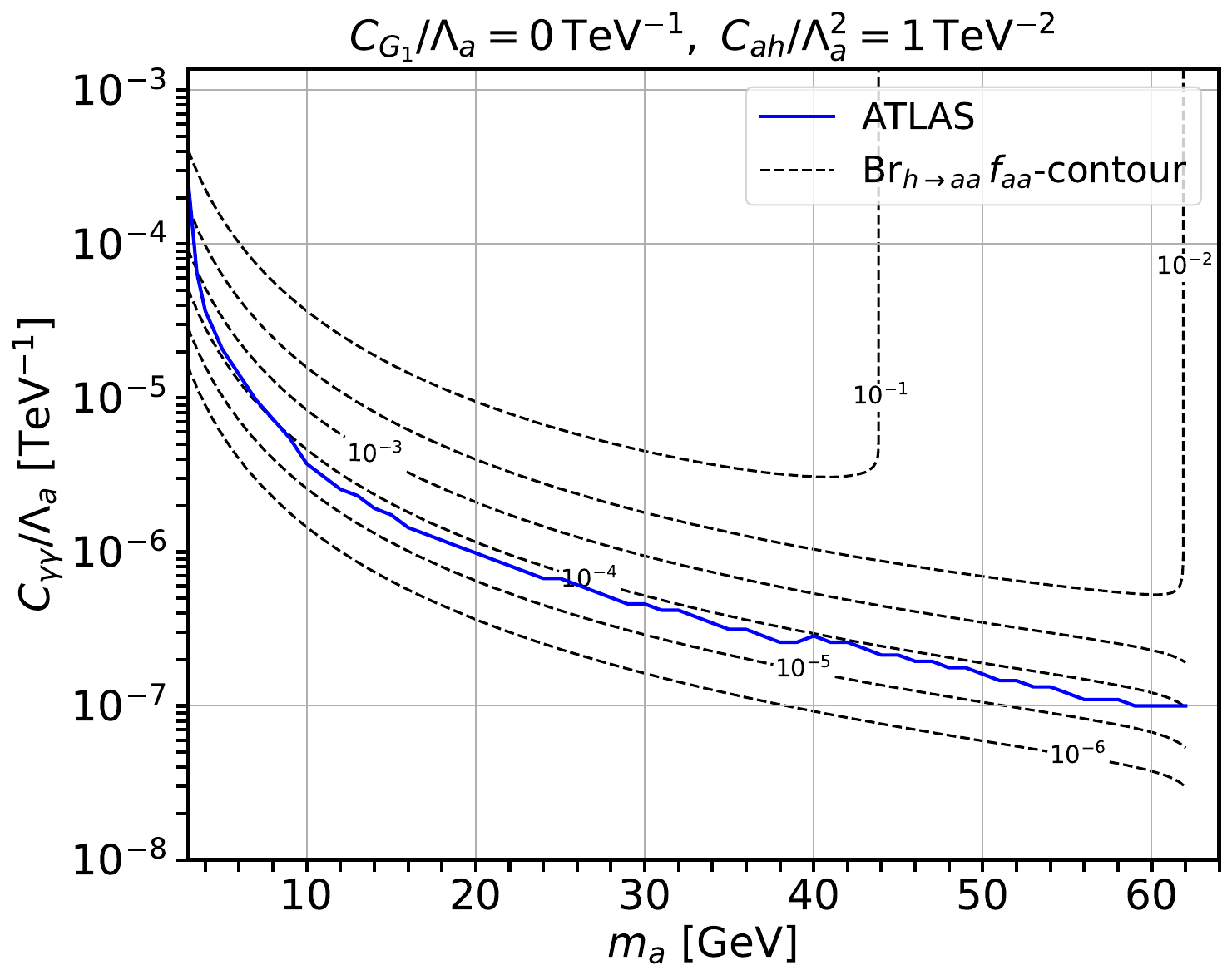}
 \hfill
 \includegraphics[width=0.48\textwidth]{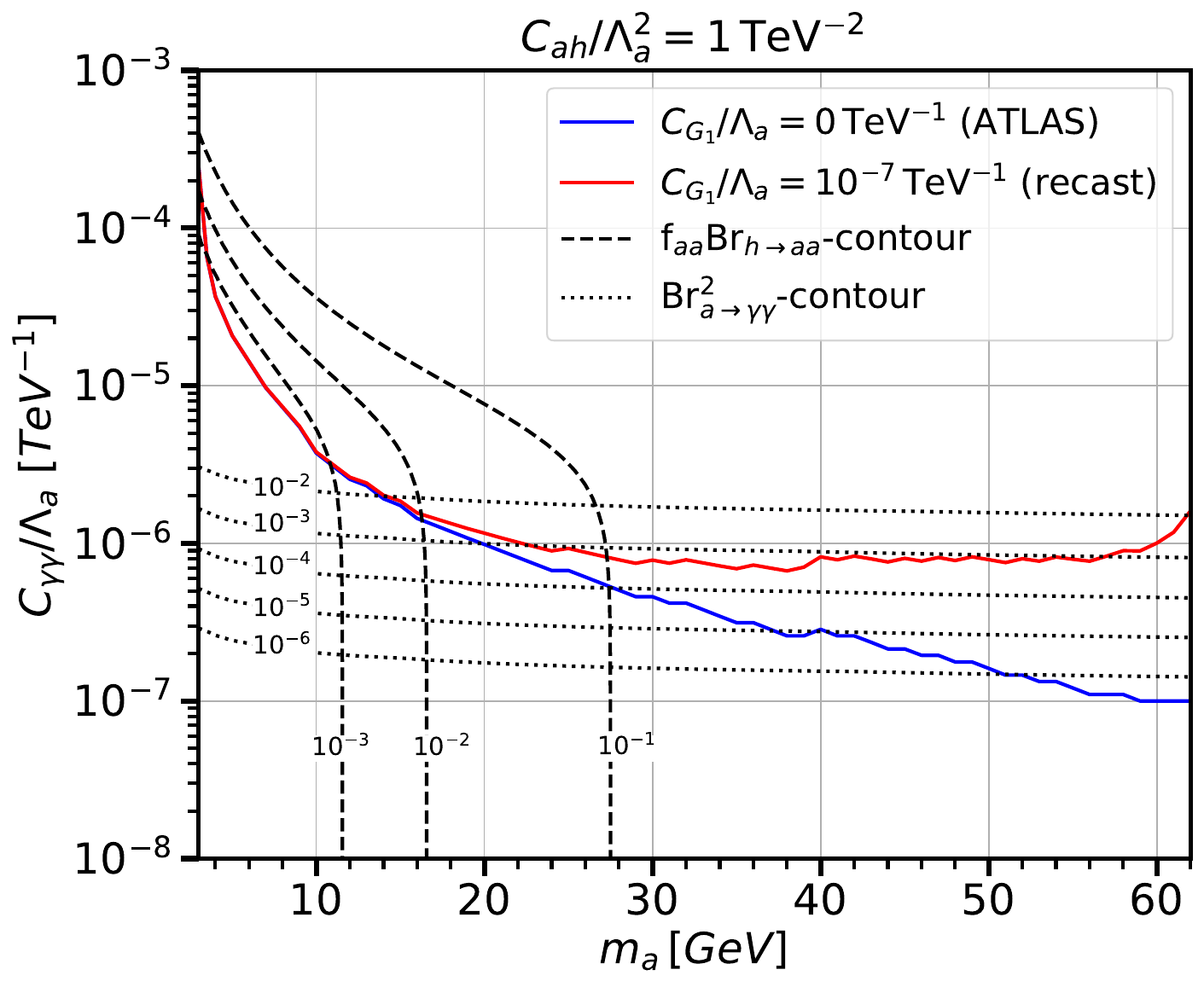}\caption{Left: Bounds on $\cgamma/\La$ extracted in Ref.~\cite{ATLAS:2023ian} from the measured limits on $\Br(h\to aa\to 4\gamma)_{\rm eff}$,  under the assumptions $\cgoh/\La^2=\unit[1]{TeV^{-2}}$ and $\cgo=0$ (blue). Right: comparison of the same bounds (blue) with their reinterpretation for $\cgoh/\La^2=\unit[1]{TeV^{-2}}$ and $\cgo/\La=\unit[10^{-7}]{TeV^{-1}}$ (red). In both panels, the gray dashed lines show contours of constant $\Br(h\to aa) f_{aa}$. In the right plot, the dotted grey lines show contours of constant $\Br(a\to\g\g)^2$. 
 }
  \label{fig:ATLAS_vs_contours}
 \end{figure}

Fig.~\ref{fig:ATLAS_vs_contours} (left) shows the upper limits on $\cgamma/\La$ reported in Ref.~\cite{ATLAS:2023ian}, together with iso-contours of the product $\Br(h\to aa)f_{aa}$, as a function of $m_a$ and for fixed $\cgoh/\La^2=\unit[1]{TeV^{-2}}$. Over most of the ALP mass range, the ATLAS bound is consistent with the simple requirement $\Br(h\to aa)f_{aa}\lesssim 10^{-4}$. Small deviations from this behavior can only be observed for very small masses, where the analysis loses sensitivity due to the more challenging event reconstruction, and  close to the $m_h/2$ threshold, where the Higgs branching ratio drops sharply. For intermediate ALP masses and in the couplings region close to the ATLAS bound, the iso-contours dominantly scale as $\Br(h\to aa)f_{aa}\sim f_{aa}\sim (\Gamma_a^{\rm tot})^{2}\sim (m_a^6\cgamma^4)$ as a function of $m_a$ and $\cgamma$, which determines the bound's slope. The isocontours for $\Br(h\to aa)f_{aa} =0.1$ and $0.01$ bend upwards in correspondence of the maximum $m_a$ for which $\Br(h\to aa) =0.1 (0.01)$ can be achieved (see Eq.~\eqref{Eq:Br_h_aa}).

The upper bounds on $\cgamma/\La$ reported in Ref.~\cite{ATLAS:2023ian} can be easily re-interpreted in scenarios with $\cgo\neq 0$ and $\cgoh/\La^2\neq \unit[1,\,0.1]{TeV^{-2}}$. The bound at a generic parameter space point  $(\cgamma, \cgo, \cgoh)$ can be evaluated by requiring 
\begin{align}
\Br(h\to aa\to 4\gamma)_{\rm eff} \,(m_a, \cgamma, \cgo, \cgoh)
 \leq \Br(h\to aa\to 4\gamma)_{\rm eff}\, (m_a, \overline\cgamma,\cgo= 0, \cgoh= 1) \,,
 \label{eq.resonant_recast}
\end{align}
where $\overline\cgamma$ is the upper limit on $\cgamma$ provided in Ref.~\cite{ATLAS:2023ian} for the corresponding $m_a$ and $\cgoh=1$.\footnote{This equation recasts the reported bounds on $\cgamma/\La$ for $\cgoh/\La^2=\unit[1]{TeV^{-2}}$. Alternatively, one could use the bounds on $\cgamma/\La$ for $\cgoh/\La^2=\unit[0.1]{TeV^{-2}}$ as a starting point, or re-derive the constraint starting from the bounds on $\Br(h\to aa\to4\gamma)_{\rm eff}$. Because our parameterization of $f_{aa}$ is only an approximation to the one adopted by ATLAS, the results obtained with the three methods are not identical, but we have verified that they agree to very good numerical accuracy.}

\paragraph{Reinterpretation varying $\cgo$.} Fig.~\ref{fig:ATLAS_vs_contours} (right) shows the result obtained varying only $\cgo$, which is set to a very small but non-zero value. In this case, the inequality is solved numerically for $\cgamma$ at each value of $m_a$, with fixed $\cgo/\La=\unit[10^{-7}]{TeV^{-1}}$, $\cgoh/\La^2=\unit[1]{TeV^{-2}}$. 
The figure also shows, for reference, iso-contours of the product $\Br(h\to aa) f_{aa}$ (gray dashed lines) and of $\Br(a\to \g\g)^2$ (gray dotted lines), which depend on the fixed value of $\cgo$. The recast bound (in red) is controlled by the product of these two quantities. The plot shows that, while at small $m_a$ the bound is dominated by the former factor, at larger $m_a$ it is dominated by the latter one.
Indeed, introducing $\cgo$ has two physical effects: it increases $\Gamma_a^{\rm tot}$ and therefore $f_{aa}$, and it reduces $\Br(a\to \g\g)$.
Since the contribution to $\Gamma_a^{\rm tot}$ scales with $\cgo^2 m_a^3$, turning on a small $\cgo$ brings $f_{aa}$ close to 1 at larger masses, while leaving $f_{aa}<1$ at small $m_a$. In the region where $f_{aa}\simeq 1$, $\Br(h\to aa\to4\gamma)_{\rm eff}$ depends on $\cgamma$ only through $\Br(a\to \g\g)^2$, so the bound becomes approximately independent of $m_a$. 

\begin{figure}[t]\centering
\includegraphics[width=11cm]{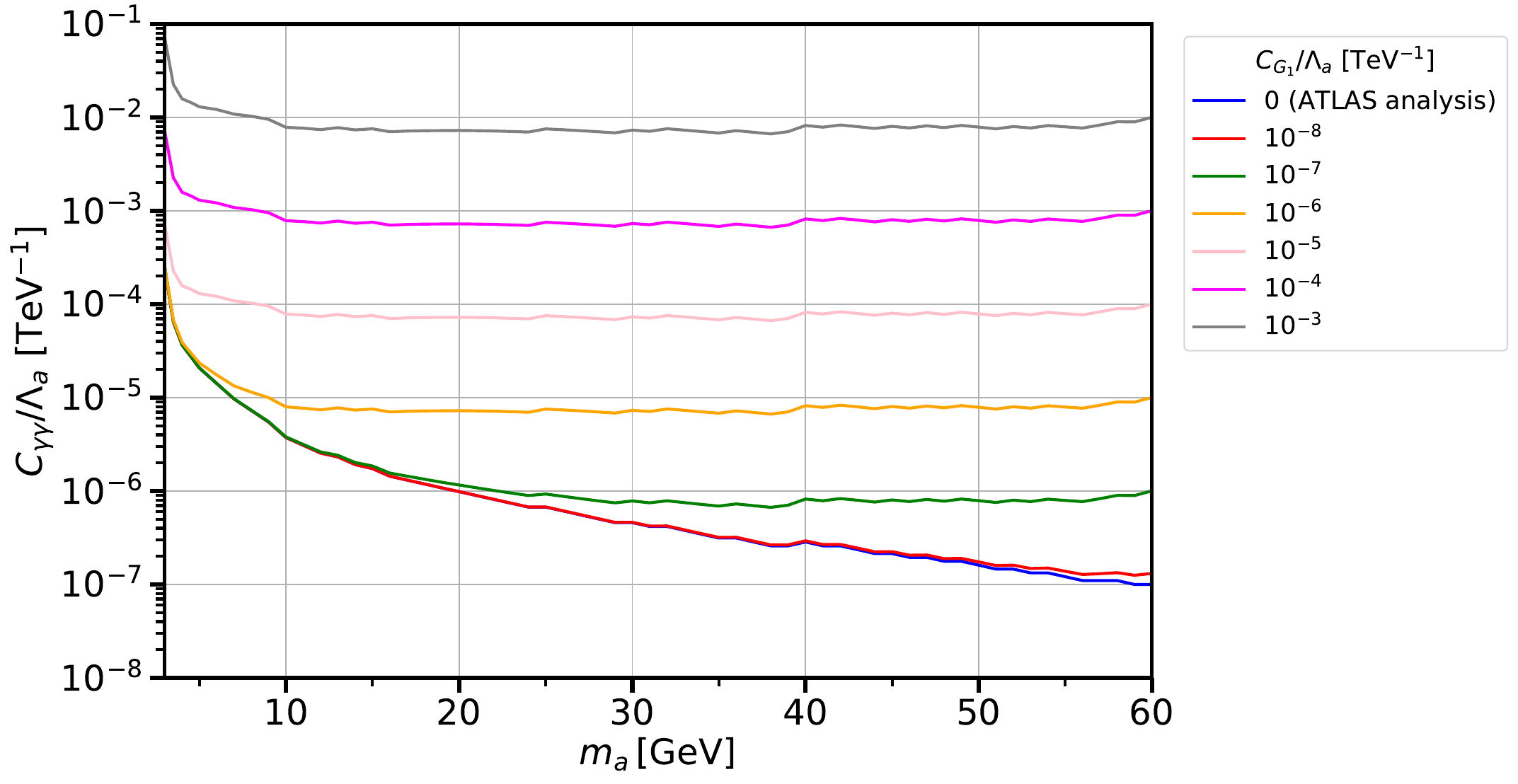}
\caption{Upper limits on $\cgamma/\La$ as a function of $m_a$, obtained recasting the limits reported in Ref.~\cite{ATLAS:2023ian} by varying $\cgo/\La\neq 0$, while keeping $\cgoh/\La^2$ fixed to $\unit[1]{TeV^{-2}}$.
}
\label{fig.resonant_recast_cG1}
\end{figure}

Fig.~\ref{fig.resonant_recast_cG1} shows the bounds on $\cgamma/\La$ obtained with the same procedure, for various values of $\cgo/\La$, while keeping $\cgoh/\La^2=\unit[1]{TeV^{-2}}$. Here one observes how, going towards larger $\cgo$, $f_{aa}$ saturates to 1 at smaller and smaller masses,  eventually reaching a point where $f_{aa}\simeq 1$ across the entire mass spectrum. Correspondingly, the bound on $\cgamma$ weakens and becomes more and more dominated by the $\Br(a\to\g\g)^2$ suppression.
In this regime, the upper bound on $\cgamma$ worsens linearly with $\cgo$, due to the fact that $\Br(a\to\g\g)$ is a function of $r_\gamma = \cgo/\cgamma$.
For reference, for $\cgo/\La=\unit[10^{-3}]{TeV^{-1}}$, which is well within the current experimental bounds (see Sec.~\ref{sec:currentbounds}), the constraint on $\cgamma/\La$ weakens by  3--5 orders of magnitude compared to the $\cgo=0$ limit, across the entire mass spectrum.

\begin{figure}[t]
\centering
\includegraphics[height=6cm]{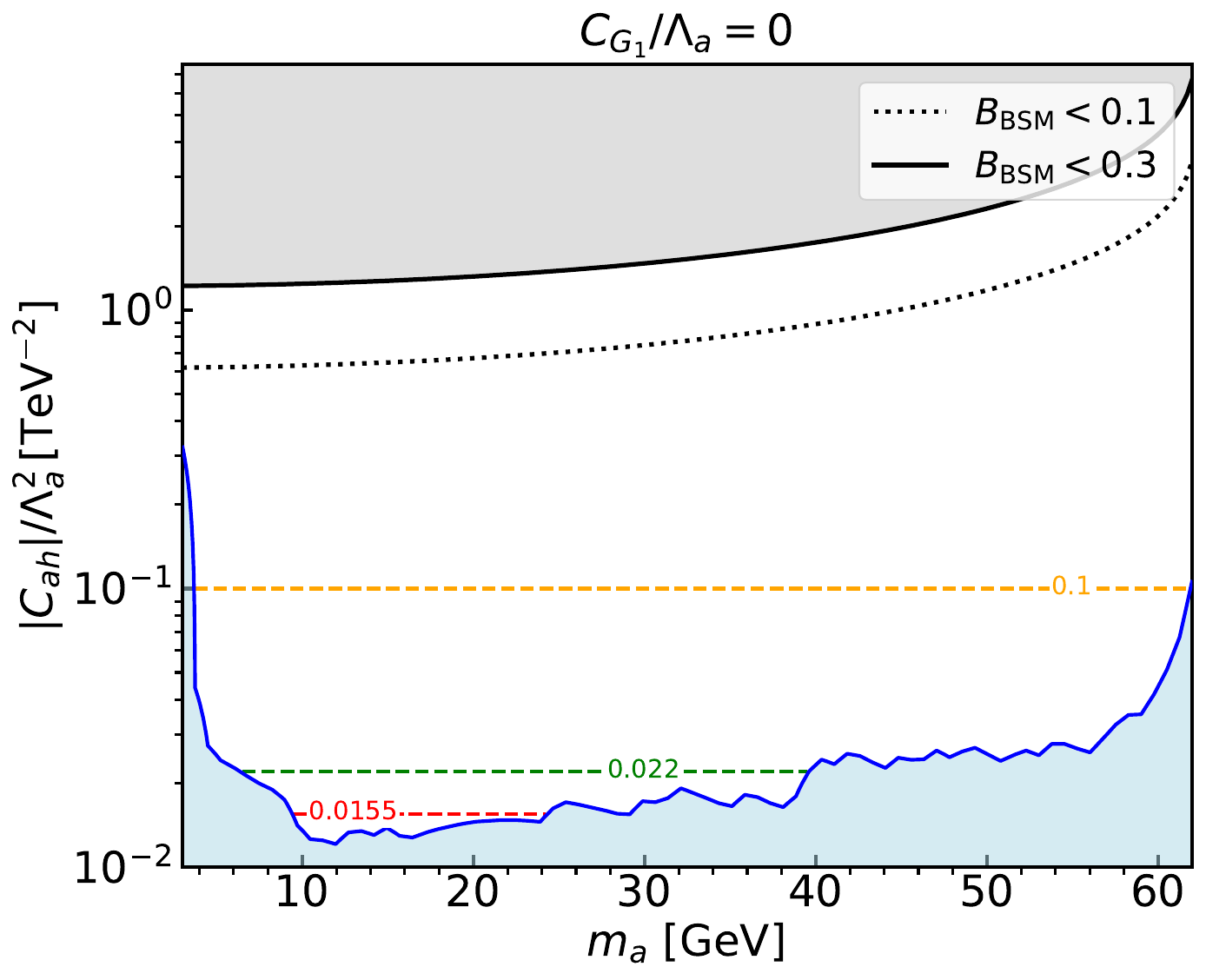}
\hfill
\includegraphics[height=6cm]{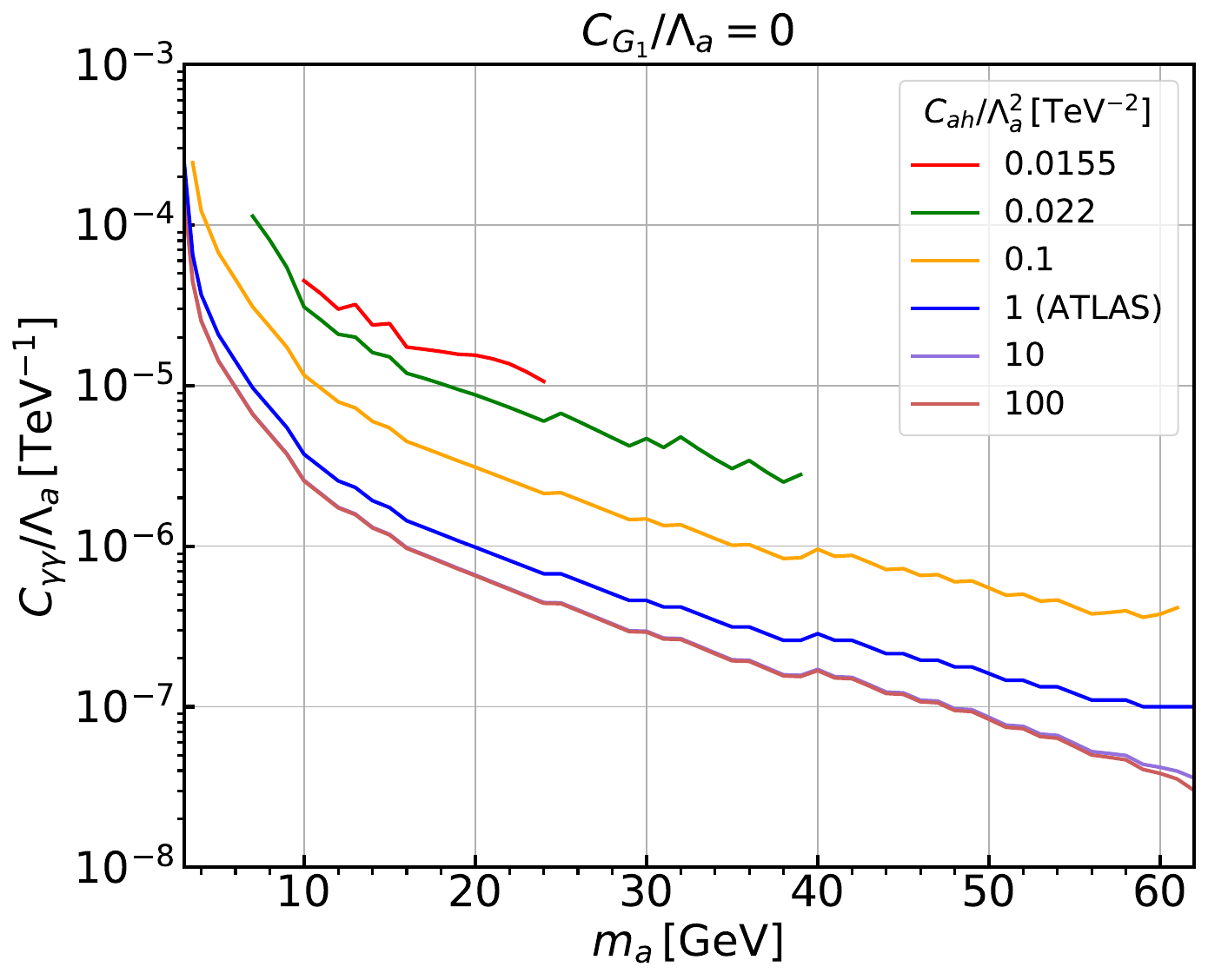}
\caption{Left: values of $\cgoh/\La^2$, as a function of $m_a$, for which it is possible to infer constraints on the ALP parameter space from the limits on $\Br(h\to aa\to 4\gamma)_{\rm eff}$ measured in Ref.~\cite{ATLAS:2023ian}.
The gray region is excluded by $\Br(h\to aa)\leq 30\%$. Within the light blue region $N_{\rm signal}$ is too small to set bounds on $\cgamma$ and $\cgo$.
Right: upper limits on $\cgamma/\La$ as a function of $m_a$, obtained recasting the limits of Ref.~\cite{ATLAS:2023ian} by varying $\cgoh/\La^2$ while keeping $\cgo=0$. The red, green and orange curves are derived for the values of $\cgoh$ marked with the corresponding color in the left panel: in all three cases, limits on $\cgamma$ can only be inferred for $m_a$ in a limited range. 
}
\label{fig.min_cah}
\end{figure}

\paragraph{Reinterpretation varying $\cgoh$.} Varying $\cgoh$ in Eq.~\eqref{eq.resonant_recast}, rather than $\cgo$,  has a qualitatively different impact on the bounds obtained from the Higgs-resonant search. The Higgs-ALP interaction $\cgoh$ controls the production rate of the ALP pairs through $\Br(h\to aa)$: if $\cgoh$ is too small, $N_{\rm signal}$ remains below the observation threshold for all values of $\cgamma$ and $\cgo$, and therefore it is not possible to infer any meaningful constraint on the ALP parameter space. Fig.~\ref{fig.min_cah} (left) shows the minimum value of $|\cgoh|/\La^2$ for which the search in Ref.~\cite{ATLAS:2023ian} constrains the $d=5$ ALP interactions (blue curve).  The values shown are extracted by solving numerically Eq.~\eqref{eq.resonant_recast} for $\cgoh$, with $\cgo=0$ and $\cgamma$ fixed to a large number such that $f_{aa}\Br(a\to\g\g)^2\equiv 1$ for all values of $m_a$. 

The figure shows that for $\cgoh/\La^2 \lesssim \unit[0.011]{TeV^{-2}}$, no bound can be set on $\cgamma$, for any value of the ALP mass.
Interestingly, because the minimum varies with $m_a$, there exist values of $\cgoh/\La^2$ for which the ATLAS measurement yields constraints only over a limited range of ALP masses. This condition is exemplified by the values of $\cgoh$ marked by the green ($\cgoh=0.022$) and red ($\cgoh=0.0155$) horizontal lines in Fig.~\ref{fig.min_cah} (left). 
The corresponding bounds on $\cgamma/\La$, extracted via the recasting procedure, are shown in the right panel: where the bound line stops, the number of produced $aa$ events drops below observability and the constraint on $\cgamma$ vanishes.
Moving towards larger values of $\cgoh$, the ATLAS constraint becomes relevant over a larger ALP mass range,  with $\cgoh/\La^2=\unit[0.1]{TeV^{-2}}$ being the smallest coupling for which the bound applies for \emph{all} the masses shown. At the same time, the bound on $\cgamma/\La$ becomes stronger, as one would expect.  In the limit of very large $\cgoh$, the Higgs branching ratio $\Br(h\to aa)$ eventually saturates to 1, and the constraints on $\cgamma$ cannot be improved further. This condition is visualized  by the two curves at $\cgoh/\La^2=10$ and~$\unit[100]{TeV^{-2}}$ in Fig.~\ref{fig.min_cah} (right), which indeed overlap. 

As one might suspect, such large values of $\cgoh$ are already excluded by current Higgs measurements, as reviewed in Sec.~\ref{sec:currentbounds}.
The solid (dotted) black line in Fig.~\ref{fig.min_cah}  (left) shows the upper limit on $\cgoh/\La^2$ corresponding to the condition $\Br(h\to aa)\leq 30\%$ (10\%) introduced in Eq.~\eqref{eq.BRhaa_bounds}. It is worth reminding that these constraints only depend on the ALP mass, and not on $\cgamma$ and $\cgo$.  For reference, they require
\begin{align}
  |\cgoh|/\La^2 &\leq \unit[1.22\;  (0.62)]{TeV^{-2}}
  &&\text{for }m_a=\unit[3]{GeV}\,,
  \\
  |\cgoh|/\La^2 &\leq \unit[1.47\; (0.75) ]{TeV^{-2}}
  &&\text{for }m_a=\unit[30]{GeV}\,.
\end{align}
Thus, in order to allow for the extraction of meaningful constraints on $\cgamma$ \emph{and} remain compatible with current Higgs measurements, $\cgoh$ must take values within a relatively limited window, which depends on $m_a$ and corresponds to the white region in Fig.~\ref{fig.min_cah} (left).

\begin{figure}[t]
\centering
\includegraphics[height=6cm]{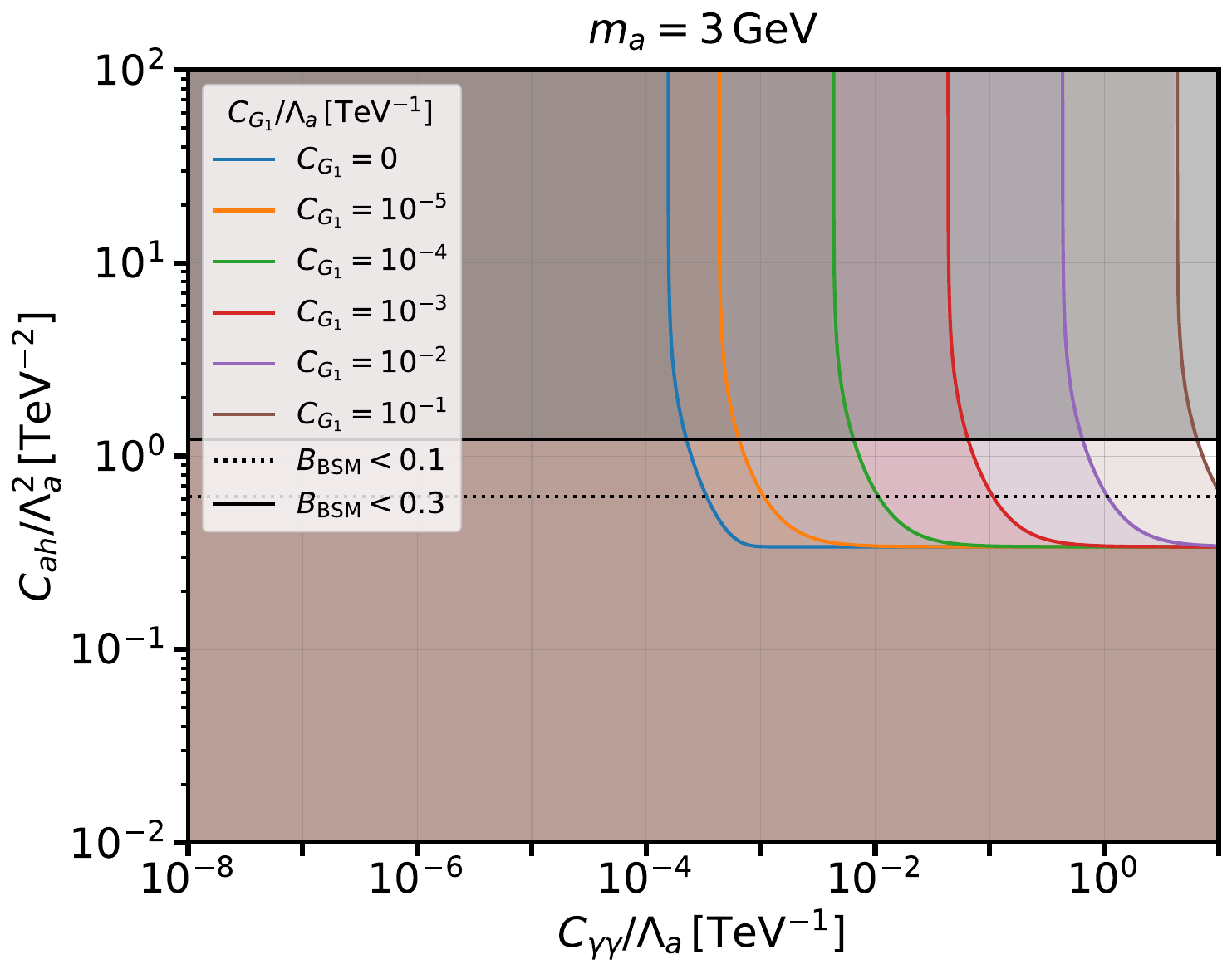}
\hfill
\includegraphics[height=6cm]{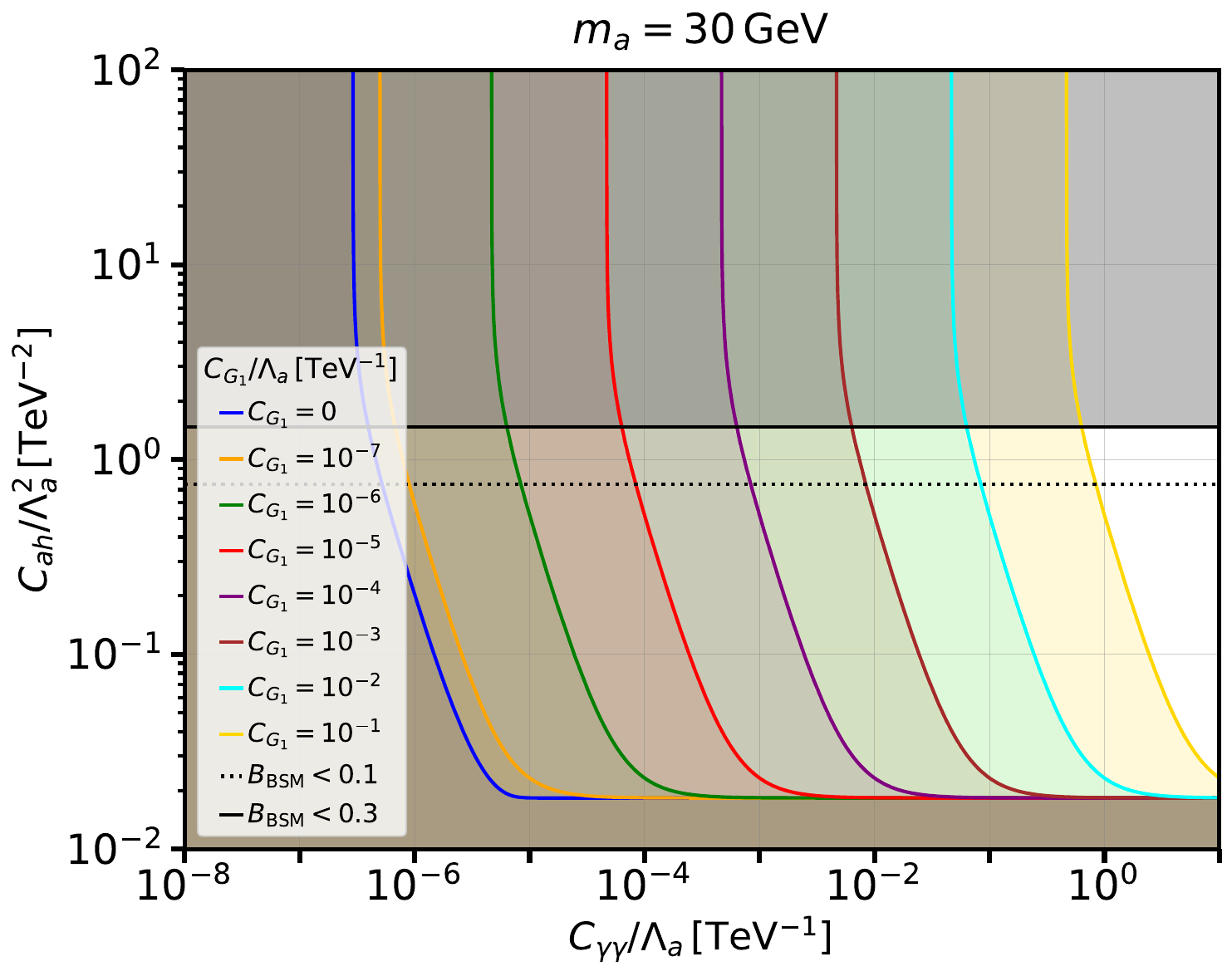}
\\[3mm]
\includegraphics[height=6cm]{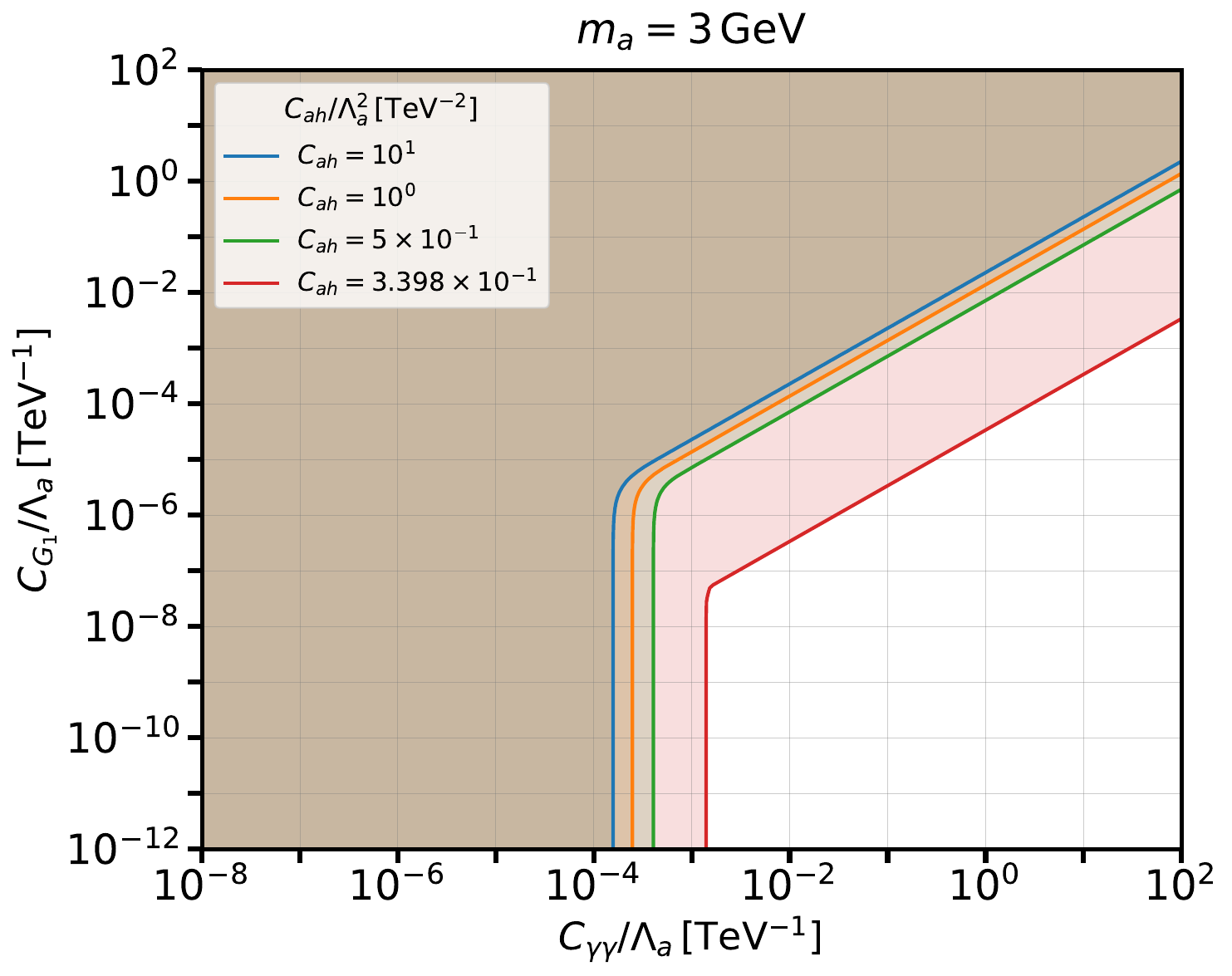}
\hfill
\includegraphics[height=6cm]{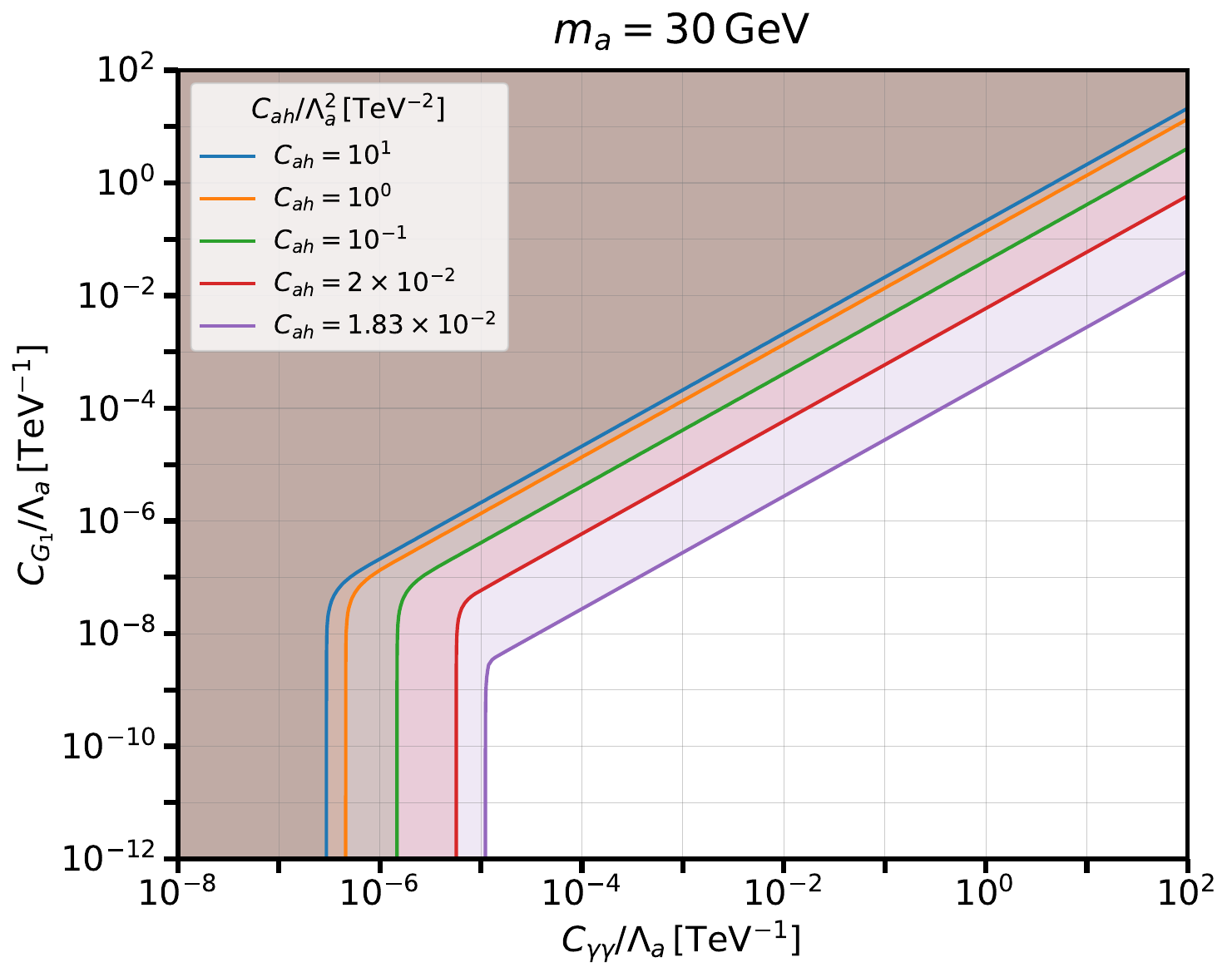}
\caption{
Upper panels: 
two-dimensional constraints in the $(\cgamma,\cgoh)$ plane, derived by recasting the upper bounds from the $pp\to h\to aa\to 4\gamma$ search reported in Ref.~\cite{ATLAS:2023ian}, for fixed values of $\cgo$ and of the ALP mass $m_a=\unit[3]{GeV}$ (left) and $\unit[30]{GeV}$ (right). The solid (dotted) black line marks the upper bound on $\cgoh/\La^2$ obtained by requiring $\Br(h\to aa)\leq 30\%$ (10\%).
Lower panels: two-dimensional constraints in the $(\cgamma,\cgo)$ plane, derived for fixed values of $\cgoh$ and of $m_a$.
}
\label{fig.2Dpanels_resonant}
\end{figure}

\paragraph{Bounds in 2D planes.}
Solving numerically Eq.~\eqref{eq.resonant_recast}, two-dimensional bounds on pairs of Wilson coefficients can be derived for fixed values of the ALP mass and of the third coefficient, as done for the non-resonant analysis.
Fig.~\ref{fig.2Dpanels_resonant} shows 95\%CL-allowed regions in the $(\cgamma,\cgoh)$ plane for fixed values of $\cgo$ (upper plots) and in the  $(\cgamma,\cgo)$ plane for fixed values of $\cgoh$ (lower).

A visible feature in the upper panels of Fig.~\ref{fig.2Dpanels_resonant} is the existence
of a minimum value of $\cgoh$ (horizontal boundary) below which $\cgamma$ is unconstrained for any value of $\cgo$. Such minimum value is different for the two chosen masses, consistently with Fig.~\ref{fig.min_cah} (left). 
The saturation of the Higgs branching ratio at very large values of $\cgoh$ is also visible, and it appears as vertical boundaries. For intermediate values of $\cgoh$, the branching ratio can be approximated as $\Br(h\to aa)\simeq \Gamma_{h\to aa}/\Gamma_h^{\rm SM}$, so the bounds scale approximately with $\cgoh^2$. 
The upper bounds imposed by current Higgs measurements are indicated, for reference, by the solid and dotted black lines, corresponding to $\Br(h\to aa)\leq 30\%$ and 10\%  respectively.
Let us note that, for most values of $m_a$, the two benchmarks $\cgoh/\La^2=0.1$~and~$\unit[1]{TeV^{-2}}$ chosen in Ref.~\cite{ATLAS:2023ian}  lie in the intermediate regime where the bound scales with $\cgoh^2$. 
Within this limited region, the bounds scaling  with $\cgoh$ can be reliably estimated by interpolating between the two benchmarks. However,  away from it, the scaling is dramatically different and it cannot be extrapolated in that way.

The behavior as a function of $\cgo$ can be understood as follows:
for sufficiently large $\cgo$, $f_{aa}\simeq 1$ and the bounds on $\cgamma$ are dominated by $\Br(a\to\g\g)^2$. By moving towards smaller $\cgo$, bounds improve following a linear scaling with $\cgo$, until the bounds become dominated by $f_{aa}$. At this point, as $\cgo$ approaches 0, its contribution to $\Gamma_a^{\rm tot}$ becomes negligible compared to that of $\cgamma$, and the bounds remain unchanged. The value of $\cgo$ at which this transition occurs depends on the ALP mass, as is manifest from the comparison of the left and right plots. These results are qualitatively consistent with the projections provided in Ref.~\cite{Bauer:2017ris} for $m_a\leq \unit[10]{GeV}$.

The patterns discussed so far can be visualized  in the lower panels of Fig.~\ref{fig.2Dpanels_resonant} as well. Here, the diagonal boundaries correspond to the regime in which $\Br(a\to\g\g)^2$ controls the bound, while the vertical boundaries are induced once $f_{aa}$ becomes the dominant effect.
Comparing the lower panels of  Figs.~\ref{fig.2Dpanels_resonant} and~\ref{fig:cg1vscgamma}, we can observe that the allowed regions take very similar shapes and are also numerically very similar. Indeed, they are essentially induced by the same physics: in both cases, the production cross section is fixed to a large value (determined respectively by $\cgoh$ and $\cgt$), and the constraint on $\cgamma$ and $\cgo$ stems from requiring that $N_{\rm signal}$ is depleted enough to fall below the observation threshold, via a long ALP lifetime and/or small branching ratio to photons.
The fact that meaningful constraints in this plane can only be derived for $\cgoh$ within a limited window is reflected in the finite ranges spanned by the curves shown in Fig.~\ref{fig.2Dpanels_resonant}. The right-most curve in both panels is derived for a value of $\cgoh$ very close to the minimum threshold (see Fig.~\ref{fig.min_cah}). This value was tuned to the 4th decimal digit: a downward or upward variation by $10^{-4}$ in the shown value of $\cgoh$ makes the bound disappear or move very close to the other curves, respectively. The left-most curve in both panels assumes $\cgoh/\La^2=\unit[10]{TeV^{-2}}$, which is excluded by Higgs measurements, but is representative of the maximal sensitivity achievable in the $(\cgamma,\cgo)$ plane with current Higgs-resonant searches.

\paragraph{Bounds on $\cgoh$ as a function of $m_a$, for fixed $\cgamma$ and $\cgo$.} We conclude this section by recasting the results of the ATLAS search into constraints on $\cgoh/\La^2$ as a function of $m_a$, for fixed values of $\cgo$ and $\cgamma$. These results are presented in the same spirit as Fig.~\ref{fig:Cg2_vs_ma_log_lin}, \ie\ noting that, in a potential global analysis of the ALP EFT, $\cgamma$ and $\cgo$ are likely to be constrained by a number of processes, while $\cgoh$ would be dominantly bounded by $pp\to h\to aa$ searches, as this is the simplest process to which $\Ocah$ contributes at tree level.

\begin{figure}[t]
\centering
\includegraphics[width=0.48\textwidth]{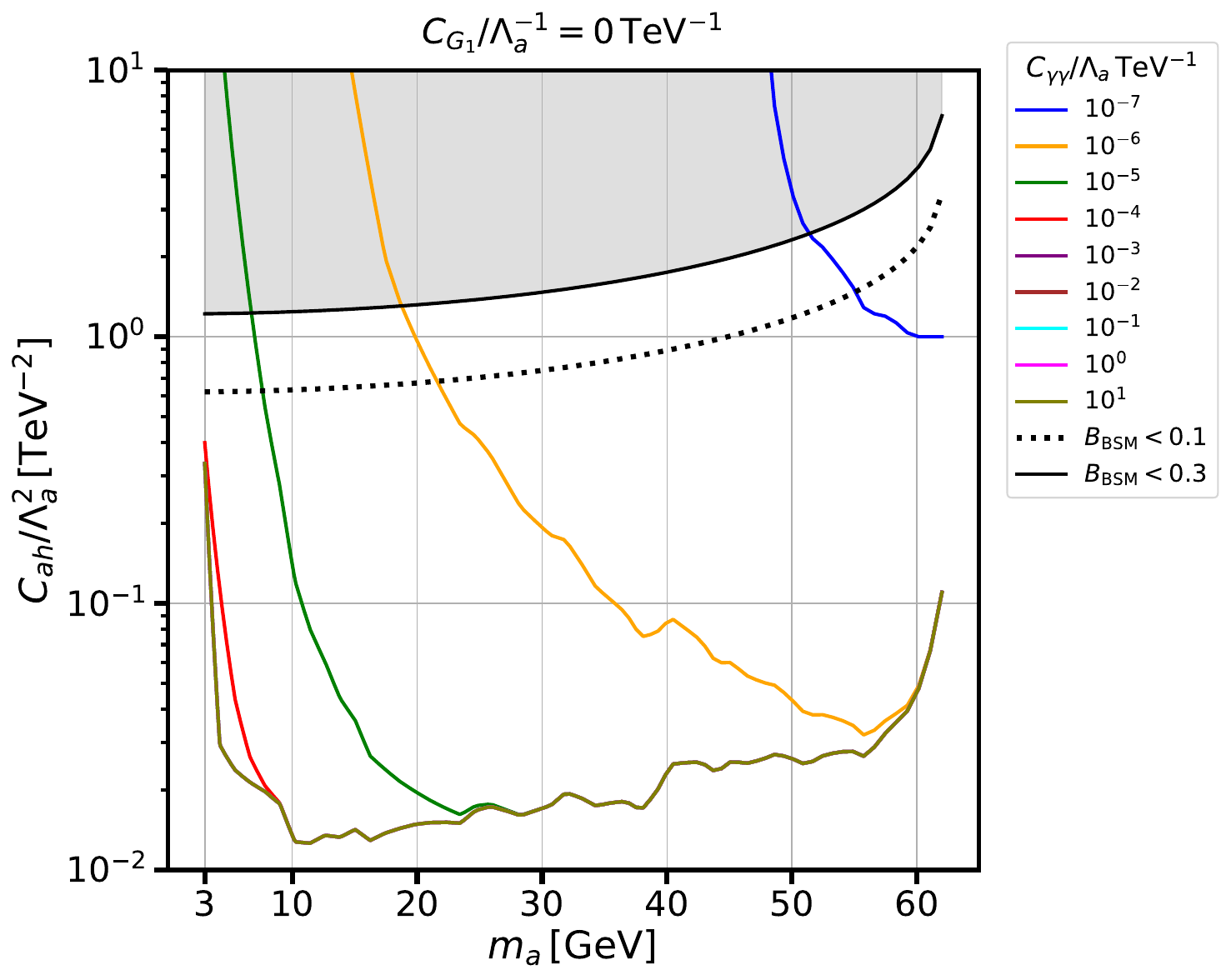}
\hfill
\includegraphics[width=0.48\textwidth]{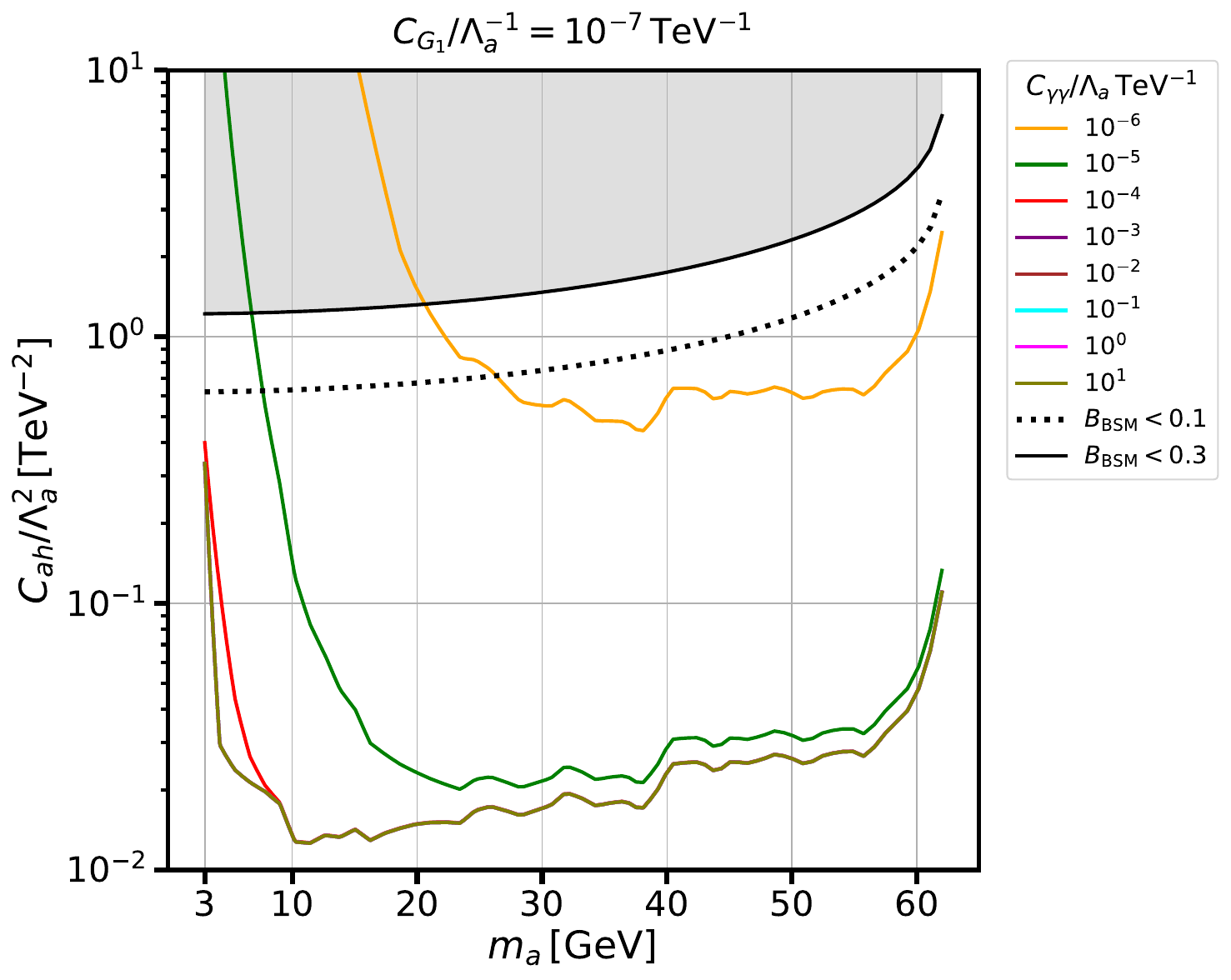}

\vspace{0.5cm}

\includegraphics[width=0.48\textwidth]{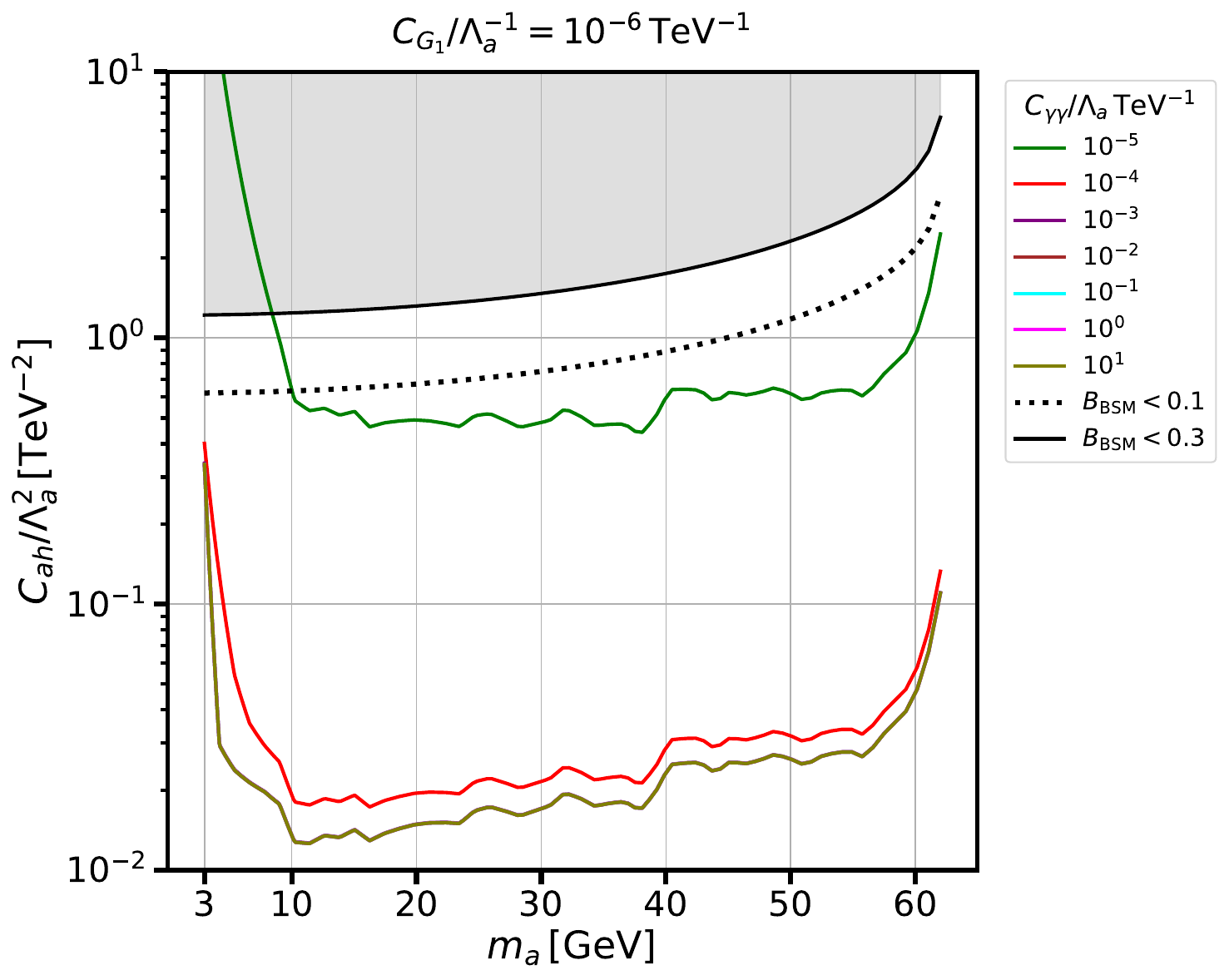}
\hfill
\includegraphics[width=0.48\textwidth]{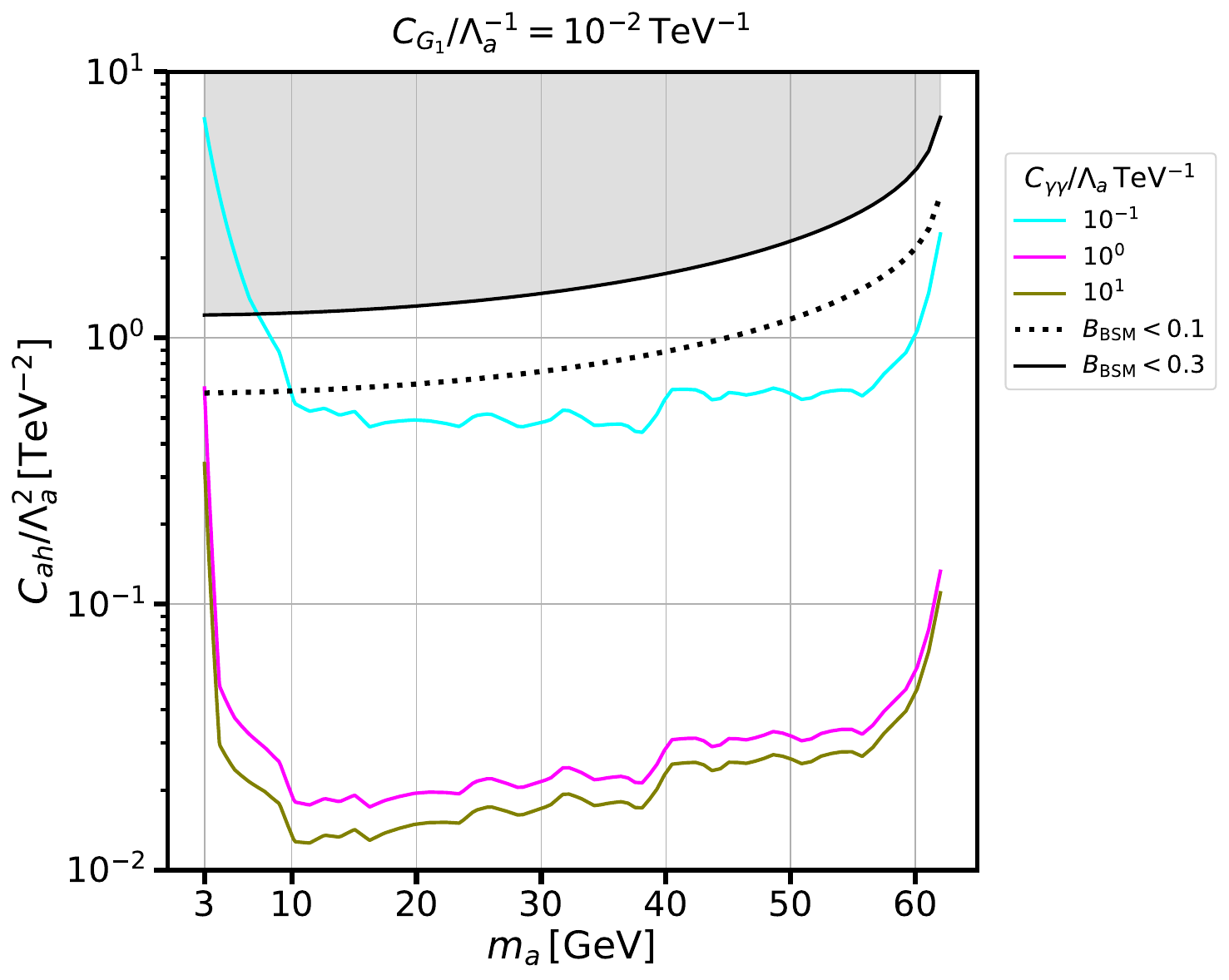}

\caption{95\%CL upper bounds on $\cgoh/\La^2$ obtained recasting the limits from the $pp\to h\to aa\to4\gamma$ search in Ref.~\cite{ATLAS:2023ian}, as a function of $m_a$ and for fixed values of $\cgo/\La$ (panels) and $\cgamma/\La$ (curves).
In the first two plots, the curves for $\cgamma/\La\in [10^{-4} -1]~\unit{TeV^{-1}}$ mostly overlap. The same occurs for  $\cgamma/\La\in [10^{-3} -1]~\unit{TeV^{-1}}$ in the third plot.
}
\label{fig:cah_vs_ma}
\end{figure}

Fig.~\ref{fig:cah_vs_ma} shows the 95\%CL upper limits on $\cgoh/\La^2$ obtained for various $\cgamma$ and $\cgo$ hypotheses. The solid and dotted black lines mark, for comparison, the exclusion limits derived by requiring $\Br(h\to aa)\leq 30\%$ and 10\% respectively. 
The top left panel assumes $\cgo=0$, and therefore $\Br(a\to\g\g)= 1$. In this case, a dependence on $\cgamma$ can only enter through FSDE, which are relevant for $\cgamma/\La\lesssim \unit[10^{-4}]{TeV^{-1}}$. Above this threshold, any value of $\cgamma$ yields the same constraint $\cgoh/\La^2\lesssim \unit[10^{-2}]{TeV^{-2}}$, which also coincides with the minimum value of $\cgoh$ for which a constraint on $\cgamma$ could possibly be obtained, see  Fig.~\ref{fig.min_cah} (left).
When $\cgamma$ is small enough that $f_{aa}<1$  (blue, orange, green curves), the sensitivity to $\cgoh$ drops. As noted above, for small enough $\Gamma_a^{\rm tot}$, we have that  $f_{aa}\sim (\Gamma_a^{\rm tot})^{2}\sim (\cgamma^4m_a^6)$. Therefore this effect is much more pronounced at lower ALP masses. Within this regime, the bounds from $h\to aa\to 4\gamma$ searches are complementary to those from $\Br(h\to aa)$: the former dominate at large $m_a$, while the latter dominate at smaller~$m_a$. 

We have verified that increasing $\cgo/\La$ up to $  \unit[10^{-8}]{TeV^{-1}}$ leaves the constraints on $\cgoh$ approximately unchanged.
Once we reach $\cgo/\La=\unit[10^{-7}]{TeV^{-1}}$ (second panel), we have $f_{aa}\simeq 1$ for $m_a\gtrsim\unit[30]{GeV}$ and any $\cgamma$, \ie\ FSDE disappear from the right side of the plot. At the same time, for $\cgamma\lesssim 10\cgo$ (orange curve),  $\Br(a\to\g\g)$ drops below its maximum. This new suppression weakens the bound, compensating for the lack of FSDE.
The behavior of the upper bounds on $\cgoh$ shown here is very similar to the one observed for the bounds on $\cgt$ in Fig.~\ref{fig:Cg2_vs_ma_log_lin}: the reduction of $\Br(a\to \g\g)$ acts approximately as a rigid upward translation of the constraints, while FSDE change their shape, worsening them more significantly at lower $m_a$.

For $\cgo/\La\gtrsim\unit[10^{-6}]{TeV^{-1}}$, FSDE are irrelevant for all masses shown. Then the dependence on $\cgamma$ only enters through $\Br(a\to\g\g)$. The picture is now identical to the one observed in the $\cgt$ case: for $\cgamma\gtrsim 100\,\cgo$ ($r_\gamma\lesssim 10^{-2})$, the constraint on $\cgoh$ is always the same, and it coincides with the maximum one attainable. For smaller values of $\cgamma$, the bound worsens by a factor given approximately by $\Br(a\to\g\g)^{-1}$, \ie\ by a factor $20$ for $\cgamma = 10 \,\cgo$ ($r_\gamma=0.1$), a factor $2000$ for $\cgamma=\cgo$ ($r_\gamma=1$) {\it etc}. Unlike the $\cgt$ case, here this behavior can in principle be extrapolated up to arbitrarily large $\cgo$ and $\cgamma$, as $N_{\rm signal}$ remains sensitive only to the $r_\gamma$ ratio.

\subsection{Discussion of the results}
\label{sec:discussion}

We conclude this section by discussing the main findings that emerge from the results presented in Sec.~\ref{sec:nonresresults} and~\ref{sec:resresults}.

\paragraph{Features of the multi-dimensional parameter space.}
A general conclusion of our study is that a search for $pp\to aa\to 4\gamma$ alone can never 
fully constrain the ALP parameter space,  \ie\ an upper limit on the cross section of this process always identifies an \emph{unbounded} region in $(\cgo,\cgt,\cgamma,\cgoh)$. 
This statement remains true even if $\Lag_{\rm ALP}$ is truncated to dimension-5, \ie\ in the limit $\cgt=\cgoh=0$,\footnote{In this case the Higgs-resonant process is absent at tree level.}
and it occurs because of a number of reasons:
(a) the allowed region will always contain values of $\cgo,\cgt$ and $\cgoh$ for which the $aa$ production rate is too small to cast limits on $\cgamma$; 
(b) $N_{\rm events}$ becomes independent of $\cgamma$ in the limit ${\cgamma\to\infty}$, due to the branching ratio $\Br(a\to\g\g)$ saturating to its maximum allowed value; (c) in non-resonant production, $N_{\rm events}$ becomes independent of $\cgo$ in the limit $\cgo\to\infty$, due to a cancellation in the dependence on this parameter between the ALP production and decay; (d) again in non-resonant production, a flat direction in the $(\cgo,\cgt)$ plane is present for $\cgt<0$ due to negative interference in the production cross section. 
The only way to extract robust constraints on the ALP Wilson coefficients would then be to incorporate a measurement of $pp\to aa\to 4\gamma$ into a broader global analysis of the ALP EFT which, however, is beyond the scope of the present work. Instead, we provided representative 2D-slices of the multidimensional parameter space by estimating the projected sensitivity of di-ALP searches for fixed values of two of the four relevant parameters,  namely the ALP mass $m_a$ and the three Wilson coefficients $\cgo,\cgamma$ and $\cgt$ (for the non-resonant case) or $\cgoh$ (for the Higgs-resonant one).

Let us comment first on Figs.~\ref{fig:cg1vscgamma}--\ref{fig:cg2vscgamma} and~\ref{fig.2Dpanels_resonant}, which show constraints for fixed $m_a$. These figures indicate that, in general, the upper bounds attainable on a pair of Wilson coefficients depend significantly on the value of the third.  The allowed region can take qualitatively very different shapes depending on the regime considered: for instance, in some cases it extends up to arbitrarily large values of (one of) the Wilson coefficients, as a consequence of the effects listed above. In some others, a Wilson coefficient is constrained within a finite range of non-zero values: this is the case, for instance, of the cusps and band shapes visible in the upper panels of Fig.~\ref{fig:cg1vscgamma}, and in Fig.~\ref{fig:cg2vscgamma}. Both of these features typically occur for $\cgt<0$ and large values of $\cgamma$.

There are also cases in which  it is possible to set very stringent bounds on the ALP couplings: 
the most notable example are the constraints on $\cgamma$ that can be extracted in the parameter space region dominated by FSDE. In the Higgs-resonant case, bounds of order $\unit[10^{-5}-10^{-7}]{TeV^{-1}}$ can be placed on $\cgamma/\La$ for $m_a=\unit[30]{GeV}$, $\cgoh/\La^2\gtrsim \unit[10^{-2}]{TeV^{-2}}$ and $\cgo/\La\lesssim\unit[10^{-7}]{TeV^{-1}}$ (lower right panel of Fig.~\ref{fig.2Dpanels_resonant}). This result is consistent with those found in Refs.~\cite{Bauer:2017ris,ATLAS:2023ian} and it weakens only by a couple of orders of magnitude for $m_a=\unit[3]{GeV}$ (lower left panel in the same Fig.). Even stronger limits can be placed in the non-resonant case, with bounds $\cgamma/\La\lesssim\unit[10^{-9}]{TeV^{-1}}$ being reached for $m_a=\unit[1]{TeV}$, $\cgt/\La^2=\unit[1]{TeV^{-2}}$ and $\cgo/\La\lesssim\unit[10^{-9}]{TeV^{-1}}$ (lower right panel of Fig.~\ref{fig:cg1vscgamma}). As highlighted in the previous sections, these conditions are quite peculiar, as they physically correspond to assuming a sizable $aa$ production cross section driven by a large dimension-6 coupling, and subsequently requiring the ALP lifetime, controlled by dimension-5 interactions, to be long enough to deplete the number of visible signal events below the observation threshold.

\paragraph{Interplay with EFT power counting and other bounds.}
The situation pointed out at the end of the previous paragraph is potentially unnatural from the EFT point of view, because it occurs when the dimension-6 coefficients are orders of magnitude larger than the dimension-5 ones. Indeed, not all regions of the parameter space considered in Secs.~\ref{sec:nonresresults} and \ref{sec:resresults} are expected to be equally plausible in a realistic realization of the ALP EFT. 
At the same time, arguments about which regions of parameter space are better motivated should be formulated with care. In this paragraph we examine some theoretical considerations that can be used for this purpose.
Note that, while these arguments can provide guidance in identifying the most interesting regions, some of them are necessarily subject to assumptions about the ALP EFT or its UV completion, and they can be potentially evaded in specific setups.
For this reason, in the presentation of our results we deliberately refrained from introducing theory priors that would penalize specific regions of parameter space.

\begin{table}[t]\centering
\renewcommand{\arraystretch}{1.8}
\begin{tabular}{l|*4{>{$}c<{$}}}
\hline
{\bf Upper bounds}& \cgo/\La& \cgamma/\La& \cgt/\La^2& \cgoh/\La^2 \\
\hline
Other measurements (Sec.~\ref{sec:currentbounds})&
\unit[0.01-1]{TeV^{-1}}&
\unit[0.1-1]{TeV^{-1}}&
\text{none}&
\unit[1]{TeV^{-2}}
\\\hline
EFT validity \& perturbativity&
\dfrac{1}{g_s^2}\dfrac{4\pi}{E}& 
\dfrac{1}{e^2}\dfrac{4\pi}{E}& 
\dfrac{1}{g_s^2}\dfrac{(4\pi)^2}{E^2}&
\dfrac{(4\pi)^2}{E^2}\\
\hfill Higgs-resonant ($E=m_h$)& 
\unit[68]{TeV^{-1}}&
\unit[1000]{TeV^{-1}}&
-&
\unit[10^4]{TeV^{-2}}
\\[-2mm]
\hfill Non-resonant ($E=m_{aa}$)& 
\unit[2]{TeV^{-1}}&
\unit[26]{TeV^{-1}}&
\unit[6]{TeV^{-2}}&
-
\\\hline
Unitarity~\cite{Brivio:2021fog,Bresciani:2025ojh}&
0.31\dfrac{1}{g_s^2}\dfrac{1}{E}&
1.8\dfrac{1}{e^2}\dfrac{1}{E}&
\text{none}&
\dfrac{8\pi}{E^2}
\\
\hfill Higgs-resonant ($E=m_h$)& 
\unit[2]{TeV^{-1}}&
\unit[146]{TeV^{-1}}&
-&
\unit[1.6 \times 10^3]{TeV^{-2}}\\[-2mm]
\hfill Non-resonant ($E=m_{aa}$)& 
\unit[0.04]{TeV^{-1}}&
\unit[4]{TeV^{-1}}&
\text{none}&
-
\\\hline
\end{tabular}
\caption{Representative upper limits on the absolute size of the Wilson coefficients, imposed by current measurements and by the theoretical arguments illustrated in the main text. The unitarity bounds adopted are the most stringent among those reported in Refs.~\cite{Brivio:2021fog,Bresciani:2025ojh}. The numerical values for the non-resonant case are computed taking $m_{aa}\simeq\unit[5]{TeV}$ as the maximum $m_{aa}$ reached by signal events and taking a fixed $\alpha_s=0.118$. With these conditions, EFT validity additionally requires $\La\geq 10$ and $\unit[400]{GeV}$ respectively for the two processes. No experimental or unitarity bounds on $\cgt$ are currently present in the literature.
}\label{tab.other_bounds}
\end{table}

We begin by examining the conditions required for the validity and perturbativity of the ALP EFT, which read respectively
\begin{align}
\left(\frac{E}{\La}\right)&\leq 4\pi\,,
&&\text{and}&
(e^2 \cgamma),\, (g_s^2 \cgo),\, (g_s^2 \cgt) ,\,\cgoh &\leq 1\,,
\label{eq.EFTvalidity}
\end{align}
where $E$ stands for the typical energy flowing through the EFT vertex, which is a process-dependent quantity.
Combining these two inequalities, one finds upper bounds on the combinations $(\cgamma/\La),(\cgo/\La),(\cgt/\La^2),(\cgoh/\La^2)$, which are reported in the second block of Table~\ref{tab.other_bounds}. For instance:
\begin{align}
  \frac{\cgo}{\La} = \cgo\frac{E}{\La}\frac{1}{E}\;\leq\; 
  \frac{1}{g_s^2}\times (4\pi)\times  \frac{1}{E}\,.
\end{align} 
The inequalities in Eq.~\eqref{eq.EFTvalidity} can be easily derived using Naive Dimensional Analysis (NDA)~\cite{Manohar:1983md} (see also~\cite{Brivio:2025yrr} for a recent review):
$\La$, the characteristic scale of ALP interactions, has the dimensions of a mass $\times \sqrt{\hbar}$, the same as a scalar field. A well-known NDA result is that $\La$ must satisfy the condition $4\pi\La\geq M$, where $M$ is the mass scale such that the EFT expansion parameter is the ratio $(E/M)\leq 1$. The first condition in Eq.~\eqref{eq.EFTvalidity} follows trivially from this argument,  namely $(E/\La) \leq 4\pi (E/M)\leq 4\pi$. 
NDA provides a very convenient way to derive perturbativity conditions as well, which are simply obtained by requiring that the pre-factor the effective operators written in NDA normalization be less than 1. For the operators defined in Sec.~\ref{sec:theo}, those coincide with the products $ (e^2\cgamma),\,(g_s^2\cgo),\, (g_s^2\cgt),\, \cgoh$ respectively, hence the second set of conditions in Eq.~\eqref{eq.EFTvalidity}.

The upper bounds on the ($C_i/\La^{(2)}$) combinations listed in Tab.~\ref{tab.other_bounds} can be interpreted numerically  for fixed processes.
Tab.~\ref{tab.other_bounds} reports the numerical values obtained fixing $E=m_h=\unit[125]{GeV}$ for the Higgs-resonant production process, and $E=m_{aa}$ for the non-resonant one. In the latter case,  $m_{aa}=\unit[5]{TeV}$ was taken as representative of the maximum invariant mass of the ALP pair reached by signal events. Note that, with the same values, the first inequality in Eq.~\eqref{eq.EFTvalidity} imposes $\La\geq\unit[10]{GeV}$ for the Higgs-resonant channel and $\La\geq\unit[400]{GeV}$ for the non-resonant one. As one would expect, the EFT-validity bounds from the non-resonant channel are significantly more stringent than those from Higgs-resonant production, as the former tests the EFT at much higher center-of-mass energies. Nevertheless, none of them are particularly strong: in fact, they are satisfied in nearly all the parameter space shown in Secs.~$\ref{sec:nonresresults}$ and $\ref{sec:resresults}$.
Indeed, these constraints should be understood as minimal, conservative requirements: their violation would indicate a severe breakdown of the ALP EFT formalism and, in practice, one would expect the Wilson coefficients to take values well within these boundaries. 

Similar considerations apply to unitarity bounds that are inferred by requiring all the possible $2\to2$ scattering amplitudes produced in the ALP EFT to fulfill partial-wave unitarity up to a certain center-of-mass energy $E = \sqrt{s}$. Unitarity bounds on ALP operators were computed in Refs.~\cite{Brivio:2021fog,Bresciani:2025ojh}, which adopted different techniques for the calculation of the relevant matrix elements (see also~\cite{Bresciani:2025toe}). Tab.~\ref{tab.other_bounds} reports the most stringent among the bounds reported in the two references,\footnote{The unitarity bound on $\cgamma/\La$ was derived by composing the bounds reported on $C_{W}, C_{B}$. } indicating both the analytic expressions and the numerical values that apply to Higgs-resonant and non-resonant ALP pair production, computed in same way described for EFT validity bounds.  To our knowledge, unitarity bounds on the dimension-6 operator $\Ocgt$ have not been derived in the literature. By naive dimensional arguments, we would expect it to scale with $(8\pi/g_s^2 E^2)$ times some numerical factor. If this guess is correct, requiring unitarity to be preserved up to $E=\unit[5]{TeV}$ would translate into a unitarity bound of the order of $\cgt/\La^2\lesssim \unit[0.7]{TeV^{-2}}$. This estimate is of course far from being rigorous, but it could be taken as a reference in the absence of a proper calculation.
Similar to the constraints from EFT validity, 
unitarity bounds should not be interpreted as sharp bounds, but rather as an indication of the order of magnitude at which $C_i/\La^{(2)}$ can in principle induce unitarity violation in scattering processes.  

These constraints can be directly compared to current experimental bounds, for which we only report the order of magnitude in the first block of  Tab.~\ref{tab.other_bounds}. As discussed in detail in Sec.~\ref{sec:currentbounds}, experimental constraints are currently available only on $\cgamma,\cgo,\cgoh$ and their interpretation in terms of individual Wilson coefficients is typically very sensitive to the modeling assumptions. This is particularly true for the constraints on the dimension-5 parameters. The absolute value of the constraints additionally depends significantly on the ALP mass. The ranges indicated should then be understood as indicative of the current sensitivity.

All the constraints discussed so far define box-like\footnote{Unitarity bounds are derived by varying multiple Wilson coefficients simultaneously so they technically can have non-trivial shapes in the parameter space. However, bosonic operators do not show significant correlations among them, so the bounds can be approximately considered as  on one operator at a time.} boundaries for the space spanned by the Wilson coefficients, but do not provide information about their relative size. 
To make considerations about the latter, one can for instance resort to assumptions about the UV completion: in scenarios in which it is weakly coupled, the UV-EFT matching is perturbative and we can note that the operator $\Ocah$ can be potentially generated at tree-level, while all the remaining operators considered here can only arise at 1-loop. This implies that the Wilson coefficient $\cgoh$ could naturally carry a loop factor enhancement relative to $\cgo,\cgamma$ and $\cgt$. Interestingly, a tree-level $\Ocah$ would be produced by integrating out a heavy scalar, while the remaining operators can only be produced via loops of particles with spin. 
In the simplest class of UV completions, often referred to as KSVZ-like constructions, they are heavy vector-like fermions~\cite{Kim:1979if,Shifman:1979if,DiLuzio:2020wdo,Palavric:2026vej}. This distinction suggests that the mechanism generating a non-zero $\cgoh$ could be potentially different from the one generating $\Ocgamma,\,\Ocgo,\,\Ocgt$. 
In ``pure'' KSVZ-like models, it is also reasonable to expect that $\cgamma/\cgo\sim $ an $\mathcal{O}(1)$ factor related to the ratio of electromagnetic and color anomalies in the UV model. Something similar occurs in models in which only fermionic operators are produced at LO in the UV-matching, which then generate bosonic ALP interactions via loops.  For instance Ref.~\cite{Blasi:2023hvb} examined a top-philic scenario, in which the ALP couples only to top quarks at LO. In this case, one has $\cgamma/\cgo\sim \mathcal{O}(1)$. 
Ref.~\cite{Bauer:2017ris} considered a similar scenario, assuming tree-level couplings only to up-type SM quarks via a diagonal and universal $C_u/f_a =\unit[0.1-10]{TeV^{-1}}$. They examined in particular the 1-loop contributions to $\cgamma$ and $\cgoh$, finding a fixed ratio $\cgamma/\sqrt{\cgoh}\simeq 10^{-2}$.

All in all, in the absence of ad-hoc suppressions, one would expect $(\cgo/\La),(\cgamma/\La)$ to be of comparable size, perhaps within a couple of orders of magnitude of each other. A comparison between dimension-5 and dimension-6 parameters is more delicate: the matching-based arguments above concern only the Wilson coefficients $C_i$, so applying them to the parameter space in Figs.~\ref{fig:cg1vscgamma}-\ref{fig:cg2vscgamma},~\ref{fig.2Dpanels_resonant} requires an additional specification of $\La$.  With this caveat, the discussion above suggest that a quite large $\cgoh$, (giving $\cgoh/(\cgo,\cgamma)\gtrsim 10^2$) is natural in a number of constructions. The size of $\cgt$, on the other hand, strongly depends on how the ALP's shift symmetry is realized: in the presence of large violations of this symmetry (\ie\ in the limit in which the ALP is a generic pseudoscalar), $\cgt$ is unsuppressed, and it can be expected to be of a similar size as $\cgoh$, or a loop factor smaller. Since the structure of the $\Ocgt$ operator is very similar to that of $\Ocgo$, it is also reasonable to expect the corresponding Wilson coefficients to take values that are not too far apart. If the shift symmetry is preserved to a good approximation, though, then $\cgt$ would carry an additional suppression and it could be naturally much smaller than the remaining coefficients. 

Parameter space regions that lie very far from these conditions could be regarded as disfavored \emph{a priori}. This includes for instance the purple and black benchmarks defined in Tab.~\ref{tab.stars}, which are characterized by $(\cgt/\La^2)/(\cgo/\La)\gtrsim\unit[10^5]{TeV^{-1}}$: even for the lowest possible $\La\simeq\unit[400]{GeV}$, realizing this condition requires $\cgt/\cgo\gtrsim 10^5$.

\paragraph{Sensitivity to dimension-6 operators.}
Coming back to discussing our results, one of the main goals of this study was to assess the sensitivity to dimension-6 ALP interactions. The projected upper bounds on $\cgt,\cgoh$ were displayed in Figs.~\ref{fig:Cg2_vs_ma_log_lin} and~\ref{fig:cah_vs_ma} respectively, as a function of the ALP mass and for different assumptions on $\cgamma,\cgo$.

Let us discuss first the case of $\cgt$: Fig.~\ref{fig:Cg2_vs_ma_log_lin} shows that the maximal sensitivity to this operator is reached in the limit of small $\cgo/\La\lesssim\unit[10^{-3}]{TeV^{-1}}$ and for $\cgamma\gtrsim 100 \cgo$, such that $\Br(a\to\g\g)$ is maximized. In this scenario, which is compatible with current bounds and with the considerations at the previous paragraph, the upper limit is $\cgt/\La^2\lesssim\unit[10^{-3}]{TeV^{-2}}$ for $m_a\lesssim\unit[300]{GeV}$ and independently of the exact values of the dimension-5 coefficients. Physically, the constraint is driven by the upper bound on the $gg\to aa$ production cross section, which is dominated by $\cgt$. Going towards higher masses weakens the constraint only slightly: we find $\cgt/\La^2\lesssim\unit[10^{-2}]{TeV^{-2}}$ for $m_a=\unit[1]{TeV}$.
Moving away from this regime, we observe that the upper bound on $\cgt$ is most sensitive to variations in $\Br(a\to\g\g)$: moving towards larger $\cgo/\La\simeq\unit[1]{TeV^{-1}}$ while keeping a maximized $\Br(a\to\g\g)$ weakens the constraints only up $\cgt/\La\lesssim\unit[1]{TeV^{-2}}$. On the other hand, reducing the branching ratio can weaken it by several orders of magnitude: as discussed in Sec.~\ref{sec:nonresresults}, the bound on $\cgt$ would scale with $r_\gamma^2$. Sensitivity to the sign of $\cgt$ can only be achieved if $\cgo$ is large enough for interference effects to be relevant. In this case, the interference causes the bounds on $\cgt$ to take alternative shapes, which can identify finite preferred ranges (rather than upper limits) for $\cgt<0$.

In the case of $\cgoh$, Fig.~\ref{fig.2Dpanels_resonant} shows that the strongest constraint achievable on this parameter through a reinterpretation of the ATLAS search in Ref.~\cite{ATLAS:2023ian} is $\cgoh/\La^2\lesssim \unit[10^{-2}-10^{-1}]{TeV^{-2}}$, with the smaller values being reached for $m_a\simeq\unit[10]{GeV}$, and the larger ones for $m_a\simeq m_h/2$ and $m_a\simeq\unit[3]{GeV}$. The sensitivity is maximized for $\Br(a\to\g\g)=1$ regardless of the exact values of the dimension-5 Wilson coefficients, and drops proportionally to $\Br(a\to\g\g)^{-1}$ away from this limit. 

Overall, we find that the sensitivity of $aa$ searches to dimension-6 coefficients is very good, both in the Higgs-resonant and non-resonant channels, already with integrated luminosities of the order of $\unit[100]{fb^{-1}}$.

\paragraph{Truncating at dimension-5.}
One could also ask whether searches for ALP-pair production could be an interesting probe for the ALP Lagrangian truncated at dimension-5, \ie\ when $\cgt=\cgoh=0$. In this limit, Higgs-resonant production is absent, but bounds on $\cgamma$, $\cgo$ can in principle be set from searches for the non-resonant process. 

Numerically, the results would essentially coincide with those obtained for  $\cgt/\La^2=\unit[10^{-7}]{TeV^{-2}}$, shown in red in the upper-left panel of Fig.~\ref{fig:cg1vscgamma}:  the search could in principle exclude the region where $\cgamma/\La\gtrsim \unit[1]{TeV^{-1}}$ and $\cgo/\La\gtrsim\unit[10^{-2}]{TeV^{-1}}$. Although a di-ALP measurement alone would not be able to disentangle the two operators, nor to set individual bounds on each of them, it is interesting to note that this sensitivity is quite close to current experimental bounds. This suggests that di-ALP searches at the LHC could play a complementary role to those for single-ALP production, even in a dimension-5-only analysis.

\paragraph{Comparison of Higgs-resonant and non-resonant channels.}
Finally, it can be interesting to ask how non-resonant di-ALP production compares to Higgs-resonant production that had been previously studied in the literature. 
Besides the obvious fact that the two processes probe different dimension-6 operators, the main differences between them are the following: (i) the non-resonant channel is a $2\to2$ scattering process, while the Higgs-resonant one is a 2-body decay. This implies that the non-resonant channel probes the ALP EFT up to higher energies (in fact, over a spectrum of energies, although this feature was not exploited in our analysis) with the production cross section growing with $s^2$, and also that it is in principle sensitive to arbitrarily large values of $\cgt/\La^2$. By contrast, $\Br(h\to aa)$ saturates to 1 in the limit of large $\cgoh/\La^2$, making the measurement insensitive to variations of the coupling within this regime. 
(ii) The production cross section of non-resonant ALP pair production depends simultaneously on a dimension-5 ($\cgo$) and a dimension-6 ($\cgt$) operator, while $\Br(h\to aa)$ is proportional solely to $\cgoh$. This can generate a non-trivial interplay between $\cgo$ and $\cgt$ in the non-resonant channel, which can provide sensitivity to the sign of $\cgt$.
(iii) The non-resonant $gg\to aa$ channel is a QCD process, while $h\to aa$ is a purely EW decay. This implies that the non-resonant cross section could receive sizable higher-order corrections, and that this process could also be searched for in alternate final states with photons plus jets. Although an exploration of $gg\to aa$ at NLO is beyond the scope of this work, this feature makes non-resonant ALP pair production a potentially richer playground to test ALP interactions.

In terms of numerical sensitivity to dimension-6 operators, we find that a search for non-resonant ALP pair production would yield constraints that are certainly competitive and potentially stronger than those from current measurements. In particular, the sensitivity to dimension-6 interactions is approximately an order of magnitude stronger in the non-resonant case.

\section{Conclusions}\label{sec:conclusion}
We have studied ALP-pair production at the LHC in final states with four isolated photons, exploring its sensitivity to dimension-5 and dimension-6 ALP interactions. In particular, we considered for the first time the dimension-6 operator $\Ocgt$, which induces a $ggaa$ contact interaction, and performed the first phenomenological study of the non-resonant process $gg \to aa$. We also recast an ATLAS search for the Higgs-resonant process $pp \to h \to aa$~\cite{ATLAS:2023ian}, projecting its exclusion limits onto the three-dimensional parameter space spanned by $(\cgamma,\cgo,\cgoh)$.

Our results, discussed in detail in Sec.~\ref{sec:results}, demonstrate that ALP-pair production is a very promising probe of the ALP EFT. We find that, already with an integrated luminosity of $\unit[300]{fb^{-1}}$, a search for non-resonant ALP pairs could be sensitive to values of $\cgt/\La^2$ as small as $\unit[10^{-3}]{TeV^{-2}}$. For comparison, current searches for Higgs-resonant ALP pairs, based on $\unit[140]{fb^{-1}}$ of data, probe values of $\cgoh/\La^2$ down to about $\unit[10^{-2}]{TeV^{-2}}$. Depending on the values of the dimension-6 Wilson coefficients, both searches could also constrain dimension-5 ALP interactions competitively with existing bounds. Altogether, these results indicate that a meaningful exploration of the ALP EFT beyond dimension-5 is already possible at the LHC.

A particular emphasis of this work has been the multidimensional nature of the ALP parameter space, accounting as much as possible for the simultaneous dependence of the signal on multiple ALP interactions. A general conclusion is that searches for ALP-pair production alone cannot fully constrain the relevant parameter space, which remains unbounded along certain directions. This feature, together with several properties of the allowed regions discussed in Secs.~\ref{sec:nonresresults} and \ref{sec:resresults}, originates from the interplay between ALP production and decay, as well as from the dependence of the signal on the ALP lifetime. As such, it is not specific to the processes or final states considered here, but is expected to arise more generally in collider searches for ALPs. Indeed, as reviewed in Sec.~\ref{sec:currentbounds}, similar degeneracies appear in nearly all collider probes of ALPs. A global analysis of the ALP EFT would therefore be essential to properly account for these effects and extract robust information on ALP interactions. We hope that the methodology developed in this work can provide a useful blueprint for broader studies of ALP production and decay processes. 

Finally, the analysis presented here could be improved in several respects. Our statistical analysis was kept simple, assuming negligible backgrounds and relying exclusively on the total expected number of signal events. Although these assumptions are well motivated, the analysis could be refined through dedicated studies of the relevant backgrounds, more accurate estimates of the detection and reconstruction efficiencies, and a kinematic characterization of the signal. On the theory side, the predictions could be improved by performing the simulation at NLO in QCD, possibly matched to parton-shower effects, which would allow an extension of the study to final states with additional jets. The theoretical framework could also be generalized to include the dependence on $C_W$ and on fermionic ALP couplings, which were fixed in this work to maximize $\Br(a\to\g\g)$. These operators primarily affect ALP decays, modifying both $\Br(a\to\g\g)$ and $\Gamma_a^{\rm tot}$. ALP couplings to quarks --particularly the top quark-- can also contribute to the production cross section, depending on the perturbative order considered. CP-odd operators (particularly the trilinear ALP self-coupling) and dimension-6 fermionic interactions giving $q \bar q aa$ contact vertices could also be retained in future studies, extending the scope of di-ALP searches. Finally, the search strategy could be extended to lower ALP masses by considering final states with collimated photons originating from the decay of highly boosted ALPs, and incorporating a proper modeling of hadronic decays. We leave the exploration of these directions to future work.

\acknowledgments{
We thank K.~Schmieden and M.~Schott for clarifications on the ATLAS analysis in Ref.~\cite{ATLAS:2023ian} and K. Mimasu for helpful discussions.
S.M.~thanks the University of Southampton for hosting him during the period of development of this project. 
D.P.~acknowledges financial support by the MUR through the PRIN2022 Grant 2022EZ3S3F, funded by the European Union – NextGenerationEU. 
The authors acknowledge support from the COMETA COST Action CA22130.

}

\appendix

\section{Analytic minimization of ALP decays into EW bosons}
\label{sec:decayanalytic}

In this Appendix we report analytic formulas for the value $\bar C_{WW}$ of $C_{WW}$ that
minimizes $\Gamma_{a\to {\rm EW}}$, as explained in Sec.~\ref{sec:decay}, and for the function $F(m_a)$ defined in Eq.~\eqref{eq:pheno_minimum_ew_decay_width}.
\\
Depending on the ALP mass, we have
\begin{equation}\label{eq:CW_min}
\bar C_{WW} = \cgamma \times  
\begin{cases} 
s_w^2 
&m_Z \leq m_a < 2m_W 
\\[8pt]
\displaystyle 
s_w^4 \, 
\frac{\lambda^{3/2}_{Z \g}}
{ c_w^2 \,  \lambda^{3/2}_{WW} 
+ s_w^2 \,  \lambda^{3/2}_{Z \g} }
&2m_W \leq m_a < 2m_Z 
\\[1cm]
s_w^4\, 
\dfrac{ 2c_w^2 \lambda^{3/2}_{Z \g} 
- c_{2w} \, \lambda^{3/2}_{Z Z} }
{ 2 c^4_w\, \lambda_{WW}^{3/2}
+ c_{2w}^2\, \lambda_{ZZ}^{3/2}
+ 2 c^2_w s^2_w\, \lambda_{Z \g}^{3/2} }
&m_a \geq 2m_Z\,,
\end{cases}
\end{equation}
where $\lambda_{ij} = (1-x_i-x_j)^2-4x_ix_j$ with $x_i = m_{V_i}^2/m_a^2$.
Correspondingly:
\begin{equation}\label{eq:F_ma}
F(m_a)=\begin{cases}
0
& m_a<2m_W 
\\[2mm]
2s^2_w\, 
\dfrac{
\lambda_{WW}^{3/2}\, \lambda_{Z \g}^{3/2}
}{  
s^2_w\, \lambda_{Z \g}^{3/2} 
+ c^2_w\, \lambda_{WW}^{3/2} 
} 
& 2m_W\leq m_a < 2m_Z 
\\[1cm]
2 s_w^2\,
\dfrac{
s_w^2\, \lambda_{WW}^{3/2} \lambda_{ZZ}^{3/2}
+ c_w^2\,\lambda_{Z \g}^{3/2}\,
(2 \lambda_{WW}^{3/2}
+ \lambda_{ZZ}^{3/2} )
}{
2\, c_w^4\, \lambda_{WW}^{3/2}
+ c_{2w}^2\, \lambda_{ZZ}^{3/2}
+ 2\, c_w^2\, s_w^2\, \lambda_{Z \g}^{3/2}
} 
& m_a\geq 2 m_Z  \,.
\end{cases}
\end{equation}

\section{ALP transverse momentum distributions}\label{app.pT_dist}

\begin{figure}[t]
    \centering
    \setlength{\abovecaptionskip}{2pt}
    \setlength{\belowcaptionskip}{-5pt}
    \begin{subfigure}{0.33\textwidth}
        \centering
        \includegraphics[width=\linewidth]{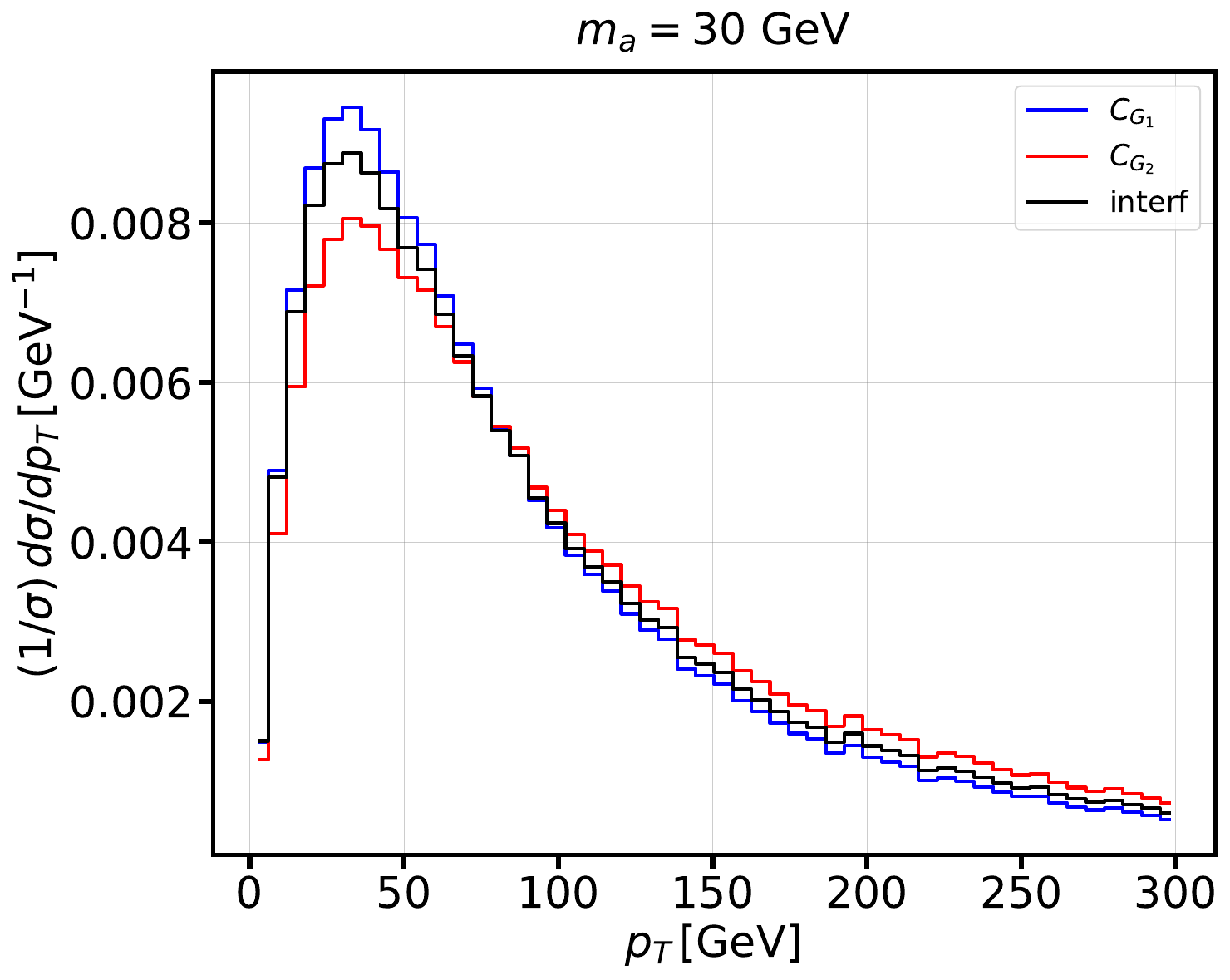}
        \caption{}
    \end{subfigure}
    \begin{subfigure}{0.33\textwidth}
        \centering
        \includegraphics[width=\linewidth]{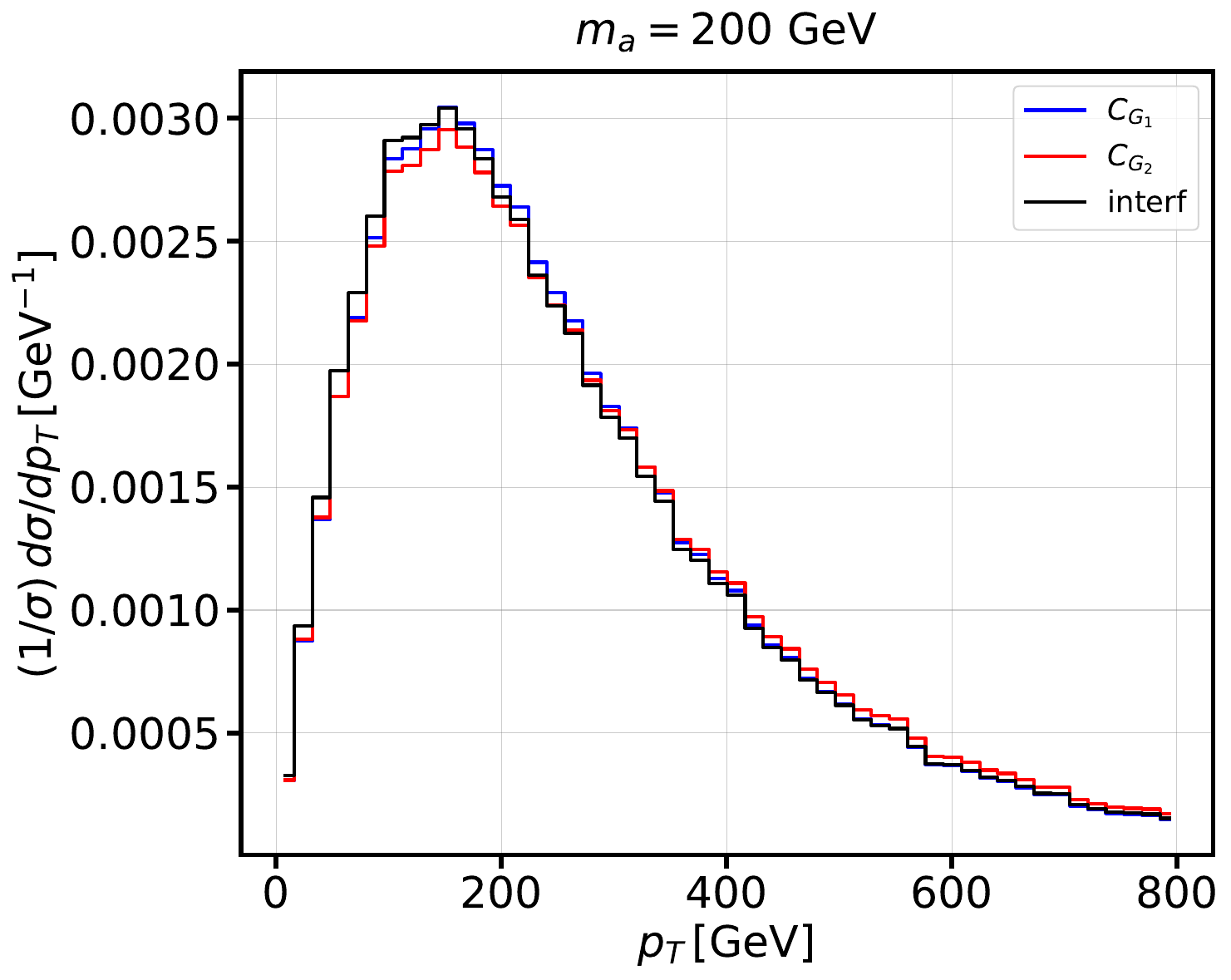}
        \caption{}
    \end{subfigure}\hfill
    \begin{subfigure}{0.33\textwidth}
        \centering
        \includegraphics[width=\linewidth]{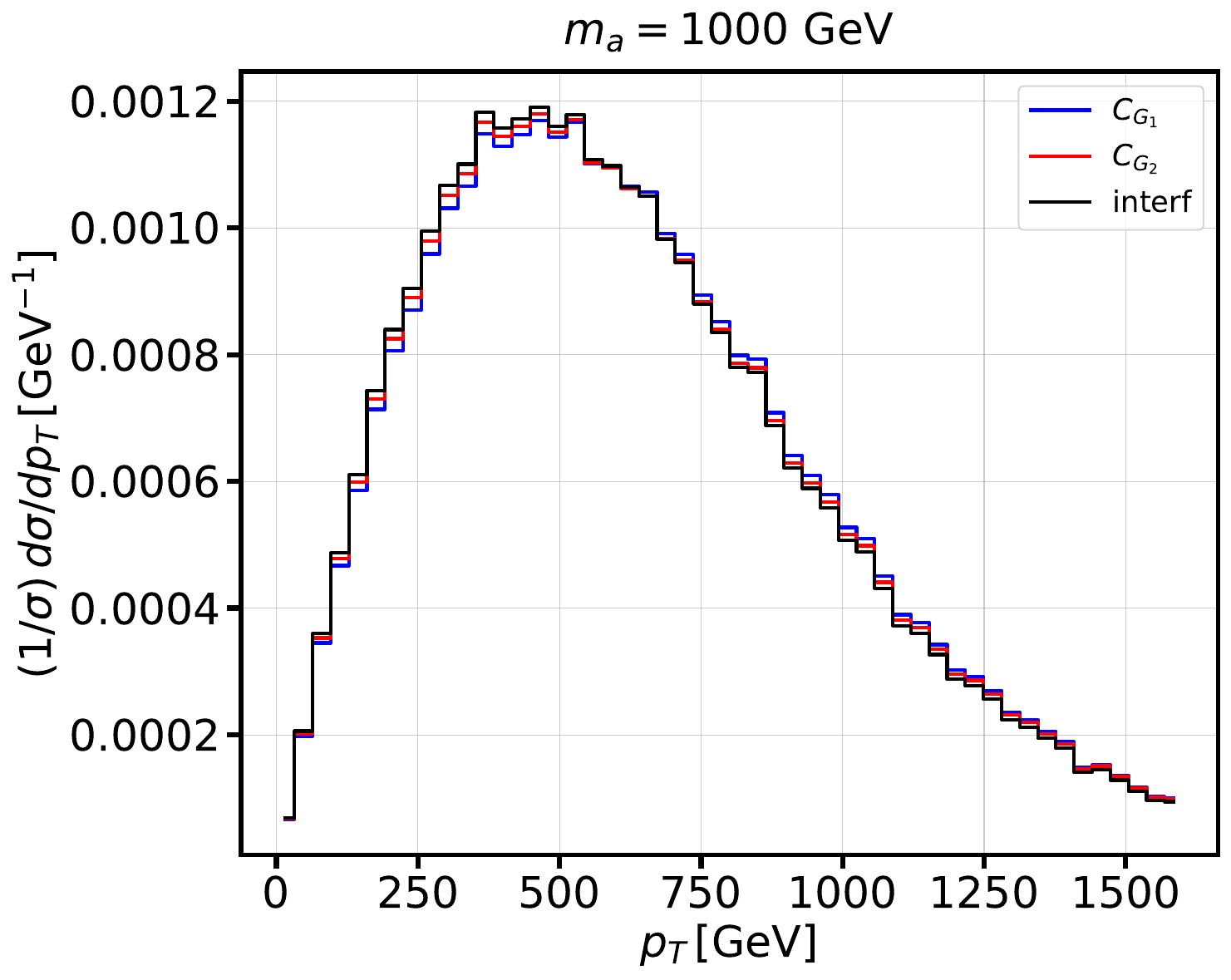}
        \caption{}
    \end{subfigure}

    \vspace{1em}
    \caption{
        Normalized distribution of the ALP transverse momentum in non-resonant $pp\to aa$ production, for various ALP masses and for the individual components $\sA (\cgo),\,\sB (\cgt)$ and $\sC$ (interf).
    }\label{fig:alp_pt_dists}
\end{figure}

Fig.~\ref{fig:alp_pt_dists} shows the normalized distributions $(d\sigma_{ij}/dp_{T,a})/\sigma_{ij}$ for representative values of the ALP mass, 
obtained with a simulation of non-resonant $pp\to aa$ production at $\sqrt{s}=\unit[13]{TeV}$ performed as described in Sec.~\ref{sec:nonres}. No selection cuts or FSDE are implemented at this stage, as the decay of the ALPs to photons is not included in the modeling.

Although $|\mathcal{M}_1|^2$ explicitly depends on $t$ and $u$ at the partonic level, while $|\mathcal{M}_2|^2$ and the $\mathcal{M}_1\mathcal{M}_2^\dag$ interference only depend on $s$ (see Eqs.~\eqref{eq:A_Feynman}--\eqref{eq:C_Feynman}), we observe only very mild differences in the shape of the corresponding distributions: $d\sA/dp_{T,a}$, \emph{vs.} $d\sB/dp_{T,a}$ and $d\sC/dp_{T,a}$.  
The discrepancies only amount to a few percent for $m_a=\unit[30]{GeV}$ and essentially vanish at higher ALP masses.
We have verified that these differences are actually due to the dynamical scale dependence of $\alpha_s$, which enters with different powers in $\sA,\sB$ and $\sC$. If the renormalization scale is kept fixed, only negligible differences remain, even at $m_a=\unit[30]{GeV}$.

\bibliographystyle{JHEP.bst}
\bibliography{bibliography}
\end{document}